\documentclass[11pt]{article}

\usepackage[skip=14pt plus2pt, indent=0pt]{parskip}
\usepackage{verbatim}
\usepackage{epsfig,graphics}
\usepackage {graphicx}
\usepackage {epsfig}
\usepackage {subfigure}
\usepackage {tabularx} 
\usepackage{rotate}	
\usepackage{slashed}
\usepackage{enumerate}
\usepackage{amsmath}

\usepackage{amsfonts}
\usepackage{amssymb}
\usepackage{graphicx}
\usepackage{pifont}
\usepackage{xcolor}
\usepackage{cite}
\usepackage{mathrsfs}

\usepackage{fancyhdr}
\usepackage[hidelinks,linktocpage=true]{hyperref}

\usepackage{braket}
\usepackage{cancel}

\hypersetup{colorlinks=true,linkcolor=nicecolor,citecolor=nicecolor,urlcolor=nicecolor}
\definecolor{nicecolor}{rgb}{0.30392  0.35294  0.57647}

\def\beq{\begin{equation}}
\def\eeq{\end{equation}}
\def\beqa{\begin{eqnarray}}
\def\eeqa{\end{eqnarray}}
\newcommand{\cmark}{\textcolor{green!60!black}{\ding{51}}}
\newcommand{\xmark}{\textcolor{red}{\ding{55}}}

\def\Mpd{M_{\text{Pl;}\, d}}
\def\Mpf{M_{\text{Pl;}\, 4}}
\def\Mpfive{M_{\text{Pl;}\, 5}}
\def\Mps{M_{\text{Pl;}\, 6}}

\def\Mpe{M_{\text{Pl;}\,  11}}

\def\LQG{\Lambda_{\text{QG}}}

\def\a{\alpha}
\def\b{\beta}

\newcommand{\bZ}{\mathbb{Z}}
\newcommand{\bC}{\mathbb{C}}

\newcommand{\bP}{\mathbb{P}}
\newcommand{\bR}{\mathbb{R}}

\newcommand{\cI}{\mathcal{I}}
\newcommand{\cN}{\mathcal{N}}
\newcommand{\cM}{\mathcal{M}}
\newcommand{\cF}{\mathscr{F}}
\newcommand{\cG}{\mathcal{G}}
\newcommand{\cZ}{\mathcal{Z}}
\newcommand{\cD}{\mathcal{D}}
\newcommand{\cO}{\mathcal{O}}

\newcommand{\cC}{\mathcal{C}}
\newcommand{\cK}{\mathcal{K}}
\newcommand{\cS}{\mathcal{S}}

\newcommand{\cT}{\mathcal{T}}
\newcommand{\cV}{\mathcal{V}}
\newcommand{\cR}{\mathcal{R}}
\newcommand{\fF}{\mathfrak{F}}

\newcommand{\td}{\text{d}}

\def\tr{\mbox{Tr}}

\catcode`\@=11   
\newdimen\@rotdimen
\newbox\@rotbox  

\def\@vspec#1{\special{ps:#1}}
\def\@rotstart#1{\@vspec{gsave currentpoint currentpoint translate
		#1 neg exch neg exch translate}}
\def\@rotfinish{\@vspec{currentpoint grestore moveto}}
\def\@rotr#1{\@rotdimen=\ht#1\advance\@rotdimen by\dp#1%
	\hbox to\@rotdimen{\hskip\ht#1\vbox to\wd#1{\@rotstart{90 rotate}%
			\box#1\vss}\hss}\@rotfinish}
\def\@rotl#1{\@rotdimen=\ht#1\advance\@rotdimen by\dp#1%
	\hbox to\@rotdimen{\vbox to\wd#1{\vskip\wd#1\@rotstart{270 rotate}%
			\box#1\vss}\hss}\@rotfinish}%
\def\@rotu#1{\@rotdimen=\ht#1\advance\@rotdimen by\dp#1%
	\hbox to\wd#1{\hskip\wd#1\vbox to\@rotdimen{\vskip\@rotdimen
			\@rotstart{-1 dup scale}\box#1\vss}\hss}\@rotfinish}%
\def\@rotf#1{\hbox to\wd#1{\hskip\wd#1\@rotstart{-1 1 scale}%
		\box#1\hss}\@rotfinish}%
\def\rotate{\@ifnextchar[{\@rotate}{\@rotate[l]}}
\def\@rotate[#1]#2{\setbox\@rotbox=\hbox{#2}\@nameuse{@rot#1}\@rotbox}

\catcode`\@=12

\topmargin
-1.5cm
\textwidth
15.5cm
\textheight
23.5cm
\oddsidemargin
0.7cm
\evensidemargin
0.7cm

\begin{document}
\makeatletter
\@addtoreset{equation}{section}
\makeatother
\renewcommand{\theequation}{\thesection.\arabic{equation}}
\pagestyle{empty}
\vspace{-0.2cm}
\rightline{IFT-UAM/CSIC-25-61}
\rightline{EFI-25-8}
\vspace{1.2cm}
\begin{center}
		
\LARGE{ \bf Laplacians in Various Dimensions \\
        and the Swampland }
\vspace{1.2cm}

\large{ C. Aoufia$^\heartsuit$, A. Castellano$^{\diamondsuit,\,\clubsuit}$ and L.E. Ib\'a\~nez$^\heartsuit$
			\\[12mm]}
\small{
$^\heartsuit$ {Departamento de F\'{\i}sica Te\'orica
and Instituto de F\'{\i}sica Te\'orica UAM/CSIC,\\
Universidad Aut\'onoma de Madrid,
Cantoblanco, 28049 Madrid, Spain}  \\[5pt]
$^\diamondsuit$  {Enrico Fermi Institute and Kadanoff Center for Theoretical Physics, \\ University of Chicago, Chicago, IL 60637, USA} \\[5pt]
$^\clubsuit$ {Kavli Institute for Cosmological Physics, \\ University of Chicago, Chicago, IL 60637, USA}
\\[20mm]
		}
\small{\bf Abstract} \\[6mm]
\end{center}
\begin{center}
\begin{minipage}[h]{15.60cm}

The species cutoff is a moduli-dependent quantity signaling the onset of quantum gravitational phenomena, whose form can be oftentimes determined from higher-derivative and higher-curvature corrections within low-energy gravitational EFTs. In this work, we point out that these Wilson coefficients are eigenfunctions of an appropriate second-order elliptic operator defined over moduli space in theories with more than four supercharges. This was already known to be the case for the leading ${\cal R}^4$-correction to the two-derivative (bosonic) action of maximal supergravity in $d\leq 10$. Here, we reconsider this fact from the Swampland point of view and show how, in $d=10,9,8$, solving a Laplace equation imposes non-trivial restrictions on the species hull vectors. We further argue that this property is also satisfied in settings with less supersymmetry. In particular, we focus on the $\cR^4$-operator in minimal supergravity theories in $d=10,9$, and on the leading $\cR^2$-term in setups with 8 supercharges in $d=6,5,4$. Finally, we provide a symmetry-based criterion for determining when the relevant elliptic operator should be the Laplacian. A bottom-up rationale for this constraint remains to be fully understood, and we conclude by outlining some compelling possibilities.

\end{minipage}
\end{center}
\newpage
\pagestyle{empty}
\renewcommand{\thefootnote}{\arabic{footnote}}
\setcounter{footnote}{0}
	
\tableofcontents
\pagestyle{empty}
\newpage
\setcounter{page}{1}
\pagestyle{plain}

\section{Introduction and Summary of Results}
\label{s:intro}

The past decade has seen significant progress in identifying universal properties that any theory of quantum gravity should possess. Remarkably, from a low-energy perspective, requiring a consistent UV-completion imposes non-trivial constraints on the structure of allowed gravitational Effective Field Theories (EFTs). The charting and exploration of these criteria is one of the main goals of the Swampland program \cite{Vafa:2005ui} (see \cite{Brennan:2017rbf,Palti:2019pca,vanBeest:2021lhn,Grana:2021zvf,Harlow:2022ich,Agmon:2022thq,VanRiet:2023pnx} for detailed reviews). 

In supersymmetric theories, one often encounters a number of exactly massless fields, commonly referred to as moduli. In the context of the Swampland program, the topological space $\cM$ they span has been extensively studied within string-theoretic constructions and has proven to be a valuable setting for gaining insights about the nature of the UV-complete framework, owning to its rich geometrical structure. Of distinguished importance are asymptotic boundaries of infinite distance, which are typically associated to weak coupling regimes in the theory. The Distance Conjecture \cite{Ooguri:2006in} posits that, when venturing towards such regions, towers of particle states must become light exponentially fast with the traversed geodesic distance. In particular, their masses become parametrically lower than the (naive) cutoff scale of the EFT, leading ultimately to its breakdown. Furthermore, in those cases where $\cM$ is multi-dimensional, different asymptotic limits typically yield distinct towers. Whenever this happens, it becomes useful to introduce \textit{convex hull} diagrams spanned by certain vectorial quantities \cite{Calderon-Infante:2020dhm,Etheredge:2022opl,Calderon-Infante:2023ler}, which allow one to encode in a simple yet powerful way the spectrum of leading towers as we move towards each of the infinite distance boundaries of moduli space.

As already mentioned, the appearance of light towers of states at these asymptotic loci prematurely invalidates the use of the original effective theory. Indeed, any such gravitational EFT is provided with a number of physically relevant scales. A prominent example would be the one associated to the onset of genuine quantum gravity phenomena, which we denote by $\Lambda_{\rm QG}$.\footnote{For an in-depth discussion about this and other relevant gravitational scales, see \cite{Bedroya:2024uva,Caron-Huot:2024lbf,Castellano:2024bna, Calderon-Infante:2025ldq}.} At generic points in the interior of the moduli space, this scale coincides roughly with the $d$-dimensional Planck scale $\Mpd$. However, in the presence of many light species, a parametric separation can arise between these two. The original one-loop graviton propagator estimate,\footnote{See also \cite{ValeixoBento:2025iqu} for a recent thorough discussion of this point.} dubbed species scale, gives an upper bound to $\Lambda_{\rm QG}$ as follows \cite{Dvali:2007hz,Dvali:2007wp, Dvali:2009ks,Dvali:2010vm, Castellano:2022bvr}
\begin{equation}
    \Lambda_{\rm sp}(\phi^i) = \dfrac{\Mpd}{N(\phi^i)^{\frac{1}{d-2}}}\, ,
\end{equation}
where $N(\phi)$ is the moduli-dependent number of light, weakly coupled species below $\Lambda_{\rm sp}$. Notice that, precisely when a tower of states becomes light in Planck units, this relation implies that $\Lambda_{\rm sp} \ll \Mpd$. The above scale must be actually regarded as a function depending on the vacuum expectation values (vevs) of the scalar fields, given that the number of species is itself moduli-dependent. This suggests that it should be possible to extend its definition to the interior of moduli space, even when a suitable account of $N$ in terms of weakly-coupled fields is not strictly speaking available anymore \cite{Long:2021jlv,vandeHeisteeg:2022btw}. For instance, one can bound $N$ from below by the number of BPS species $N_{\rm BPS}$ \cite{Long:2021jlv} in the theory, or either via black hole microstate counting through their classical and quantum-corrected entropies \cite{Dvali:2007hz,Dvali:2012uq, Cribiori:2023ffn, Basile:2024dqq}. 

Alternatively, one may estimate the species scale by the UV-cutoff, $\Lambda_{\rm UV}$, of the effective theory suppressing higher-derivative corrections to the classical two-derivative action in the gravitational sector \cite{vandeHeisteeg:2022btw, vandeHeisteeg:2023dlw, Castellano:2023aum, Calderon-Infante:2025ldq}. In the present paper, we shall adopt this latter approach,\footnote{The various perspectives are all linked in an intricate manner. For instance, some higher-derivative corrections are index-like in nature\cite{vandeHeisteeg:2022btw} and count BPS states. Relatedly, some of these coefficients determine the entropy of certain (minimal) black holes and are therefore related to the species scale, as in \cite{Cribiori:2022nke, Cribiori:2023ffn, Calderon-Infante:2023uhz,Basile:2023blg, Basile:2024dqq, Bedroya:2024ubj,Aoufia:2024awo,Herraez:2024kux,Castellano:2025ljk,Calderon-Infante:2025pls}.} and even though all these scales are different a priori, we still refer to the quantity extracted from Wilson coefficients involving higher powers of the Riemann tensor as the species scale. We moreover denote it by $\LQG$ to emphasize its relation to quantum gravitational effects. Its behavior in different string vacua and their infinite distance limits has been the subject of many recent investigations, see e.g.,\cite{Grimm:2018ohb,  Corvilain:2018lgw, Castellano:2021mmx, Castellano:2022bvr, Castellano:2023qhp, Calderon-Infante:2023ler, vandeHeisteeg:2022btw, Cribiori:2023ffn, Cribiori:2023sch, Blumenhagen:2023yws, Blumenhagen:2023tev, Blumenhagen:2023xmk, vandeHeisteeg:2023ubh, Andriot:2023isc, Cota:2022yjw, Cota:2022maf, Calderon-Infante:2023uhz,Bedroya:2025ris}. In particular, let us mention that analogously to the definition of the convex hull generated by tower scalar charge-to-mass vectors, one can also introduce a similar notion of \textit{species hull} as originally done in \cite{Calderon-Infante:2023ler}, which becomes especially useful in multi-moduli setups. In fact, these two kinds of diagrams are not unrelated to each other, and indeed a very precise link between them both was formulated in the purportedly universal pattern observed in \cite{Castellano:2023stg, Castellano:2023jjt}. The latter implies, in turn, that the two polytopes---when available---must be dual to each other with respect to a ball of radius $1/\sqrt{d-2}$ in $d$ spacetime dimensions.\footnote{For further works exploring the underlying structure of these hulls as well as their physical consequences for string vacua, see \cite{Etheredge:2023odp,Etheredge:2024tok,Grieco:2025bjy,Etheredge:2025ahf}.}

The full knowledge of the moduli-dependence of these Wilson coefficients, particularly so in the bulk of scalar field space, is in general quite challenging due to the presence of a plethora of quantum corrections. However, it was pointed out first in \cite{vandeHeisteeg:2022btw} and subsequently studied in \cite{vandeHeisteeg:2023ubh, vandeHeisteeg:2023dlw, Castellano:2023aum,Aoufia:2024awo, Castellano:2024bna, Calderon-Infante:2025ldq}, that in the case of string vacua with 32, 16 and 8 supercharges one may sometimes be able to extract the \emph{asymptotic} functional form of the species scale in a reliable manner.\footnote{Importantly, these coefficients may also encode even parametrically lower field-theoretic scales---such as the Kaluza-Klein mass---via massive threshold effects, as recently emphasized in \cite{Calderon-Infante:2025ldq}.} Indeed, we can consider BPS-protected operators of the qualitative form 
 \begin{equation}
    S_{{\rm BPS}} = \frac{1}{2 \kappa^2_d}\int \td^dx \sqrt{-g}\ {\cF }^{(d)}_n (\phi^i)\, \frac{  {\cal R}^{2n}}{ \Mpd^{4n-2}}\, ,
\end{equation}
with $n=2$ in the case of 32 and 16 supercharges, and $n=1$ for the case of 8 supercharges. These protected terms appearing in the quantum effective action may be computed exactly, i.e., also including (non-)perturbative corrections \cite{Green:1997tv}, by different methods, and they can contribute non-trivially to scattering amplitudes involving (at least) $2n$ external
gravitons. The associated Wilson coefficients ${\cF}^{(d)}_n$ have thus been determined for maximal supergravity in diverse dimensions, ranging from $d=3, \dots ,10$ \cite{Green:2010kv,Green:2010wi}, as well as in theories preserving 8 supercharges in $d=4,5,6$, see e.g., \cite{vandeHeisteeg:2022btw, vandeHeisteeg:2023ubh, vandeHeisteeg:2023dlw, Castellano:2023aum} and references therein. The species scale can then be retrieved asymptotically as the one suppressing these corrections via the relation
 \begin{equation}\label{eq:spsc}
    \Lambda_{\rm QG}\, \sim\, \Mpd\, {\cF}_n^{-\frac{1}{4n-2}}\, ,
\end{equation}
where in our conventions $\kappa^2_d=\Mpd^{2-d}\,$. As a consistency check, this definition has been shown to yield---for the cases mentioned before---the correct dependence expected to arise from decompactification and weakly coupled string limits in many top-down examples, see \cite{vandeHeisteeg:2023ubh,Castellano:2023aum}. 
  
Therefore, given all the explicit examples currently available of global species scale functions, it is interesting to investigate the general properties that the corresponding Wilson coefficients ${\cF}^{(d)}_n$ exhibit. One of the main purposes of the present work is to argue that these functions satisfy a `Laplace-like' differential equation\footnote{The role of Laplacians and their interplay with towers of states has been also discussed recently in \cite{Etheredge:2023zjk}.} of the schematic form
\beq\label{eq:eigenvalueeqintro}
\cD^2_{\bf {\cal M}} {\cF}^{(d)}_n   = \eta_d\,  {\cF}^{(d)}_n\, .
\eeq
Here, $\eta_{d}$ are constant eigenvalues and  $\cD^2_{\bf {\cal M}}$ is a second-order, elliptic differential operator defined on the moduli space of the theory. As we will show explicitly, in many examples the latter will turn out being the Laplace-Beltrami operator, namely $\cD^2_{\cM} = \Delta_\cM$. This is true for both the $n=2$ and $n=1$ higher-curvature terms described above.\footnote{In the marginal cases, i.e., $d=8$  for ${\cal R}^4$ and $d=4$ for ${\cal R}^2$, there is an additional constant piece on the right-hand side due to regularized logarithmic divergences. As will be discussed later, one can ignore these extra contributions for our purposes herein, since they correspond to non-analytic threshold effects due to the massless modes within the theory.} However, in those cases where these two necessarily differ, string dualities link $\cD_\cM^2$ to Laplacians across various dimensions in a way that is very reminiscent of the relation that exists between the corresponding $\Delta_D$ and $\Delta_{D+1}$ operators in toroidal compactifications of maximal supergravity \cite{Green:2010wi}. Table \ref{tab:summary} summarizes the various operators $\cD^2_{\bf {\cal M}}$ and associated eigenvalues $\eta_d$ discussed at length throughout the paper.

\begin{table}[]\renewcommand{\arraystretch}{1.4}
    \centering
    \begin{tabular}{|c|c|c|c|} \hline
        \textbf{Theory} & \textbf{Coefficient} & \textbf{Operator} & \textbf{Eigenvalue}  \\ \hline
         Maximal supergravity* (10d, 9d, 8d) & $\cR^{4}$ & $\Delta_\cM$ & $ 3\, \frac {(11-d)(d-8)}{d-2}$ \\ \hline
         Type IIA & $\cR^{4}$ & $\partial_\phi^2 + \partial_\phi$ & $ \frac34$  \\ \hline
         Half-maximal supergravity (10d, 9d)  &$\cR^{4}$ & \eqref{eq:16superchargesop1}, \eqref{eq:16superchargesop2}& $ 3\, \frac {(11-d)(d-8)}{d-2}$ \\ \hline
         6d $\cN = (1,0)$ ($T$ tensor multiplets) & $\cR^{2}$ & $\Delta_{\cM}$ & $T$ \\ \hline
         5d $\cN = 2$ ($n_V$ vector multiplets) & $\cR^{2}$ & \eqref{eq:5doperator} & $\frac16 n_V$ \\ \hline
         4d $\cN =2$ &  $\cR^{2}$ & $\Delta_{\cM}$ & 0 \\ \hline
         
    \end{tabular}
    \caption{\small Summary of the various operators, Wilson coefficients and associated eigenvalues discussed throughout the paper. The * refers to the fact that Type IIA is dealt with separately: the eigenvalue is the same as Type IIB, but the operator is not the Laplacian $\Delta_\cM$ over moduli space. For the half-maximal Heterotic/Type I theories, the elliptic operator is the same as the Laplacian for the maximal theories.}
    \label{tab:summary}
\end{table}

Given this state of affairs, we argue that the Laplace condition here proposed constrains the asymptotic rates $\lambda_{\rm sp}$ of the species scale, and is closely related to its exponential damping, as expected from the Distance Conjecture. Indeed, in maximally supersymmetric theories, the exponential behavior found is related to the mathematical fact that all the relevant eigenfunctions are automorphic forms which admit some Fourier decomposition whose zero modes (`constant terms') depend only on the non-compact real moduli. These moreover correspond to the perturbative contributions to ${\cF}_n^{(d)}$ which dominate asymptotically. We take advantage of this property to extract non-trivial information about the infinite distance limits by solving equation \eqref{eq:eigenvalueeqintro} in terms of power-like ansatze for their moduli-dependence.

The fact that the protected Wilson coefficients ${\cF}_n^{(d)}$ are eigenfunction of a Laplace-Beltrami operator was in fact known to be the case in maximal supergravity for the ${\cal R}^4$-operator 
\cite{Green:2010kv, Green:2010wi, Green:1999pv, Pioline:1998mn, Green:1999pu, Green:2005ba}. After reviewing the relevant material (cf. Section \ref{s:speciesreview}), in Section \ref{s:maxsugra} we reconsider this fact from the point of view of the Swampland and show how, in $d=8,9,10$, solving the proposed Laplace equations implies certain consistency conditions on the species convex hull. A first constraint, due to the elliptic nature of the operator and the non-negative definiteness of the eigenvalues (for high enough spacetime dimensions), concerns the length of all possible species vectors in the theory, which is bounded from above in a manner that is compatible with---yet independent of---the bound put forward in \cite{vandeHeisteeg:2023ubh}. Furthermore, one can set up an inductive procedure, where upon solving the Laplace equation in each dimension it is possible to reconstruct the species hull completely. Type IIA supergravity in $d=10$ is the first instance where the relevant differential operator is not given by the Laplacian, but needs to be modified via the addition of a linear derivative term in the dilaton. This operator can be motivated by the duality with Type IIB string theory compactified on a circle, as explained in Section \ref{sss:typeIIAop}.

We further argue that the eigenvalue condition is also present in settings with less amount of supersymmetry. We first consider in Section \ref{ss:16sugra} the case with 16 supersymmetry generators corresponding to both $d=10$ heterotic strings, which have the dilaton as the only non-compact modulus. For the $SO(32)$ case, the form of the $\cR^4$-coefficient is, not surprisingly, quite similar to the Type IIB case---with the $C_0$ axion fixed to zero, as computed in \cite{Green:2016tfs}. Thus, the associated ${\cF}_{SO(32)}^{(10)}$ function is a solution of an equation of the type \eqref{eq:eigenvalueeqintro} with the same eigenvalue as in Type IIB. Relatedly, the $E_8\times E_8$ coefficient exhibits the same functional dependence as the Type IIA $d=10$ one. We repeat the computation for the 9d $SO(16)\times SO(16)$ theories, finding eventually identical results. In all of the above cases, the relevant differential operator is not the full Laplacian defined on the underlying moduli space.

Subsequently, in Section \ref{s:8supercharges} we proceed by analyzing the case of 8 supercharges in $d=4,5,6,$ spacetime dimensions, where the focus is now placed on the Wilson coefficient of the $\cR^2$-operator. The latter is also $\frac12$-BPS and only depends on the corresponding vector or tensor multiplet moduli spaces. This includes 4d ${\cal N}=2$ theories obtained from Type II Calabi--Yau compactifications. There, the appropriate Wilson coefficient ${\cF}_1^{(4)}$ solves a Laplace equation with zero eigenvalue, and we show that this condition, together with requiring the function to be periodic---which is also motivated by our earlier observations in maximally supersymmetric setups---is enough to determine the general form of the species scale. Moreover, at large moduli, the discrete periodicity enhances to a continuous one, such that asking for the latter is akin to extracting the zero mode of the Fourier decomposition with respect to the axionic fields, revealing certain similarities with the constant terms within maximal supergravity. In particular, condition \eqref{eq:eigenvalueeqintro} forces the function $ {\cF}_1^{(4)}$ to be linear in the saxions belonging to the vector multiplet sector. We also find a relation between the above constraint and certain versions of the Scalar Weak Gravity \cite{Palti:2017elp} conjectures put forward in
\cite{Lee:2018spm,Gonzalo:2019gjp,Gonzalo:2020kke,DallAgata:2020ino,Benakli:2020pkm,Andriot:2020lea,Etheredge:2022opl,Benakli:2022shq, Dudas:2023mmr, Etheredge:2023usk}. On the other hand, in six dimensions the relevant moduli space is instead parametrized by the tensor multiplet scalars. We find, once again, that the Laplace condition is satisfied from the top-down, and that is moreover compatible with $ {\cF}_1^{(6)}$ being linear in the (constrained) tensor moduli, as computed in \cite{Bonetti:2011mw,vandeHeisteeg:2023dlw}. In this case, the eigenvalue is given just by the total number of tensor multiplets $T$. Finally, in $d=5$ we find that the appropriate differential operator is different from the Laplacian, while the eigenvalue is still controlled by the number of vector multiplets, similarly to the 6d case. Here, the resulting elliptic operator can be motivated by uplifting the 4d scalar Laplacian via circle (de)compactification.

Given the plethora of casuistics that is found, it is natural to wonder whether there could be some underlying principle that would allow us to deduce when the relevant differential operator should coincide with the Laplacian. In Section \ref{s:Laplace&Symm}, we argue using standard group-theoretic techniques applied to differential operators defined on symmetric (coset) spaces, that this is the case if the continuous symmetries exhibited by $\cM$ are not broken by other interactions in the two-derivative theory. Interestingly, all the instances where this does not happen are (more directly) related to M-theory. We leave our conclusions and further discussion of these results for Section \ref{s:conclusions}.

\section{Review: The Species Scale and the Swampland Program}\label{s:speciesreview}

The aim of this section is to review the concept of the quantum gravity (or species) cutoff, which is commnonly understood as the energy scale controlling the maximum regime of validity of \emph{any} effective field theory weakly coupled to Einstein gravity. As mentioned in the introduction, its original definition involves the number of light, weakly coupled species, $N$, belonging to the effective theory \cite{Dvali:2007hz,Dvali:2007wp, Dvali:2009ks,Dvali:2010vm, Castellano:2022bvr}
\begin{equation}\label{eq:Lambdaspdef}
    \Lambda_{\rm sp} = \dfrac{\Mpd}{N^{\frac{1}{d-2}}}\, ,
\end{equation}
which makes its connection with the Swampland readily apparent. Indeed, many of the these criteria---such as the Weak Gravity \cite{Arkani-Hamed:2006emk,Heidenreich:2015nta,Heidenreich:2016aqi,Montero:2016tif,Andriolo:2018lvp} or the Distance \cite{Ooguri:2006in, Lee:2019wij} conjectures---deal with extreme regimes in the parameter space of the low energy description, where some gauge couplings are taken to zero or rather we allow for infinitely large (with respect to a certain measure) vevs in the scalar fields. Furthermore, the way in which the EFTs admitting a UV-completion to quantum gravity censor this kind of limits is via the appearance of an infinite number of light states, thereby implying that the cutoff scale \eqref{eq:Lambdaspdef} must decrease accordingly. This explains, in essence, why this concept plays a major role in understanding (some of) the infrared constraints imposed by quantum gravity, and why significant effort has recently been devoted to characterizing its most basic properties (see e.g., \cite{Castellano:2024bna} and references therein).

In this chapter, we will highlight several universal features that this energy scale is expected to exhibit, most of which have been thoroughly tested in top-down string theory constructions in various dimensions and with different amounts of supersymmetry preserved. More precisely, in Section \ref{ss:infinitedecayspecies} we review the exponential decay that arises when venturing towards infinite distance limits in the moduli space, as a consequence of the Distance conjecture together with the asymptotic definition \eqref{eq:Lambdaspdef}. We will also discuss the rich geometric structure---inherited from the duality group of the theory---that emerges in those setups where the number of moduli is greater than one, which require from the introduction of certain vectorial quantities dubbed species vectors, as well as their convex hull diagrams. Finally, let us mention that the alternative method proposed to extract the quantum gravity cutoff from the known moduli-dependence of certain BPS-protected higher-derivative corrections to the two-derivative gravitational action \cite{vandeHeisteeg:2022btw, vandeHeisteeg:2023ubh}, has been shown to reproduce all these properties asymptotically \cite{Castellano:2023aum, vandeHeisteeg:2023dlw}, providing strong evidence for this picture. Accordingly, one of the goals of the present paper is to investigate which of these properties---and to what extent---can be derived from a simple condition, namely the (generalized) Laplace equation applied to such higher-derivative terms.

\subsection{Infinite distance limits and exponential decay}\label{ss:infinitedecayspecies}

To keep things general, we consider in what follows a $d$-dimensional EFT containing a set of massless (moduli) scalar fields, weakly coupled to Einstein gravity, as follows
\begin{equation}\label{eq:genericaction}
	S_{\text{EFT}} \supset \dfrac{1}{2\kappa_d^2}\int R\star 1 + g_{i j} (\phi)\, d \phi^i \wedge \star d \phi^j\, ,
\end{equation}
where $g_{ij}(\phi)$ is a 2-rank symmetric tensor over the moduli space, $\mathcal{M}_{\phi}$, spanned by the vevs of the massless scalars and which defines, in turn, a natural notion of distance over the latter. In fact, one of the most well-established Swampland principles is the Distance Conjecture \cite{Ooguri:2006in}, according to which we should have an infinite set of states becoming exponentially light at every infinite distance boundary within the moduli space. Thus, in terms of the traversed \emph{geodesic} distance, which is computed as
\begin{equation}\label{eq:modspacedist}
	\Delta_{\phi} = \int_{\gamma} \td\sigma \sqrt{g_{i j} (\phi) \frac{d \phi^i}{d \sigma} \frac{d \phi^j}{d \sigma}}\, ,
\end{equation}
for any geodesic path $\gamma$ parametrized by $\sigma \in \mathbb{R}$, there should exist a tower of particles whose mass scale decreases in Planck units as $m\sim e^{-\lambda \Delta_{\phi}}$ for $\Delta_{\phi} \gg 1$. Here, $\lambda$ denotes the asymptotic \emph{decay rate}, which is given by some $\mathcal{O}(1)$ number.\footnote{\label{fnote:sharpened}The observed values for the decay rate of the lightest tower are such that there seems to exist a lower bound for the latter, given by $\lambda_{\rm t} \geq \frac{1}{\sqrt{d-2}}$\cite{Etheredge:2022opl}.}

However, in the presence of several moduli within the theory, it is useful to define some vector-like quantity for every tower becoming light asymptotically, with components
\begin{equation}\label{eq:chargetomass}
	\zeta^i := - g^{i j} \frac{\partial}{\partial \phi^j} \log m\, .
\end{equation}
These are usually referred to as \emph{scalar charge-to-mass vectors} \cite{Calderon-Infante:2020dhm,Etheredge:2022opl,Etheredge:2023odp},\footnote{The name stems from the Scalar Weak Gravity Conjecture \cite{Palti:2017elp}, given that these vectors measure the relative strength of the scalar force induced by the moduli between two such identical particles, and their gravitational attraction.} and they encode the information about how fast the mass of the associated tower decays for every possible infinite distance direction. In particular, given an asymptotically geodesic trajectory characterized by some normalized tangent vector $\hat{n}$, the decay rate of the tower can be determined as
\begin{equation}\label{eq:decayrate}
  \lambda= g_{ij}\zeta^i \hat{n}^j\, .
\end{equation}
Furthermore, experimentally one observes that the norms of these vectors, computed with respect to the metric $g_{ij}$, are also strongly constrained and depend essentially on the microscopic nature of the tower, namely wether it corresponds to a Kaluza-Klein or a fundamental, critical string \cite{Lee:2019wij}. The precise dependence found in string theory examples is as follows
\beq\label{eq:zetavecnorm}
	|\zeta_{{\rm KK},\, p}| = \sqrt{\frac{d+p-2}{p (d-2)}}\, , \qquad |\zeta_{\rm osc}|= \frac{1}{\sqrt{d-2}}\, ,
\eeq
where $p$ counts the number of decompactifying dimensions. In practice, however, the scalar metrics $g_{ij}(\phi)$ appearing in \eqref{eq:genericaction} tend to be rather cumbersome, owing to the complicated structure of moduli space. To overstep this difficulty, one oftentimes introduces a local orthonormal frame,\footnote{In general, this can only be done locally, even if in some examples certain slices involving the non-compact fields admit a globally defined flat frame, see Section \ref{s:maxsugra} below.} $e^a_i(\phi)$, wherein the charge-to-mass vector components read 
\begin{equation} \label{eq:zeta-vectors}
  \zeta^{a} = e^{a}_{i} \zeta^i\, , 
\end{equation}
and such that the inner product \eqref{eq:decayrate} and norms \eqref{eq:zetavecnorm} are simply computed using the diagonal metric $\delta_{ab}$. Let us also stress that, since the decay rate $\lambda$ for any such tower depends on the geodesic trajectory (cf. eq. \eqref{eq:decayrate}), it does not define, per se, an intrinsic property of the states, whereas the tower (or $\zeta$-)vectors do. Therefore, given the set of all possible towers becoming light, we denote by $m_{\rm t}$ the one that does so at the fastest rate, which means that $\lambda_{\rm t} = \vec{\zeta}_{\rm t} \cdot \hat n$ is the largest exponent. Notice that determining which one is the dominant tower for each possible asymptotic limit is not a mere academic exercise, since it can inform us about the different duality frames of the theory (whose boundaries typically occur whenever the former changes), and it can restrict as well the maximum variation of the scalar fields that can be accommodated within the original EFT. 

\begin{figure}
    \centering
    \includegraphics[width=0.6\linewidth]{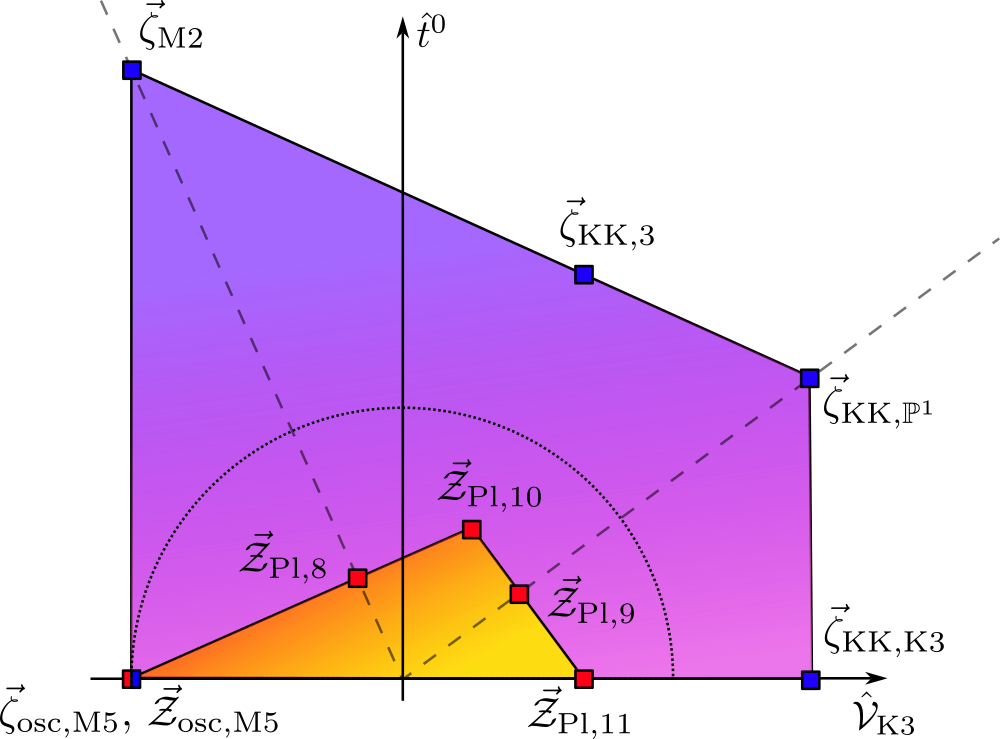}
    \caption{\small Example of convex hull generated by tower (blue) and species vectors (red). This example is realized in the string theory Landscape via M-theory compactification on an attractive K3 surface \cite{Castellano:2023jjt}. The dashed line separates different duality frames, wherein a different tower provides the dominant decay rate. Notice that the two polytopes are dual to each other with respect to a ball of radius $1/\sqrt{5}$, i.e., the black (half-)circumference centered at the origin.}
    \label{fig:hullexample}
\end{figure}

On the other hand, in practical terms, it is convenient to plot all the relevant $\zeta$-vectors (once they have been canonically normalized, cf. \eqref{eq:zeta-vectors}) in a \emph{single} diagram, see Figure \ref{fig:hullexample}. Whenever this is possible, one can draw the \emph{convex hull} determined by all such points, which is a very useful geometrical object that serves for many purposes. For instance, it allows us to verify in a simple manner whether certain lower/upper bounds on $\lambda_{\rm t}$ (see footnote \ref{fnote:sharpened}) are satisfied \cite{Calderon-Infante:2020dhm,Etheredge:2023odp}. Similarly, these diagrams often exhibit a symmetry structure that reflects the duality properties associated to the quantum UV-theory. In fact, when restricting to the non-compact directions, the resulting symmetries correspond to the Weyl group of the analogous (U-)duality one \cite{Castellano:2023jjt} (see Section \ref{s:maxsugra} for explicit examples of this).

Interestingly, a similar story can be seen to hold for the quantum gravity cutoff, whose determination---using eq. \eqref{eq:Lambdaspdef}---in the presence of multiple infinite sets of states becomes less straightforward, given that it may receive contributions from towers other than the lightest one. Indeed, as originally explained in \cite{Castellano:2021mmx, Castellano:2022bvr}, the details of the computation depend in a sensitive way on how the towers relate to each other, i.e., whether they are additive or multiplicative. Since we will refer several times to the latter case in this work, we review the calculation in the following.

Thus, consider a spectrum of mixed states with quantum numbers $(j, k) \in \mathbb{Z}^2\,$, associated to two such infinite towers. For concreteness, we take their mass dependence to be of the form
\begin{equation} \label{eq:effectivetower}
  m_{j,k}^{2} = j^{2/p_1} m_{\text{t}}^2 + k^{2/p_2} m_{\text{t}'}^2 \, ,\qquad \text{with}\quad m_{\text{t}}\leq m_{\text{t}'}\, .
\end{equation}
A useful way to think of this is as if it was coming from two distinct \emph{multiplicative} towers with mass scales $\{ m_{\text{t}},\, m_{\text{t}'} \}$ and density parameters $p_1$ and $p_2$, respectively.\footnote{A prominent example of \eqref{eq:effectivetower} would be the case of a pair of Kaluza-Klein towers corresponding to two compact internal manifolds of dimensions $p_1, p_2$ and radii $\{ R_1, R_2\}$, with masses $m_{\text{t}}=1/R_1$ and $m_{\text{t}'}=1/R_2$.} Hence, to each of these towers we can separately associate a would-be species scale as
\begin{equation}\label{eq:powerlikespecies}
  \Lambda_{\text{t}}\, \sim\, m_{\text{t}}^{\frac{p_1}{d-2+p_1}} \Mpd^{\frac{d-2}{d-2+p_1}}\, , \qquad \Lambda_{\text{t}'}\, \sim\, m_{\text{t}'}^{\frac{p_2}{d-2+p_2}}\Mpd^{\frac{d-2}{d-2+p_2}}\, ,
\end{equation}
where each of them is computed via \eqref{eq:Lambdaspdef} by accounting only for the subset of states associated to the corresponding tower, thus ignoring the remaining one.

However, to appropriately account for states with mixed quantum numbers $(j,k)$ in \eqref{eq:effectivetower}, one should consider the combined effect of the two aforementioned towers at once. Doing so, one finds \cite{Castellano:2021mmx}
\begin{equation}\label{eq:effspeciesscale}
  \Lambda_{\text{sp, eff}}\, \sim\, m_{\text{eff}}^{\frac{p_{\text{eff}}}{d-2+p_{\text{eff}}}}\Mpd^{\frac{d-2}{d-2+p_{\text{eff}}}}\, ,
\end{equation}
where we define effective (averaged) quantities as follows\footnote{Recall that such `effective' towers, together with their averaged mass scale and density parameters, are just book-keeping devices that allow us to easily compute the total number of species and quantum gravity cutoff via e.g., eq. \eqref{eq:effspeciesscale} \cite{Castellano:2021mmx, Castellano:2022bvr, Calderon-Infante:2023ler}.}
\begin{equation}\label{eq:masseffectivetower}
  m_{\text{eff}}\, \sim\, \left( m_{\text{t}}^{p_{1}}\, m_{\text{t}'}^{p_{2}} \right)^{1/p_{\text{eff}}}\, , \qquad p_{\text{eff}} = p_1 + p_2 \, .
\end{equation}
Notice that the reason why separating this computation in a two-step procedure becomes useful is because, depending on the asymptotic limit that we take, the states associated to either one of the two towers can become arbitrarily lighter than those coming from the second one. Therefore, in certain circumstances it may be enough to consider particles arising from just one of them in order to compute $\Lambda_{\rm sp}$. Most likely, one may still need to consider mixed states thereof, which happens whenever \cite{Castellano:2021mmx}
\begin{equation}\label{eq:conforeffectivetower}
  m_{\text{t}'}\, \lesssim\, \Lambda_{\text{t}} \iff \Lambda_{\text{eff}}\, \lesssim\, \Lambda_{\text{t}}\, .
\end{equation}
This simple condition implies that the true species cutoff for any given infinite distance limit is given by the smallest out of the scales $\{\Lambda_{\text{t}},\, \Lambda_{\text{t}'},\, \Lambda_{\text{eff}}\}$. Moreover, from \eqref{eq:effspeciesscale} it becomes clear that, since the asymptotic moduli-dependence of the mass of the infinite towers is exponential---as per the Distance Conjecture, the species cutoff should also exhibit this kind of behavior, namely
\begin{equation}\label{eq:speciesexponential}
  \Lambda_{\rm sp}\, \sim\, e^{-\lambda_{\rm sp} \Delta_{\phi}}\, ,
\end{equation}
with $\lambda_{\rm sp}$ being some $\mathcal{O}(1)$ factor. Hence, in analogy with the tower vectors defined in \eqref{eq:chargetomass}, one may introduce some \emph{species vectors}, which are computed from the expression \cite{Calderon-Infante:2023ler}
\begin{equation}\label{eq:defspeciesvectors}
  \mathcal{Z}^{a} = - \delta^{ab} e^{i}_{b} \,  \partial_{i} \log\Lambda_{\text{sp}}\, ,
\end{equation}
and whose projection gives, for every asymptotic direction $\hat n$, the exponential rate in \eqref{eq:speciesexponential}
\begin{equation}\label{eq:decayratespecies}
  \lambda_{\text{sp}}=\vec{\mathcal{Z}} \cdot \hat{n}\, .
\end{equation}
Similarly to the tower case, the norm of the species vectors thus defined is frequently linked with the nature of the infinite distance boundary they encapsulate. Indeed, in string theory examples one observes essentially two possibilities \cite{Lee:2019wij}: either the species cutoff corresponds to certain (possibly dual) higher-dimensional Planck mass or rather to some fundamental string scale. This yields \cite{Calderon-Infante:2023ler}
\beq\label{eq:speciesveconemodulus}
	|\mathcal{Z}_{{\rm KK},\, p}| = \sqrt{\frac{p}{(d+p-2) (d-2)}}\, , \qquad |\mathcal{Z}_{\rm osc}|= \frac{1}{\sqrt{d-2}}\, ,
\eeq
where again $p$ corresponds to the (effective) number of dimensions decompactifying. 

Concerning the allowed values for the $\mathcal{O}(1)$ parameter controlling the exponential decay of the species cutoff, there have been various upper and lower bounds proposed and studied in the literature. The most well-known ones were introduced and analyzed in \cite{Calderon-Infante:2023ler, vandeHeisteeg:2022btw,vandeHeisteeg:2023ubh}, which constrain $\lambda_{\rm sp}$ to lie in the range
\begin{equation}\label{eq:loweandupperrboundspecies}
  \frac{1}{\sqrt{(d-1)(d-2)}} \leq\lambda_{\rm sp} \leq \frac{1}{\sqrt{d-2}}\, .
\end{equation}
Particularly relevant for us will be the upper bound \cite{vandeHeisteeg:2022btw, vandeHeisteeg:2023ubh,vandeHeisteeg:2023dlw}, which we will study in relation to the Laplace equation \eqref{eq:eigenvalueeqintro} at various places in the upcoming sections.

Lastly, let us mention that, by collecting all the possible species vectors of a given quantum gravity theory into a single diagram, one also obtains a convex hull which captures in a global fashion (when possible) the different duality frames of the theory \cite{Calderon-Infante:2023ler}. An example of this kind of polytope is depicted in Figure \ref{fig:hullexample}. In fact, as originally observed in \cite{Castellano:2023stg, Castellano:2023jjt},\footnote{See also \cite{Rudelius:2023spc,Basile:2025bql,Etheredge:2024tok} for recent studies and applications of this idea.} the two diagrams resulting from plotting both the $\zeta$-vectors associated to the leading towers of states and the species hull satisfy the following mathematical constraint
\beq \label{eq:patternmass}
	\vec\zeta_{\rm t} \cdot\vec{\mathcal{Z}}= \frac{\kappa_d^2}{d-2}\, ,
\eeq
where the product is taken using the metric in the moduli space and we recall that $\kappa_d^2= \Mpd^{2-d}$ is the gravitational coupling constant. More precisely, this means that the corresponding polytopes are dual to each other with respect to a ball of radius $1/\sqrt{d-2}$ in $d$ spacetime dimensions, as illustrated in Figure \ref{fig:hullexample}.

\section{Maximal and Half-Maximal Supergravity}\label{s:maxsugra}

In this section, we examine the role of elliptic operators in determining the higher-derivative $t_8t_8\cR^4$ \cite{Green:2012oqa,Green:2012pqa} Wilson coefficient in the case of (half-)maximal supergravity in ten, nine and eight dimensions, while also emphasizing their link with the Distance Conjecture \cite{Ooguri:2006in} and the derivation of the species polytope \cite{Calderon-Infante:2023ler} in these theories.

The $\frac12$-BPS operator relevant for the analysis in maximal supergravity in $d$-dimensions, seen as M-theory compactified on a $D=11-d$ torus or Type II on $\mathbf{T}^{D-1}$, is the $\cR^4$-operator, whose computation in terms of four-supergraviton\footnote{Here, we refer to the whole massless gravity multiplet, whose operator can be written in terms of a generalized curvature invariant $\mathfrak{R}$ \cite{Green:2008bf,Green:2010wi,Boels:2012ie,Wang:2015jna}. In this work, we only focus on the purely gravitational contribution.} scattering amplitudes has been performed and thoroughly discussed in the literature \cite{Green:2010wi,Green:2010kv,Green:1997as,Green:1997tv,Green:1998by,Green:2005ba,Kiritsis:1997em,Obers:1999es,Lambert:2006ny,Obers_2000}. The part of the low-energy amplitude which is analytic in the Mandelstam variables $s,t,u$ can be split into a tree-level contribution and a radiative correction depending on the relevant supergravity moduli space $\cM_{\td}$, which in turn may be expressed in the form 
\beq
{\cal A}_{\text{1-loop}} ^{(d)} =  \sum_{p=0}^{\infty} \sum_{q=0}^{\infty }  \left[{\cF}_{(p,q)}^{(d)}\big( \cM_{\td}\big) \right]\sigma_2^p\sigma_3^q \, {\cal R}^4\, ,
\eeq
where one defines the dimensionless kinematic parameters $\sigma_n$ in terms of the $d$-dimensional Planck scale $\Mpd$ as follows
\beq
 \sigma_n =  \frac {\big(s^n+t^n+u^n\big)}{\left(2\Mpd\right)^{2n}}\, ,
 \eeq
whereas the functions $\cF_{(p,q)}^{(d)}$ are necessarily automorphic forms invariant under the U-duality group $G_{\rm{d}}(\bZ)$ of the $d$-dimensional theory. Indeed, the moduli space of maximal supergravity has the structure of a double quotient $\cM_\td = G_{\text{d}}(\bZ)\backslash G_{\text{d}}(\bR)/K$, where $K$ is the maximal compact subgroup of $G_\text{d}(\bR)$. To keep the notation simple, we will use ${\cal R}^4$ to refer to the total contraction of this object with the standard tensor $t_8t_8$ \cite{Green:2012pqa}. The $p=q=0$ piece of the above sum precisely corresponds to the Wilson coefficient of the $\cR^4$-operator in the effective action expressed in the Einstein frame
\begin{equation}
    S_{\cR^4} = \frac{1}{2 \kappa_D^2}\int \td^dx \sqrt{-g}\, \cF_{(0,0)}^{(d)} \frac{\cR^4}{\Mpd^6}\, ,
\end{equation}
so that the species scale can be defined from it using \eqref{eq:spsc} as
\cite{vandeHeisteeg:2023dlw,Castellano:2023aum}
\beq\label{eq:speciescoeffrelation}
  \frac{\Lambda_{\text{QG}}}{\Mpd} \sim \left({\cF}_{(0,0)}^{(d)}\right)^{-1/6}\, .
\eeq
For definiteness, henceforth we will focus on maximal supergravity with $d\geq 8$, where the operator is relevant/marginal (in the Wilsonian sense). This ensures that every infinite distance limit that one can take yields a suppression controlled by the species cutoff instead of, say, any other (lower) field-theoretic scale \cite{Calderon-Infante:2025ldq}. For completeness, we show the corresponding duality groups of the supergravity theories and coset space factors in Table \ref{tab:dualitygroups}.

\begin{table}[]
    \centering
    \renewcommand{\arraystretch}{1.3}
    \begin{tabular}{c|c|c|c}
        \hline
        $d$ & $G_{\rm{d}}(\mathbb{R})$ & $K$ & $G_{\rm{d}}(\mathbb{Z})$ \\
        \hline
        10A & $GL(1, \mathbb{R})$ & 1 & 1 \\
        10B & $SL(2, \mathbb{R})$ & $SO(2)$ & $SL(2, \mathbb{Z})$ \\
        9 & $GL(2, \mathbb{R})$ & $SO(2)$ & $SL(2, \mathbb{Z})$ \\
        8 & $SL(3, \mathbb{R}) \times SL(2, \mathbb{R})$ & $SO(3) \times SO(2)$ & $SL(3, \mathbb{Z}) \times SL(2, \mathbb{Z})$ \\
        \hline
    \end{tabular}
    \caption{\small Duality groups and coset space factors of the moduli spaces $G_{\text{d}}(\bZ)\backslash G_{\text{d}}(\bR)/K$ of maximal supergravity theories in $d=10,9,8$ \cite{Hull:1994ys}. }
    \label{tab:dualitygroups}
\end{table}

It was shown in \cite{Green:2010wi,Green:2010kv} that the function ${\cF}^{(d)}_{(0,0)}$ obeys a second order differential equation written in terms of the Laplacian on $G_\td(\bR)/K$, which schematically reads\footnote{The only exception being 10d Type IIA, which will be discussed in detail in Section \ref{ss:10dIIA} below.}
\beq\label{eq:eigenvalueeq}
\Delta_{G_\td/K} {\cF}^{(d)}_{(0,0)} = \eta_d\,  {\cF}^{(d)}_{(0,0)} +6\pi  \delta _{0,(d-8)}\, ,
\eeq
where the dimension-dependent factor $\eta_d$ is given by
\beq
\eta_d = 3\, \frac {(11-d)(d-8)}{d-2}\, .
\label{eigenvalue}
\eeq
Notice that for $d>8$, this is an eigenvalue equation, while for $d=8$, the $\eta$-piece in \eqref{eq:eigenvalueeq} vanishes and a constant source term $6\pi$ appears on the right-hand side. This signals the presence of threshold effects due to massless modes running in the loop \cite{Obers_2000,Green:2010wi,Green:2010kv}, reproducing the typical field-theoretic logarithmic running of marginal couplings, now dressed with moduli vevs. In string theory, they come from the Rankin-Selberg-Zagier regularization\footnote{Equivalently, they can be obtained upon imposing any duality-invariant renormalization procedure in spacetime, such as dimensional regularization as used in e.g., \cite{Green:2008uj} and mentioned in Appendix \ref{ap:Massform} for the cases of $SL(2,\bZ)$ and $SL(3,\bZ)$.} of the modular integrals defining the analytic part of the amplitude\footnote{Let us note that, despite the whole amplitude being finite, the splitting into analytic and non-analytic pieces introduces divergences which must be regulated in a modular-invariant fashion.}\cite{rankin1940contributions,selberg1940bemerkungen,zagier1981rankin,Angelantonj:2011br,benjamin2021harmonic}. A detailed acount of the relevant automorphic forms, which are invariant under $G_\td(\bZ)$ and moreover define some of the $\cF_n^{(d)}$ coefficients in $d\ge8$, can be found in Appendix \ref{ap:Massform}. 
 
It is also enlightening to rewrite the eigenvalue \eqref{eigenvalue} in a perhaps more suggestive way. In any spacetime dimension $d$ there must exist an asymptotic limit corresponding to full decompactification back to 11d M-theory. Hence, there should be an effective tower of KK modes with $p=11-d$ and species scale vector (in the notation of refs. \cite{Castellano:2021mmx,Castellano:2022bvr}) with length
\beq
 |{\vec \cZ}_{\rm{M-th}}|^2  = \frac{p}{{(d-2+p)(d-2)}} = \frac{11-d}{9\, (d-2)}\, .
\label{ximtheory}
\eeq
Thus, one may also write the possible eigenvalues in the form
\beq
\eta_d  =  27\, (d-8) |{\vec \cZ}_{\rm{M-th}}|^2\, ,
\label{eigenotra}
\eeq
which can be seen as (a multiple of) the classical dimension of the $\cR^4$-operator times the norm of the M-theory species vector. In particular, the length of this vector encodes information about the number of non-compact fields (i.e., radii) in the lower-dimensional theory.

Since one of our main interests here is to make contact with towers and species hulls, which are more naturally defined in asymptotic limits, in the following a particular focus will be devoted to studying the Laplace condition precisely within these regimes. To this end, and following the spirit of \cite{vandeHeisteeg:2022btw}, we will neglect certain threshold effects in the Wilson coefficients, which oftentimes naturally combine with the non-analytic parts of the amplitude \cite{Obers_2000,Green:2010wi, Calderon-Infante:2025ldq}, since these contributions come from massless modes running in the loops, and are not due to new towers of species entering the effective description. In particular, this allows us to ignore the source term in \eqref{eq:eigenvalueeq} (see Section \ref{ss:8dmaxsugra} for more on this). This intuition is quantified by the fact that---as explored in Appendix \ref{ap:Massform}---the asymptotic behaviour of the relevant automorphic functions is insensitive to the term thus neglected. Therefore, the Laplace equation becomes an eigenvalue equation in every dimension for $d\ge8$, as far as the asymptotics is concerned. 

Let us reformulate what it means to focus on infinite distance boundaries from the perspective of the spectral properties of the Laplacian. The discussion in Appendix \ref{ap:Massform} and above shows that automorphic forms are $G_\td(\bZ)$-invariant eigenfunctions of the Laplacian over moduli space. More generally, the Laplacian admits a spectral decomposition in terms of \textit{cusp forms} and \textit{non-cusp forms}. The former consists in eigenfunctions relative to the discrete part of the spectrum on $L^2\left(G_\td(\bR)/K\right)$, while the latter, of which the Eisenstein series are the prototypical example, describe the continuous part of the spectrum, and are thus more directly related to the non-compactness of moduli space. More specifically, a relevant role in infinite distance limits is played be the `\textit{constant terms}' of these functions, which capture their polynomial growth for large values of the non-compact directions---dubbed saxions. The terminology refers to the fact that these terms appear as zero modes of a Fourier decomposition with respect to the compact---i.e., axionic---fields (cf. Appendix \ref{ap:Massform} for details). In this language, duality forces the higher-derivative corrections to be automorphic forms, whose constant terms precisely yield the expected exponential decay\footnote{In terms of properly, i.e., canonically normalized scalar fields.} of the Distance Conjecture. One of the consequences of this automorphicity property is indeed the Laplace condition \eqref{eq:eigenvalueeq}, which is thus strongly tied to dualities and therefore to the Distance Conjecture.

In what follows, we will reinterpret these results from a different point of view. Namely, assuming the condition \eqref{eq:eigenvalueeq} to hold, and given a generic polynomial ansatz written in terms of saxionic fields resembling the duality-motivated constant terms, we will analyze what type of constraints arise for the Wilson coefficient, ${\cF}^{(d)}_{(0,0)}$, as well as for the species scale, $\Lambda_{\text{QG}}$. As mentioned before, we will focus on the case of maximal supergravities in $d=10,9,8$. Finally, let us remark that throughout the paper we will always assume that the relevant duality group exhibited at the two-derivative level (or rather a discrete subgroup thereof) survives at the quantum level. From the bottom-up, these dualities are believed to be needed to be imposed so as to make the moduli-space compactifiable, in the sense of \cite{Delgado:2024skw}.

\subsection{10d Type IIB}\label{ss:10dIIB}

\begin{figure}
    \centering
    \includegraphics[width=0.7\linewidth]{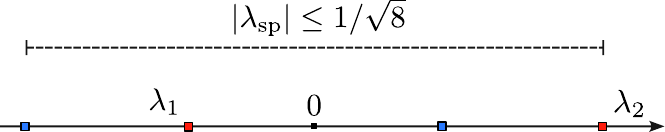}
    \caption{\small The $d=10$ Laplace equation translated as a geometric constraint imposed on the saxionic tangent space parametrized by $\lambda$. The \textit{red} points represent the solutions to $2\lambda\left(\lambda-\sqrt2\right)=3$. The duality symmetry acts on this space as a discrete $\bZ _2$ remnant sending $\lambda \to -\lambda$, which leads to the dual Laplace solutions captured by the \textit{blue} points. All solutions of both Laplace equations are characterized by leading to a decay rate of the species scale $|\lambda_{\text{sp}}| \le 1/\sqrt{d-2}$, which is indicated by the dashed interval. The outer dots reconstruct the Type IIB species polytope.}
    \label{fig:10dIIBcircle}
\end{figure}

Let us start with the case of ten-dimensional Type IIB supergravity, whose moduli space $\cM_{\text{IIB}} = SL(2,\bZ) \backslash SL(2,\bR) /SO(2)$ is parametrized by a single complex modulus $\tau=\tau_1+i\tau_2$, commonly referred to as the axio-dilaton. As usual, $\cM_{\text{IIB}}$ can also be written as $SL(2,\bZ)\backslash \mathfrak{h}$ with $\mathfrak{h}$ the complex upper-half plane, and the fundamental domain can be taken to be the standard one, namely ${\cal F} = \{ \tau \in \mathfrak{h}:\, |\tau|\ge 1 ,\, |\tau_1|\le 1/2\}$. The relevant two-derivative action in the Einstein frame reads \cite{Polchinski:1998rr}
\beq
 S_{\text{IIB}} = \frac{1}{2\kappa_{10}^2} \int \td^{10}x \sqrt{-g} \left( R - \frac{\partial \tau \cdot \partial {\overline \tau}}   {2\tau_2^2} + \dots \right)\, ,
\eeq
where $R$ denotes the Ricci scalar. Using the shorthand notation $\cF$ for the 10d coefficient $\cF^{(d)}_{(0,0)}$, the Laplace equation specialized to the case at hand is given by \cite{Green:1997tv,Green:1997as,Green:1998by,Sinha:2002zr,Green:2010wi,Green:2010kv}
\beq\label{eq:IIBlaplace}
 \Delta_{\rm sl_2} {\cal \cF}(\tau,{\overline \tau})  = \frac {3}{4} {\cF}  (\tau,{\overline \tau})\, ,
\eeq
where here the Laplace-Beltrami operator takes the form
\begin{equation}\label{eq:IIboperator}
 \Delta_{\rm sl_2}= \tau_2^2 (\partial_{\tau_1}^2  + \partial_{\tau_2}^2)\, ,  
\end{equation}
and is, by construction, invariant under $SL(2,\bZ)$. As mentioned already, the perspective that we will adopt here is to impose \eqref{eq:IIBlaplace} given an ansatz that reproduces the asymptotic contributions from the duality-motivated constant terms, and see what restrictions arise. Furthermore, imposing S-duality symmetry allows us to cut away the $|\tau|\le 1$ region. To that end, consider the generic polynomial ansatz for the only saxion field $\tau_2$, i.e.,  
\begin{equation}
{\cF} \sim \cF_{\rm const} \propto \tau_2^{\lambda/\sqrt{2}} = e^{\lambda \hat \tau}\, , 
\end{equation}
where we have introduced the canonically normalized modulus $\hat \tau = \frac{1}{\sqrt{2}}\log \tau_2$. From eq. \eqref{eq:IIBlaplace}, we thus find the constraint
\beq\label{eq:IIBlaplaceotro}
 \Delta_{\rm sl_2} {\cF}_{\rm const} = \frac12\lambda \left(\lambda -\sqrt{2}\right) {\cF}_{\rm const} \stackrel{!}= \frac {3}{4}e^{\lambda \hat \tau}\, ,
\eeq
which is a purely algebraic equation whose solutions $\lambda_{1,2}=-\frac{1}{\sqrt2},\frac{3}{\sqrt2}$ agree with the two constant terms of $E_{3/2}(\tau,\bar \tau)$ (cf. eq. \eqref{eq:nonpertexpansion}). Even if trivial in this one-dimensional case, it is useful for later purposes to give a geometric interpretation of this condition, considering the line $\lambda \in \bR$ of all possible coefficients as a vector space. The latter can be identified with the tangent space $\cT \cM_{(s)}$ of the saxionic slice of moduli space, given that for the canonically normalized saxion, $\hat \tau$, the species vector reads (as per \eqref{eq:speciescoeffrelation}) 
\begin{equation}
    \vec\cZ_{\text{sp}} =- \frac{\partial_{\hat \tau} \Lambda_{\text{QG}}}{\Lambda_{\text{QG}}} = \frac{1}{6}\lambda \equiv \lambda_{\text{sp}}\, ,
\end{equation}
independently of the base point. The $\lambda$ coefficient is thus (a multiple of) the decay rate of the tower of species, and the Laplace equation can be seen as a constraint on the possible species vectors $\vec\cZ_{\text{sp}}$, see Figure \ref{fig:10dIIBcircle} above.

Given that the full solution is the linear combination
\begin{equation}\label{eq:polynomialansatzIIB}
    \cF_{\rm const} = c_1\tau_2^{-1/2} + c_2\tau_2^{3/2}\, ,
\end{equation}
the species scale dependence on the constant terms is given by 
\begin{equation}
    \Lambda_{\text{QG}} \sim \left(c_1 \tau_2^{-1/2} + c_2 \tau^{3/2}_2\right)^{-\frac16}\, ,
\end{equation}
with $c_{1,2}$ undetermined numerical coefficients. Notice that for large $\tau_2$, one finds again the exponential damping of the species scale predicted by the Distance Conjecture, along with the appropriate decay rate $1/\sqrt{8}$. Moreover, as $\lambda=3 \cdot 2^{-1/2}$ is the bigger coefficient, the bound $|\lambda_{\text{sp}}|\le 1/\sqrt{d-2}$ is respected for all solutions of the Laplace equation. We thus find the upper bound put forward in \cite{vandeHeisteeg:2023ubh} (see also \cite{Calderon-Infante:2023ler, Castellano:2023jjt}). Before proceeding, it is important to mention that one could also choose to work directly in terms of canonical variables, where the saxionic piece of the Laplacian reads instead as (cf. eq. \eqref{eq:IIboperator})
\beq\label{eq:typeIIB-A}
\Delta_{\hat \tau} =  \partial_{\hat \tau}^2 - \partial_{\hat \tau}\, .
\eeq
This will be useful when comparing the present discussion with 10d Type IIA in Section \ref{ss:10dIIA}.

Lastly, notice that the constant terms we recover are not duality invariant, per se. This is expected from the fact that in order to perform the analysis we implicitly took the large $\tau_2$ regime to engineer our polynomial ansatz \eqref{eq:polynomialansatzIIB}, thereby selecting implicitly a certain fundamental domain. The remnant Weyl $\bZ_2$-duality symmetry acting on the saxionic slice of moduli space $\cT \cM_{(s)}$ (via $\lambda\to-\lambda$) maps the Laplace constraint \eqref{eq:IIBlaplaceotro} to\footnote{Notice that the Laplacian is not invariant under this $\bZ_2$ symmetry despite being $SL(2,\bZ)$-invariant, since this is part of $SL(2,\bZ)$ only if $\tau_1=0$.}
\begin{equation}
  \frac12\lambda\left(\lambda+\sqrt2\right) = \frac{3}{4}\, ,
\end{equation}
whose associated roots are depicted in Figure \ref{fig:10dIIBcircle}. Finally, notice that the solution with $\lambda=-1/\sqrt{2}$ does not appear to dominate any asymptotic limit, since $\tau_2 = 0 \notin {\cal F}$. Indeed, by picking the only solution in the fundamental region $\lambda >0$, and using the $\bZ_2$-symmetry one can reconstruct the entire Type IIB species polytope \cite{Calderon-Infante:2020dhm,Calderon-Infante:2023ler}, as shown in Figure \ref{fig:10dIIBcircle}. 

Moving towards the bulk of moduli space, the eigenfunctions of the Laplacian reconstruct the $SL(2,\bZ)$ automorphic forms
\cite{Green:2010wi,Green:2010kv} discussed in Appendix \ref{ap:Massform}. One finds it to be the non-holomorphic  Eisenstein series $E_{3/2}^{sl_2}(\tau,{\overline \tau})$, which, as mentioned before, admits the expansion
\beq
E_{3/2}^{sl_2} =  2\zeta (3)\tau_2^{3/2} + 4\zeta (2)\tau_2^{-1/2} +  {\cal O}(e^{-2\pi \tau_2})\, ,
\eeq
around $\tau_2 \to \infty$. Therefore, the solutions found above reproduce, barring numerical prefactors, the large-$\tau_2$ contribution of the full modular invariant $E_{3/2}^{sl_2}$ Eisenstein series.

\subsection{10d Type IIA}\label{ss:10dIIA}

Among the maximal supergravities in $d < 11$, 10d Type IIA is the only one presenting a moduli space $\cM_{\rm IIA} = \bR $ with trivial U-duality group. The four-graviton scattering amplitude at one-loop has been computed in \cite{Grisaru:1986dk,Grisaru:1986kw,Gross:1986iv}, where the Wilson coefficient $\cF$ of the $\cR^4$-term was determined to be
\begin{equation}
    \cF =2\zeta(3) e^{-\frac32\phi}+ \frac{2\pi^2}{3}e^{\frac\phi2}\, ,
\end{equation}
which matches the two perturbative contributions to the 10d Type IIB coefficient. Crucially, the D-instanton corrections \cite{Green:1997tv}, necessary so as to restore $SL(2,\bZ)$-duality, are absent here. This has the consequence of making the strong coupling limit $\phi \to \infty$ directly accessible and fundamentally different in the Type IIA theory. In 10d Type IIB, this would correspond to taking $\tau_2 \to 0$, which is cut away by imposing duality and restricting to the fundamental domain. Despite these differences, the coefficient still obeys an equation of the form
\begin{equation}\label{eq:typeIIAeq}
    \left(\partial_\phi^2 + \partial_\phi \right) \cF = \frac{3}{4}\cF\, ,
\end{equation}
which upon recalling that the IIB dilaton is defined as $\phi_B = - \hat \tau$, can be seen to be exactly the same as \eqref{eq:typeIIB-A}. This is our first example where the higher-derivative corrections obey an elliptic second order equation which does not involve the scalar field Laplacian. Notice that if it were not for the existence of a T-duality relating Type IIB/A in nine dimensions, there would be no need for the perturbative contributions to match the constant terms of the Eisenstein series $E_{3/2}^{sl_2}$ \cite{Green:1997as}. This gives us a hint that, upon compactifying the theory on $\mathbf{S}^1$ and uplifting the resulting Laplace equation---which is known to hold in that case \cite{Green:2010wi,Green:2010kv}---one indeed arrives at eq. \eqref{eq:typeIIAeq}. We will implement this procedure again in Section \ref{ss:maxsugra9d} below. 

For the time being, we will just show how the absence of U-dualities in 10d Type IIA readily explains why the relevant operator cannot be the Laplacian $\Delta_{\rm 10d} = \partial_\phi^2$. As already mentioned, having no non-perturbative self-dualities means that the strong coupling limit becomes now accessible. However, a putative solution to $\partial_\phi^2 \cF = \lambda \cF$, which reads $\cF \propto e^{\pm \lambda \phi}$, would naturally display a $\bZ_2$-invariance relating both weak and strong coupling regimes---where each of these contributions would dominate, thus persisting when computing $\Lambda_{\text{QG}}$ (or more precisely its asymptotic moduli dependence). Since we know this to not be the case here,\footnote{One cannot rule out a priori from this perspective real one-dimensional moduli spaces exhibiting such invariance, which would thus yield to a $\mathbb{Z}_2$-symmetric species polytope, see also \cite{Etheredge:2024tok}.} given that the strong coupling limit of 10d Type IIA string theory actually yields an 11d supergravity theory \cite{Witten:1995ex}, we conclude that the corresponding $\mathcal{R}^4$-operator cannot satisfy an eigenvalue equation with respect to the scalar Laplacian. Therefore, in a sense, the linear term present within \eqref{eq:typeIIAeq} can be interpreted as \emph{necessary} so as to lift this discrete symmetry.

Due to the differential equation being essentially the same as for Type IIB, the geometric constraint it implies on the species vectors living on the tangent space $\cT\cM = \bR$ is analogous to Figure \ref{fig:10dIIBcircle},\footnote{Recall the figure refers to $\hat \tau = - \phi$.} where now the dual circle cannot be drawn due to the absence of U-dualities, and the $\lambda_1$ point becomes accessible (and dominant) along the $\phi \to \infty$ limit. 

Finally, notice that a puzzle arises, as already in maximal supergravity there are cases in which the differential operator relevant to set up the eigenvalue equation is not the Laplacian. One could wonder whether some criteria exists differentiating theories whose higher-derivative coefficient obeys a Laplace equation from the ones obeying a more general second order differential equation. We will come back at this question in Section \ref{s:Laplace&Symm}, where we also take into account other examples analyzed in detail in Section \ref{ss:5dMtheory}. There, we argue that these eigenvalue equations can be related to properties of the isometry group associated to the appropriate moduli space. To this end, let us note that in Type IIA, the naive isometry group $\bR$ acting by shifts of the dilaton $\phi \to \phi + c$, gets broken by interaction in the Lagrangian, due to $e^{\phi}$ terms controlling the kinetic terms and couplings in the two-derivative theory, as well as higher-genus corrections expected to arise in the string effective action.

\subsection{Nine dimensions}\label{ss:maxsugra9d}

Let us now consider the unique maximal supergravity theory in 9d \cite{Gates:1984kr}, whose scalar manifold $\cM_{\rm 9d} = SL(2,\bZ)\backslash SL(2,\bR)/SO(2) \times \bR_+$ (cf. Table \ref{tab:dualitygroups}) can be parametrized, e.g., in the language of M-theory on $\mathbf{T}^2$, by a complex structure modulus $\tau$ and an overall real volume modulus $\cV$ (in 11d Planck units). The metric on $\cM_{\rm 9d}$ can be read from the relevant piece of the two-derivative action (see e.g., \cite{Calderon-Infante:2023ler})
\begin{equation}
 S_{{\rm 9d}} = \frac{1}{2\kappa_9^2} \int \td^{9}x \sqrt{-g} \left( R - \frac{\partial \tau \cdot \partial {\overline \tau}}   {2\tau_2^2} - \frac{9}{14}(\partial \log \cV)^2+\dots \right)\, .
\end{equation}
This theory, together with its tower and species hulls, has been extensively studied in the context of the Swampland program, and we refer the interested reader to \cite{Etheredge:2022opl,Calderon-Infante:2023ler,Castellano:2023jjt,Castellano:2023aum,vandeHeisteeg:2023dlw}. In 9d, the Laplace equation \eqref{eq:eigenvalueeq} obeyed by the higher-derivative $\mathcal{R}^4$-coefficient reads
\begin{equation}\label{eq:laplace9d}
    \Delta_{\rm 9d} \cF (\cV, \tau, \bar \tau) = \frac67 \cF(\cV, \tau,\bar \tau)\, , 
\end{equation}
where the relevant Laplacian operator $\Delta_{\rm 9d}$ in these coordinates is readily determined to be\footnote{Notice that this differs from that of $SL(2,\bZ)\backslash SL(2,\bR)/SO(2) \times \bR_+$ directly computed using the metric above by the last term. The operator \eqref{eq:Laplacian9d} is indeed the Laplacian for the symmetric space $E_{2(2)}(\bR)/K$ \cite{Obers_2000,Green:2010wi}.}
\begin{equation}\label{eq:Laplacian9d}
    \Delta_{\rm 9d} = \Delta_{\rm sl_2} + \Delta_\cV = \tau_2^2 (\partial_{\tau_1}^2+\partial_{\tau_2}^2) +\frac79 \cV \partial_{ \cV}( \cV \partial_{\cV}) + \frac13 \cV \partial_{\cV}\, . 
\end{equation}
As in the previous section, the approach adopted here will be to impose \eqref{eq:laplace9d} from the bottom-up, given a polynomial ansatz resembling constant terms. Differently from the 10d case, the Laplace equation will not fix the moduli dependence uniquely, and we will need additional top-down assumptions to derive stronger results. 

The saxions, denoted by $\tau_2$ and $\cV$, can be properly normalized by introducing a pair of coordinates $(\hat \cV$, $\hat \tau)$ as follows \cite{Calderon-Infante:2023ler}
\begin{equation}
    \hat \cV = \sqrt{\frac{9}{14}}\log \cV \ , \quad \hat \tau = \frac{1}{\sqrt{2}}\log \tau_2\, ,
\end{equation}
such that a constant term ansatz leads to the Laplace equation being rewritten as
\begin{equation}\label{eq:9dlaplacecircle}
    \cF (\cV, \tau) \sim \cF_{\rm const} (\hat \cV, \hat \tau) \propto e^{\hat\cV x +\hat \tau y} \ \ \Longrightarrow \ \ 
    \frac{1}{2}x \left(x+\frac{2}{\sqrt{14}}\right) + \frac{1}{2}y\left(y-\sqrt{2}\right) = \frac{6}{7}\, ,
\end{equation}
which again can be interpreted as a geometrical constraint on the $(x,y)$-plane, identified herein with the (saxionic) tangent space  $ \cT\cM_{(s)}$, and independent of the base point. 

\begin{figure}
    \centering
    \includegraphics[width=0.55\linewidth]{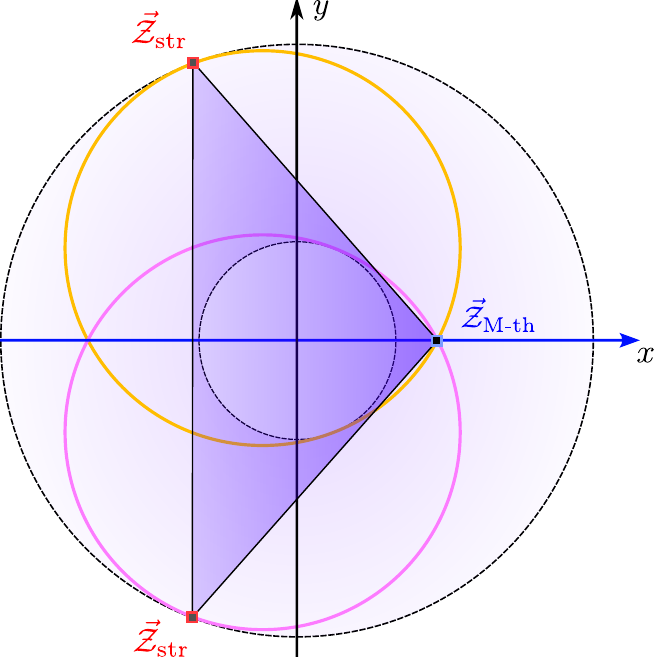}
    \caption{\small Laplace circles and species polytope for the case of maximal supergravity in $d=9$. The outer dashed circle is the upper bound \eqref{eq:upperbound9d} derived in the main text. The two circles are centered respectively in ${\vec \xi}_\pm=\left( -\frac {1}{\sqrt{14}},\pm\frac {1}{\sqrt{2}}\right)$, with radius $4/\sqrt7$.}
    \label{circulos}
\end{figure}

As anticipated before, there is now a continuous family of possible solutions to \eqref{eq:laplace9d}, which is depicted in Figure \ref{circulos}. Due to the positive definiteness of the eigenvalue and of the elliptic operator itself, the Laplace condition leads to this family being described by a circle in the $(x,y)$-plane, instead of e.g., an hyperbola. Its boundedness has an immediate consequence since, starting from the ansatz \eqref{eq:9dlaplacecircle} for the Wilson coefficient and computing the species vector through \eqref{eq:speciescoeffrelation}, one readily gets
\begin{equation}\label{eq:upperbound9d}
    |\vec{\cZ}_{\text{sp}}|^2 = \frac{1}{36}\left(x^2 + y^2\right) \le \frac{1}{7} = \frac{1}{d-2}\, ,
\end{equation}
which is the species vector obtained by assuming to fix $x,y$ and moving parallel to the $(x,y)$ direction. Of course, these $x,y$ values must obey the Laplace equation, leading to the bound. Thus, as in the Type II case, we recover the upper bound proposed in \cite{vandeHeisteeg:2023ubh}. To parallel the 10d discussion, it is also useful to notice that the remnant $\bZ_2$ Weyl subgroup duality action on the saxionic slice of moduli space leads to the `dual' Laplace circle given by the replacement $y \to -y$. We will now show how, combining the 8d Laplace equation
plus the appropriate 9d differential conditions one can reconstruct this convex hull.

To connect this discussion more directly with the species convex hull, additional assumptions will be needed. The species polytope is generated by the vertices \cite{Calderon-Infante:2023ler}
\begin{equation}\label{eq:9dspeciestowers}
    \vec{\cZ}_{\text{str}, 1} = \left( \frac{1}{2\sqrt{14}}, \frac{1}{2\sqrt{2}} \right)\,, \qquad \vec{\cZ}_{\text{str}, 2} = \left( \frac{1}{2\sqrt{14}}, -\frac{1}{2\sqrt{2}} \right)\, , \qquad \vec{\cZ}_{\text{M-th}} = \left( -\frac{\sqrt{14}}{21},0\right)\, , 
\end{equation} 
given by the two S-dual towers of string oscillators and the double KK-tower responsible for decompactifying back to 11d M-theory. The hull, together with the Laplace circles, are shown in Figure \ref{circulos}.

\begin{figure}
    \centering
    \includegraphics[width=0.6\linewidth]{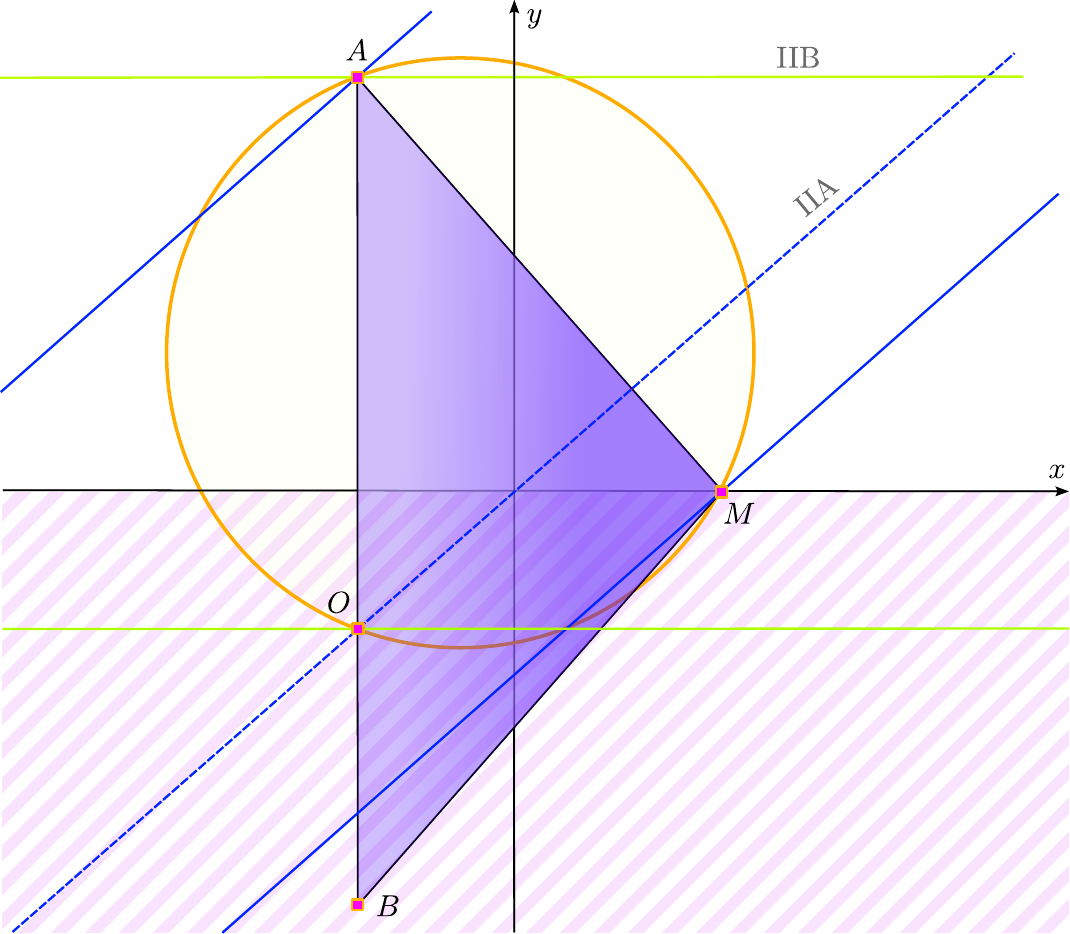}    \caption{\small 9d Laplace equation (\textit{orange}), 10d IIB (\textit{green}) and 10d IIA (\textit{blue}) equations plotted together. The $x$ axis is included in the 10d IIB solutions. In the upper half plane the intersection between all constraints is given by $M$ and $A$, which together the dual point $B$ reconstructs the species polytope. In the lower half-plane the point $O$ also solves all equations and recovers the missing constant term contribution from the $E_{3/2}^{sl_2}$ expansion in \eqref{eq:9dwilsoncoeff}. }
    \label{fig:circulos2}
\end{figure}

Motivated by duality, let us  focus on the fundamental region $y>0$. To further constrain the possible solutions, we will use the fact that the 9d theory arises as a circle compactification of both Type IIB and Type IIA. This allows us to identify directions in moduli space where the Type II circle becomes large. In terms of M-theory variables, the two radii are given by
\begin{equation}\label{eq:9ddir}
    \frac{r_B}{\ell_{10}}= \cV^{-1/3}\, , \qquad \frac{r_A}{\ell_{10}} = \cV^{9/16} \tau_2^{7/16}\, , \qquad \frac{\cV}{\tau_2} = e^{{\sqrt{\frac{14}{9}}}\hat \cV - \sqrt{2}\hat \tau} = e^{\frac43\phi_A}\, , 
\end{equation}
where $\phi_A$ is the 10d IIA dilaton. We see thus that the $\cV \to 0$ limit corresponds to a decompactification to 10d IIB, while sending $\cV\sim \tau_2 \to \infty $ with $\phi_A$ fixed, brings us back to 10d IIA. In Figure \ref{fig:circulos2} these two directions are given respectively by the $x$ axis and the dotted line perpendicular to the side of the hull.\footnote{In terms of properly normalized moduli, keeping $\phi_A$ fixed in \eqref{eq:9ddir} translates in moving along a line in the $xy$-plane.} Given these identifications, let us now impose the 10d equations satisfied by the Wilson coefficients on the 9d one,
\begin{equation}
    \left( \partial_{\phi_A}^2 + \partial_{\phi_A} \right) \cF =  \frac{3}{4}\cF\, , \qquad \tau_2^2 (\partial_{\tau_1}^2+\partial_{\tau_2}^2) \cF = \frac{3}{4}\cF\, ,
\end{equation}
whenever the dependence of the ansatz $\cF$ on the relevant variables is non-trivial. Indeed, notice that e.g., the constraint from IIB need not apply if the ansatz is written purely in terms of $\hat \cV$ i.e., if $y=0$. Next, we stress that it is natural to impose these constraints as the generic expectation from dimensional reduction is that the Wilson-coefficient in the large radius limit reads $\cF_9 \sim r^k \cF_{10}$, with $k$ some positive number and $r$ the radius---usually measured in higher-dimensional Planck units. Using the relation between the different coordinates, we find the following two equations
\begin{equation}
\begin{aligned}
   &\frac12 y(y-\sqrt{2})=\frac34\, ,\qquad \qquad \qquad\qquad\qquad \ \ \text{when}\quad y\ne 0\, ,\\
   &\frac{(\sqrt7x-3y)(8+\sqrt{14}x - 3\sqrt2 y)}{32\sqrt2} = \frac{3}{4}\, , \qquad \text{when}\quad  \sqrt{7}x \ne 3y\, ,  
\end{aligned}
\end{equation}
which are depicted in yellow and blue, respectively, in Figure \ref{fig:circulos2}. The points selected by all these constraints in the upper plane are $M, A$ and, together with its dual $B$ in Figure \ref{fig:circulos2}, they make up the hull. Notice that these are towers whose species scale appears directly in the Wilson coefficient, and which are seen by the 9d and 10d theories alike. In particular, from the top-down, we know that the triple intersection between 9d, 10d IIA and IIB equations (i.e., point A) is given by the string tower, since all theories see this limit, while the double intersection point $M$ seen by 9d and 10d IIA represents the decompactification to M-theory.

Thus, by assuming that we can impose the 10d Type II Laplace equations in the 9d theory whenever the dependence on the relevant modulus is non-trivial allows us to identify the three points \eqref{eq:9dspeciestowers}, given by $A,M$ in the fundamental region, plus their duality orbit, which reconstruct the hull. Moreover, a generic linear combination of the points $M,A,O$ which solve all of the equations and lie on the hull\footnote{Notice there would be another triple intersection point in the lower-half plane. However, it doens't lie on the hull constructed by looking at a fundamental domain, and is thus excluded.} in Figure \ref{fig:circulos2} will make up a Wilson coefficient in the $(\cV,\tau)$ variables of the kind
\begin{equation}\label{eq:9dwilsoncoeff}
    \cF \sim c_1 \overbrace{\cV^{6/7}}^{M} + c_2 \cV^{-9/14} \big( \overbrace{\tau_2^{-1/2}}^{O}+\overbrace{c_3\tau_2^{3/2}}^{A} \big)\, ,
\end{equation}
which precisely reproduces the constant terms of the full Wilson coefficient $\cF =  c_1\cV^{6/7} + c_2 \cV^{-9/14} E_{3/2}^{sl_2}$ computed in the literature \cite{Green:2010wi}. To end the discussion, notice that the point $O$ did not appear in the previous paragraph as it was not found in the fundamental domain chosen. It's appearance in the Wilson coefficient, but not in the reconstruction of the hull, mirrors the Type IIB discussion of the $\tau^{-1/2}_2$ term in Section \ref{ss:10dIIB}. 

\subsubsection{Deriving the 10d Type IIA operator}\label{sss:typeIIAop}

Now that the nine-dimensional case has been discussed, we will provide an argument to derive which differential equation should hold in 10d Type IIA supergravity. As explained in \cite{Green:2010wi}, one can set up an inductive procedure to determine the eigenvalues of the Laplace equation for maximal supergravity, using the fact that the Laplacians in $D$ and $(D+1)$ dimensions are linked via the simple relation
\begin{equation}
    \Delta_{D+1} = \Delta_D + a_D  \,r^2 \partial_r^2 + b_D\,r \partial_r\, , 
\end{equation}
where $r$ is a dimensionless radius of the circle compactification $D+1 \to D$, and $\{a,b\}$ denote some dimension-dependent constants. Thus, compactifying Type IIA on a circle of radius $r_A$, and expressing the 9d Laplacian using the set of coordinates
\begin{equation}
    r_A =\frac{R_A}{\ell_{10}} = \cV^{9/16} \tau_2^{7/16}\, , \quad  e^{\frac43\phi_A} = \frac{\cV}{\tau_2}\, , 
\end{equation}
one gets, dropping the axion $\tau_1$, the following expression
\begin{equation}
    \Delta_{\rm 9d} = \overbrace{\partial_{\phi_A}^2 + \partial_{\phi_A}}^{\cD^2} +\frac{7}{16}r_A^2 \partial_{r_A} + \frac{3}{16}r_A \partial_{r_A}\, ,
\end{equation}
which correctly identifies the elliptic operator $\cD^2$ introduced in Section \ref{ss:10dIIA}. The relevant eigenvalue depends only on the dimension, and is thus the same as in 10d Type IIB, thereby implying that the corresponding uplifted Wilson coefficient should obey
\begin{equation}
    \cD^2\cF_{\text{IIA}}  = \frac34 \cF_{\text{IIA}}\, .
\end{equation}
Compatibly, the 9d coefficient written in terms of these coordinates reads
\begin{equation}
    \cF \sim r_A^{6/7} \left( e^{-\frac{3}{2}\phi_A} + e^{\frac{1}{2}\phi_A}\right) \sim r_A^{6/7} \cF_{\text{IIA}}\, .
\end{equation}

\subsection{Eight dimensions}\label{ss:8dmaxsugra}

To end the maximal supergravity cases, let us look at the unique---up to gaugings---eight-dimensional theory given by a two-derivative action, whose scalar sector reads \cite{Salam:1984ft}
\begin{equation}
\begin{split}
    S_{8\td} &= \frac{1}{2\kappa_8^2} \int \td^8x \sqrt{-g} \Bigg[ R
    - \frac{9}{14} (\partial \log \mathcal{V}_2)^2 
    - \frac{7}{6} (\partial \log R_3)^2 
    - \frac{\partial \tau \cdot \partial \bar{\tau}}{2\tau_2^2} \\
    &\quad - \frac{\mathcal{V}_2^{-12/7} R_3^{-2}}{2} 
    \left( \partial C_{(123)} \right)^2 
    - \frac{\mathcal{V}_2^{-9/7} R_3^{+2}}{2\tau_2} 
    \Bigg| \partial \left( \frac{\text{Im} (\tau \xi_M)}{\tau_2} \right) 
    + \tau \, \partial \left( \frac{\text{Im} (\xi_M)}{\tau_2} \right) \Bigg|^2
    \Bigg]\, ,
\end{split}
\end{equation}
where we defined $\xi_M = -C_1 + iC_2 \tau_2$ and parametrized the scalar manifold in terms of coordinates $\{ \tau, R_3,\cV,C_1, C_2,C_{(123)} \}$. In the language of M-theory on a $\mathbf{T}^3 = \mathbf{T}^2\times \mathbf{S}^1$, they can be made sense of as the $\mathbf{S}^1$ radius $R_3$ and the $\mathbf{T}^2$ volume and complex structure $\cV_2,\tau$ in M-theory units. Moreover, $C_{(123)}$ arises from the reduction of the 11d 3-form on $\mathbf{T}^3$, while $C_{1,2}$ parametrize the orientation of the $\mathbf{T}^2$ inside the $\mathbf{T}^3$. We can also introduce, for future use, the canonically normalized saxions
\begin{equation}\label{eq:8dnorm}
    \hat \tau = \frac{1}{\sqrt{2}}\log \tau_2\, , \qquad \hat \rho = \sqrt{\frac76}\log R_3\, , \qquad \hat U = \frac{3}{\sqrt{14}}\log \cV_2\, .
\end{equation}
Even though these coordinates do not make it manifest, the scalar manifold is the symmetric space $\cM_{\rm 8d} = SL(3,\bZ) \times SL(2,\bZ)\backslash SL(3,\bR) \times SL(2,\bR) /SO(2) \times SO(3)$, and the theory enjoys a classical $SL(3,\bR) \times SL(2,\bR)$ invariance. Indeed, one can introduce a complex modulus $\cT = C_{(123)} + i \cV_3 \equiv C_{(123)} + i \cV_2^{6/7}R_3  $ and a $3 \times 3$ matrix $\tilde g = \cV_3^{-1/3}g$, with $g$ the $\mathbf{T}^3$ metric such that \cite[Appendix B]{Castellano:2023aum}
\begin{equation}
    S_{8\td} = \frac{1}{2\kappa_8^2}\int \td^8 x \sqrt{-g}\left[ R + \frac14 \text{tr} \left( \partial \tilde g \cdot \partial \tilde g ^{-1}\right)- \frac{\partial \cT \cdot \partial \bar \cT}{2\cT_2^2}\right]\, ,
\end{equation}
where the $SL(3,\bR)$ acts on the $\tilde g$ matrix in the adjoint representation and the $SL(2,\bR)$ on the $\cT$ modulus. Identifying the $R_3$ direction as the M-theory circle, it is also possible to introduce a set of coordinates $\{ T_A,U_A,\phi_8\}$ closely related to the language of IIA on a torus. Performing a T-duality and passing to coordinates $\{\phi_8,T_B \} \to \{\nu, \tau_B\}$ with $\tau_B$ the IIB axio-dilaton, the action can be rewritten as \cite{Castellano:2023aum} 
\begin{equation}
    S_{8\td} = \frac{1}{2\kappa_8^2}\int \td^8 x \sqrt{-g}\left[ R - \frac{1}{6} \frac{(\partial \nu)^2}{\nu^2} - \frac{\partial \tau_B \cdot \partial \bar{\tau_B}}{2 \tau_{2,B}^2} - \frac{\partial U_B \cdot \partial \bar{U_B}}{2 U_{2,B}^2} - \nu \frac{|\tau_B \partial b + \partial c|^2}{2 \tau_{2,B}}  \right]\, ,
\end{equation}
and the Laplacian in these coordinates takes the simple form
\begin{equation}
    \Delta_{\rm 8d}= \Delta_{\rm sl_2} + \Delta_{\rm sl_3} = \tau_{2,B}^2 \left( \partial_{\tau_{1,B}^2}+\partial_{\tau_{2,B}^2} \right) + \frac{1}{\nu \tau_{2,B}}|\partial_b - \tau_B \partial_c|^2 + 3 \partial_\nu(\nu^2 \partial_\nu)\, .
\end{equation}
The higher-derivative coefficient now obeys the following differential equation, cf. \eqref{eq:eigenvalueeq}
\begin{equation}\label{eq:laplace8dcsource}
    \Delta_{\rm 8d} \cF = 6\pi\, ,
\end{equation}
which is in fact not an eigenvalue equation. As discussed in \cite{Green:2006gt,Green:2008uj,Green:2010wi}, the source term on the right-hand side of the equation contributes to logarithmic dependences in the moduli typical of infrared divergences due to massless modes. In particular, as in $d=8$ the $\cR^4$ coupling appearing at one-loop is marginal, these terms arise from the regularisation of the typical field-theoretical log-divergences appearing in the one-loop graviton scattering amplitudes.\footnote{In string theory, the poles that appear in the piece of the one-loop amplitude relative to this coefficient cancel against other poles present in the non-analytic part, so that the whole amplitude is finite.} Since we are interested in making contact with Swampland conjectures and the breakdown of the effective theory due to new massive towers of states, we will ignore their contribution to the Wilson coefficient---in the spirit of \cite{vandeHeisteeg:2022btw}---when defining the species scale, and imposing instead
\begin{equation}\label{eq:8dlapl}
    \Delta_{\rm 8d} \cF  = 0\, .
\end{equation}
One should remark that the constant terms of the relevant regularized Eisenstein series appearing in the asymptotic expansion of the full coefficient as (see Appendix \ref{ap:Massform})
\begin{equation}
   \cF_{8\td} =   2E_1^{sl2}(\cT,\bar \cT) + \hat E^{sl3}_{3/2}(\cV,\tau,R_3)\, \sim\, c_1 \cV_2^{6/7}R_3 + c_2 \tau_2 + c_3 \cV^{9/7}R_3^{-2}\, ,   
\end{equation}
indeed consistently satisfy the stronger \eqref{eq:8dlapl}.

\begin{figure}
    \centering
    \includegraphics[width=1.02\linewidth]{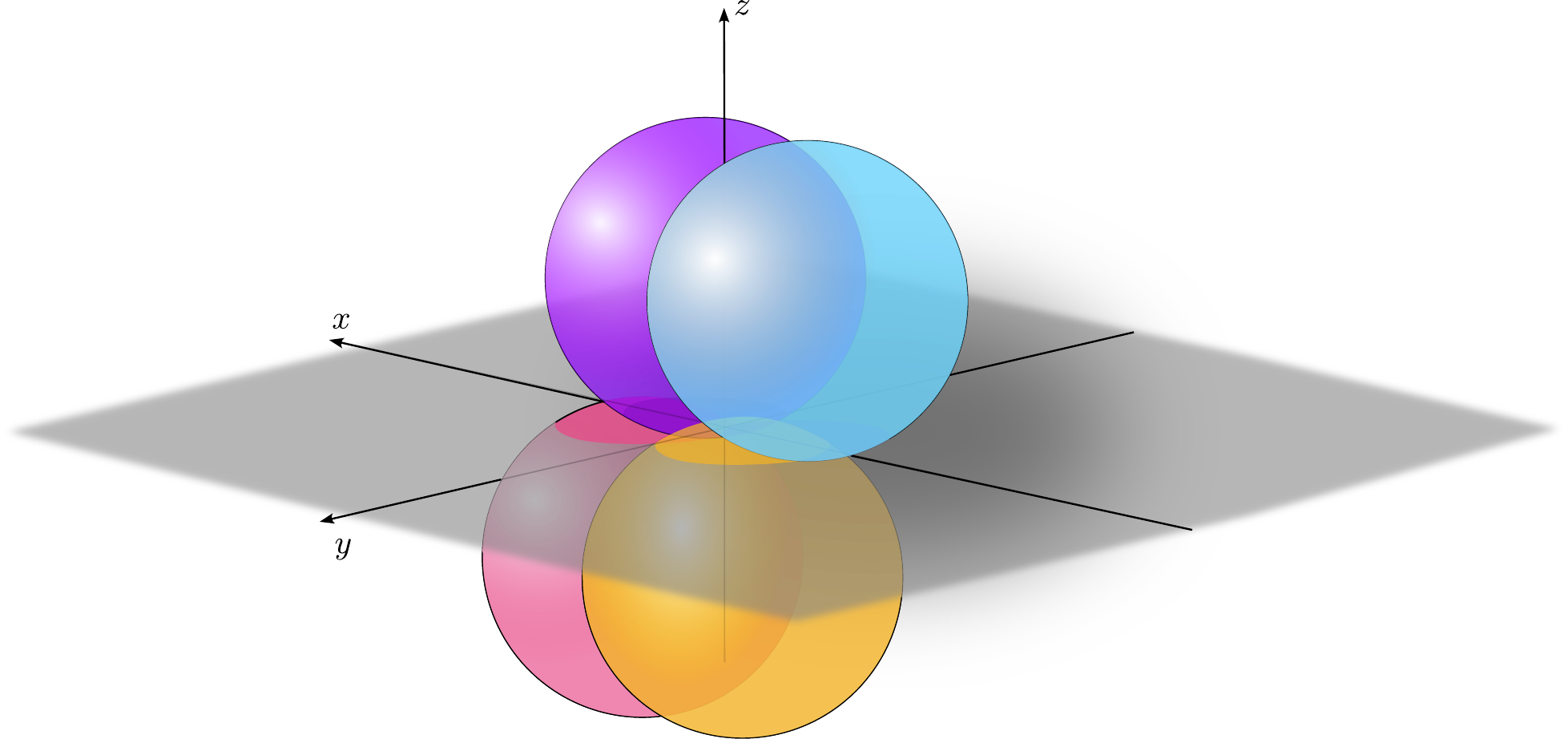}\vspace{0.05\linewidth}
    
     \includegraphics[width=1.02\linewidth]{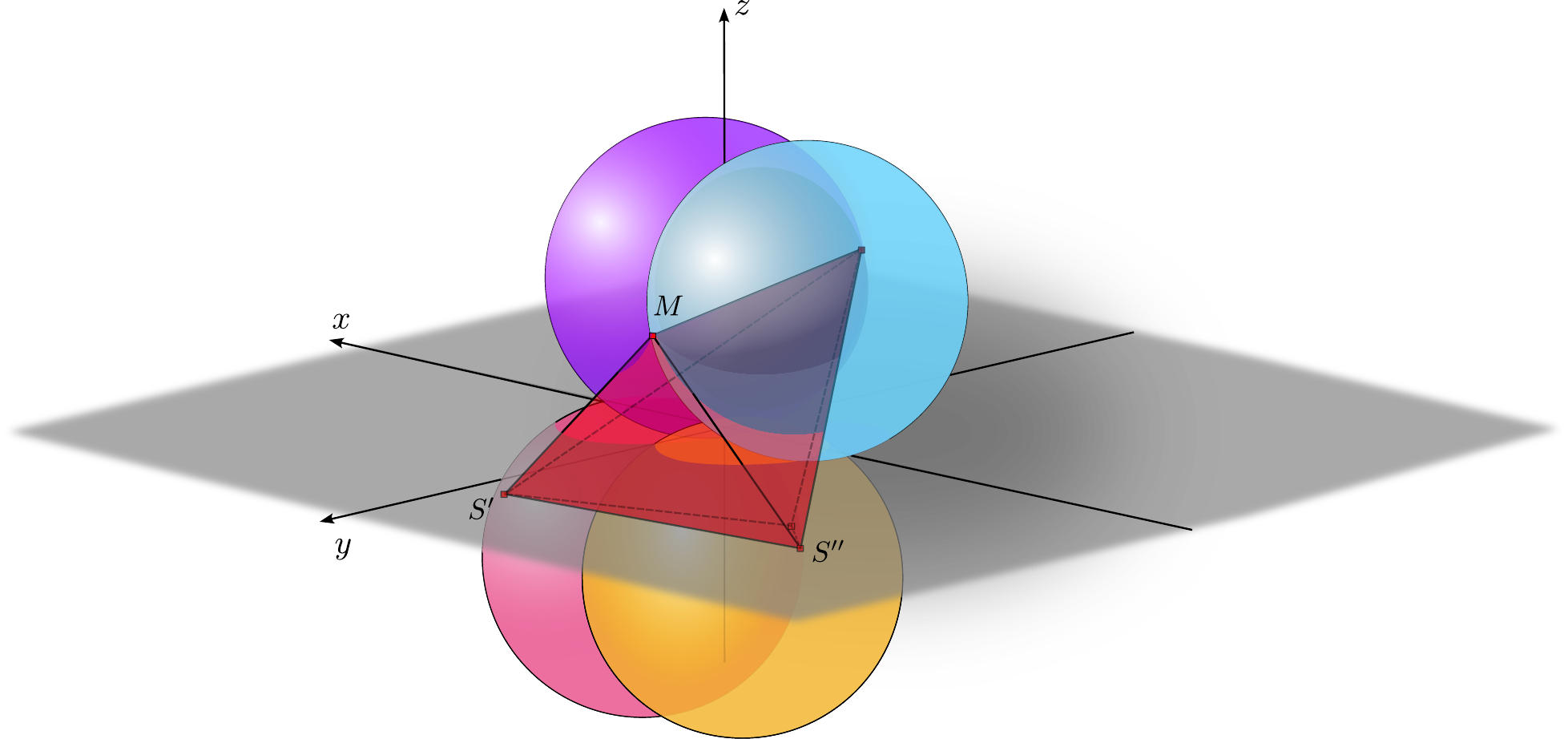}
    \caption{\small Laplace spheres in the 8d theory. The one relative to the original ansatz \eqref{eq:8dlapleq} is depicted in purple. The pink one is generated by acting with the $S_2$ premutation element of the duality remnant $S_3 \times S_2$. The other two are generated by acting with the order two element $b \in S_3$. To avoid cluttering, we did not plot the other 8 spheres generated by the order three element $a \in S_3$, which action rotates by $120^\circ$ in the plane $y-z=0$ fixed by the $S_2$ action. In the lower image we depict the species polytope, whose vertices always touch the Laplace spheres.}
    \label{fig:8dlaplballs}
\end{figure}

Following the procedure outlined in the previous sections, we once again express the duality-motivated constant term ansatz as
\begin{equation}
    \cF \sim e^{x \hat \tau + y \hat \rho + z \hat U}\, ,
\end{equation}
in terms of the canonically normalized moduli \eqref{eq:8dnorm}. Once more, one can see the space spanned by $x,y,z$ as the tangent space to the saxionic slice $\cT \cM_{(s)}$ of moduli space, and the Laplace equation as leading to the constraint
\begin{equation}\label{eq:8dlapleq}
    x^2-\sqrt{2} x+y^2+\sqrt{\frac{6}{7}} y+z^2-5 \sqrt{\frac{2}{7}} z =0\, ,
\end{equation}
which is an equation for a sphere of center $\left( \frac{1}{\sqrt2},-\sqrt{\frac{3}{14}},\sqrt{\frac{25}{14}}\right)$ and radius $r=\sqrt{\frac52}$.
Accordingly, we can use the remnant duality symmetry acting on this space to plot other Laplace spheres. This corresponds to computing the action of the Weyl subgroup of the U-duality group, which in 8d is given by $S_3 \times S_2$ \cite{Castellano:2023jjt}. The $S_2$ factor can be identified with the reflection symmetry related to the $SL(2,\bZ)$ duality group, while the $S_3$ remnant from the $SL(3,\bZ)$ group is given by the presentation
\begin{equation}
    S_3 : \ \langle b,a \, : \, b^2 =a^3=e, \, b a b=a^{-1}\rangle\, .
\end{equation}
More concretely, the action of the $S_2$ on $\cT_2 \to \cT^{-1}_2$ sends $(x,y,z) \to (x,-y,-z)$, while the order two element $b$ in $S_3$ is related to the action $\tau_2 \to \tau_2^{-1}$ and thus sends $x \to -x$. This generates the four spheres in Figure \ref{fig:8dlaplballs}. 
The order-three element $a$ finally generates a total of 12 Laplace spheres, corresponding to a $120^\circ$ rotation of the previous four spheres in the plane fixed by the $S_2$ action $y-z=0$.

As for the 9d case, by imposing the the Wilson coefficient obeys the Laplace equation \eqref{eq:8dlapl} we can get an upper bound on the decay rate of the species scale as, given the ansatz, we have 
\begin{equation}
    |\vec \cZ_{\text{sp}}|^2 = \frac{1}{36}(x^2+y^2+z^2)\, ,
\end{equation}
which corresponds to the decay rate of the tower moving in the direction $(\hat \tau,\hat \rho,\hat U)$ aligned with $(x,y,z)$. In this case we have that 
\begin{equation}
    |\vec \cZ_{\text{sp}}|^2 \le \frac{5}{18} \, ,
\end{equation}
which in particular is weaker than the upper bound $(d-2)^{-1}$ proposed in \cite{vandeHeisteeg:2023ubh}. 
Up to this point  the Laplace equation alone does not give us more  information. To proceed,  in what follows we will assume to inherit the top-down knowledge of the Laplace equation in 9d to reconstruct the species polytope, showing that the strategy described in Section \ref{ss:maxsugra9d} can be repeated, while setting up an induction procedure for lower dimensions.

We now turn to the species hull, which outer vertices can be identified in terms of $p \to \infty$ towers of string excitations and $p=3$ towers decompactifying to 11d M-theory. They are given by an $SL(3,\bZ)$ triplet of string oscillators which gets permutated under the $S_3$ symmetry, and an $SL(2,\bZ)$ doublet given by \cite{Castellano:2023jjt} 
\begin{equation}\label{eq:8dhull}
    \begin{split}
    &\vec \cZ_{\text{str},1} = \bigg( \frac{1}{2\sqrt2}, \frac{1}{\sqrt{42}}, - \frac{1}{2\sqrt{14}}\bigg)\, , \quad  \vec \cZ_{\text{str},2} = \left( -\frac{1}{2\sqrt2}, \frac{1}{\sqrt{42}}, - \frac{1}{2\sqrt{14}}\right)\, , \\ &\vec \cZ_{\text{str},3} = \left( 0, -\sqrt{\frac{2}{21}}\, \frac{1}{\sqrt{14}}\right)\,, \qquad \quad \ \vec \cZ_{\text{KK},3} = \left( 0, \frac{1}{\sqrt{42}}, \frac{2}{3\sqrt{14}}\right)\, ,\\ &\vec \cZ_{\text{M},3} = \left( 0, -\frac{1}{\sqrt{42}}, -\frac{2}{3\sqrt{14}}\right)\, ,
    \end{split} 
\end{equation}
General expectations from the analysis in the previous sections allow us to deduce that the species polytope should have vertices lying on the Laplace spheres. We see in Figure \ref{fig:8dlaplballs} that this is indeed the case. To identify one face of the hull, let us focus on the fundamental region $y>z, \ x>0$. \footnote{This is a fundamental domain with respect to the subgroup $S_2 \times \bZ_2 \subset S_2 \times S_3 $.} The Laplace equation \eqref{eq:9dlaplacecircle} for the nine-dimensional theory can be here plotted as a cylinder orthogonal to the $xz$ plane.  Doing so, one uniquely identifies the $M-S'-S''$ face by intersecting the 8d Laplace equation with the projected 9d hull inside the cylinder in the chosen duality frame. One can then use the $S_3\times S_2$ permutations to reconstruct the entire hull.

Thus, we have seen that knowledge of the Laplace equation in higher dimensions allows us to reconstruct the species polytope, corroborating the fact that the procedure can be extended to lower dimensions inductively. In the following, we will refrain from doing this, and instead turn to the case of half-maximal supergravity in $d=10,9$.

\subsection{The case of 16 supercharges}\label{ss:16sugra}

To end the section, we will briefly discuss some known results for minimal supergravity in ten and nine dimensions, closely following \cite{Green:2016tfs, vandeHeisteeg:2023dlw}. As will become clear shortly, the relevant Wilson coefficients resemble the maximal supergravity ones, allowing us to engineer proper differential equations for them, similarly to the rest of the chapter. 

In ten dimensions, at the eight-derivative level, there are three possible contributions to the effective action, given by $t_8t_8 \cR^4, \ t_8 {\rm{tr}}\cR^4$ and $ t_8({\rm{tr}}\cR^2)^2$,\footnote{Again, we will use $\cR$ to denote the Riemann curvature. The trace is taken over the tangent space $SO(9,1)$. } due to the more relaxed constraints on the kinematical structure of four-graviton scattering amplitudes in theories with 16 supercharges. As stressed in \cite{Green:2016tfs, vandeHeisteeg:2023dlw}, the Wilson coefficients of these operators suffer from an ambiguity due to the relation 
\begin{equation}
    t_8t_8\cR^4 - 24t_8 {\rm{tr}}R^4 + 6 ({\rm{tr}}R^2)^2 = 0.
\end{equation}
A natural choice of basis, which we will adopt in the following, is the one given by
\begin{equation}
    t_8t_8 \cR^4\, , \quad t_8 ({\rm{tr}}\cR^2)^2\,  , \quad \cI_{\rm anom}=24 t_8 {\rm{tr}}\cR^4 + 18 t_8 ({\rm{tr}}\cR^2)^2 + \frac{1}{4}\epsilon_{10}\epsilon_{10}\cR^4\, ,
\end{equation}
where $\epsilon_{10}$ denotes the 10d Levi-Civita symbol. This last operator is related by supersymmetry to the the anomaly cancelling term, $-12\epsilon_{10}Y_8$, through a combination of superinvariants $\cS$:
\begin{equation}
    \cS = \cI_{\rm anom}  -12\epsilon_{10}Y_8\ .
\end{equation}
Thus, because of anomaly cancellation and supersymmetry, the whole combination $\cS$ is expected to not receive corrections beyond one-loop, and neither is $\cI_{\rm anom}$. The operator $t_8({\rm tr}\cR^2)^2$ was instead chosen as it is part of the expression $t_8({\rm tr}{\cal F}^2 - {\rm tr}\cR^2)^2$ when the gauge fields are included. These are related by supersymmetry to the tree-level term $t_8({\rm tr}{\cal F}^2 - {\rm tr}\cR^2)$, which again receives no corrections. The analysis in \cite{Green:2016tfs} shows that even though there is an ambiguity in the coefficients of the single operators, there is no ambiguity in the effective action expressed in terms of superinvariants. It is interesting to notice that with minimal supersymmetry, the $t_8t_8\cR^4$ term is not $1/2$-BPS, a property which guaranteed no corrections beyond one-loop in the maximal supergravity case. Even though this protection is absent here, there is evidence that no higher-loop corrections are present \cite{Green:2016tfs}. The cause of this can be traced back to the amplitude related to this operator not being sensitive to the Ho\v{r}ava-Witten walls, i.e., to the boundaries of the $\mathbf{S}^1/\bZ_2$ interval in M-theory, thus effectively restoring the same cancellations present in the 32 supercharges computation. This, together with the fact that \cite{vandeHeisteeg:2023dlw} argued this operator to properly capture the species scale in these theories, motivates us to focus on its Wilson coefficient.

\subsubsection{10d Heterotic and Type I}

In every minimal supergravity in ten dimensions, the moduli space is simply given by the dilaton $g_s$ of each theory. The effective action of the HO theory is the same as Type I, allowing us to consider only the string coupling of the HE and HO theories, which we will call respectively $g_{\rm e}$ and $g_{\rm o}$. Denoting by $c_i$ order one factors, the Wilson coefficients $\cF$ of the $t_8t_8 \cR^4$ operators computed in \cite{Green:2016tfs}---including non-perturbative contributions---read
\begin{equation}\label{eq:16sugrawilson}
\begin{split}
    \cF_{{\rm HE}} &= c_1\, g_{\rm e}^{-3/2} + c_2\, g_{\rm e}^{1/2}\, ,\\
    \cF_{{\rm HO}} &= c_3 \,E_{\frac32}(ig_{\rm o}^{-1})\, .
\end{split}
\end{equation}
Notice that they are functionally analogous to the Type IIA and IIB coefficients, respectively.\footnote{Upon restricting $\tau_{\small\rm IIB} =ig_{\small\rm IIB}^{-1}$.} This is a reflection of the fact that the strong coupling limit of the HE theory is a decompactification limit to Ho\v{r}ava-Witten M-theory,  while the strong coupling limit of the HO theory is a weak coupling limit of Type I, making the presence of the instanton terms in $E_{3/2}$ required. This, together with the independence of the computation from the Ho\v{r}ava-Witten walls themselves, gives us an intuitive understanding of this result. Thus, as in the maximal supergravity case, one has that these coefficients satisfy
\begin{equation}\label{eq:16superchargesop1}
    g^2 \partial_g^2\cF = \frac34\cF\, , 
\end{equation}
which involves a second order differential operator that is different from the scalar Laplacian.

\subsubsection{$SO(16)\times SO(16)$ in 9d}

The compactification of the HE and HO-Type I theories on a circle involves the introduction of Wilson lines, complicating the analysis with respect to the maximal supergravity case. The resulting nine dimensional theory has a moduli space given by
\begin{equation}\label{eq:9d16modulisp}
    \cM = \bR \times SO(17,1;\bZ)\backslash SO(17,1)/SO(17)\, ,
\end{equation}
where the $\bR$ factor is parametrized again by the dilaton, while the rest is given by the radius of the circle and the $16$ Wilson lines for the heterotic gauge fields. The computation of the higher-derivative coefficients, and in particular their moduli-dependence, is thus hard, and it has been done only for specific slices of moduli space. For simplicity, and to match the discussion in \cite{Green:2016tfs,vandeHeisteeg:2023dlw}, we will focus on the case where the Wilson lines break the $SO(32)$ and $E_8\times E_8$ groups in $SO(16)\times SO(16)$, wherein the two parent theories can be related via T-duality \cite{Ginsparg:1986bx} along the extra $\mathbf{S}^1$. S-duality then relates the HE theory to Type I with Wilson lines again breaking the gauge group to $SO(16)\times SO(16)$. From this, another T-duality along $\mathbf{S}^1$ brings us to Type I', completing the duality chain. In this special case, the nine-dimensional supergravity theory thus captures the intricate web of string dualities. Focusing on this example also allows to ignore decompactifications to running solutions or the phenomenon of sliding in the tower and species polytopes \cite{Etheredge:2023odp}.

Let us focus on the HO parametrization of the amplitude. Denoting by $r$ the $\mathbf{S}^1$ radius in string units, the Wilson coefficient, again computed in \cite{Green:2016tfs}, depends on the moduli as 
\begin{equation}
    \cF_{{\rm SO(16)\times SO(16)}} = c_1 \left( \frac{r}{g_{{\rm o}}}\right)^{6/7} \left( E_{\frac32}(i g_{{\rm o}}^{-1}) + c_2\, g_{{\rm o}}^{1/2}\right)\, ,
\end{equation}
with $c_i$ order one coefficients. Remembering the 9d maximal supergravity expression 
\begin{equation}
\cF_{9\td, {\rm max}} = \hat c_1 {\cal V}^{-9/14} \left(E_{3/2}(\tau) +\hat c_2 \,{\cal {V}} ^{21/14} \right)\, ,
\end{equation}
one can see that it agrees with the 16 supercharges computation upon identifying
\begin{equation}
    \hat \tau=ig_{{\rm o}}^{-1}\, , \quad  {\hat \cV}^{-3/4} =  r \, g_{{\rm o}}^{-1/4}\,.
\end{equation}
Thus, in terms of these variables, the coefficient obeys a differential equation given by 
\begin{equation}\label{eq:16superchargesop2}
    \cD^2 \cF =  \frac67 \cF , \quad \cD^2 = \hat \tau^2 \partial_{\hat \tau}^2 +\frac79 \hat \cV \partial_{\hat \cV}(\hat \cV \partial_{\hat \cV}) + \frac13\hat \cV \partial_{\hat \cV}\, .
\end{equation}
Again, notice that the operator is not the Laplacian, at least on this slice of moduli space. The knowledge of the full Laplacian and higher derivative coefficients away from the latter is rather complicated due to the large amount of moduli, but as will be discussed in Section \ref{s:Laplace&Symm}, we do not expect this Wilson coefficient to obey an eigenvalue equation in terms of it. This is due to the fact that, similarly to the Type IIA case, the naive shift symmetry associated to the dilaton factor in \eqref{eq:9d16modulisp} is broken already within the two-derivative action.

\subsection{General comments}\label{ss:maxsugracomments}

Before continuing to analyze further lower-dimensional examples, let us make a few comments on the results obtained so far regarding the maximal and  half-maximal supergravity setting. As we just showed, the relevant Wilson coefficient in all these cases satisfies an eigenvalue equation \eqref{eq:eigenvalueeqintro} in terms of an appropriate elliptic second order operator. This turned out being the Laplacian whenever the moduli space $\cM$ is a coset space with a group of isometries that can be read off from the kinetic terms in the scalar sector, which are moreover realized as exact symmetries of the classical low-energy action. The exceptions always displayed, instead, an $\bR \subset \cM$ subset not associated with any such symmetry, which was ultimately identified with the dilaton direction in both Type IIA and the other examples with 16 supercharges. As will be explored further in Section \ref{s:Laplace&Symm}, one can translate this property into a precise group-theoretic statement by recalling that the Laplacian can be regarded as the Casimir of the isometry group in field space.

The inductive procedure from which one is able to reconstruct the species hull by using the Laplace equation in higher dimensions sets up a natural generalization of the reasoning of Section \ref{s:maxsugra} to those cases with $d\le 7$. One would then expect, similarly to $d\geq8$, that the species vectors describe a polytope whose outer vertices lie precisely on the $|{\cal W} (G_d(\bZ))|$ Laplace surfaces that define the constraint across duality frames.\footnote{$|{\cal W} (G_d(\bZ))|$ denotes the order of the Weyl subroup of the U-duality group, which connects the Laplace constraint between different duality frames, as previously seen.} However, since there the operator becomes irrelevant and the eigenvalue negative, one is presumably not able to extract any upper bound to the norm of the $\mathcal{Z}-$vectors. Moreover, in these regimes it is generically expected that in $d<7$ \textit{field-theoretic}, rather than genuine quantum gravitational effects, dominate the Wilson coefficient\cite{Castellano:2023aum,Calderon-Infante:2025ldq} for some of its limits. For these reasons, we stopped at the marginal $d=8$ case. Remarkably, a differential equation still holds in lower dimensional maximal supergravity, as per U-duality \cite{Green:2010wi}. It is also worth pointing out that after having identified the species hull using the eigenvalue equation, it is possible to use the pattern \cite{Castellano:2021mmx, Castellano:2022bvr} to derive the tower hull. In fact, as discussed in Section \ref{s:speciesreview}, this relation can be seen as the geometrical statement that the polytopes defining these two hulls are dual to each other with respect to a ball of radius $1/\sqrt{d-2}$.

At this point, it instructive to look at the dimension-dependent eigenvalues and ask whether one is able to distinguish which kind of ingredients determine their structure. Given that we restricted ourselves in this section to toroidal compactifications, it is interesting to notice that $(11-d)$ is precisely the number of circle---or interval, in the case of Ho\v{r}ava-Witten theory---moduli of the M-theory compactification, while $(d-8)$ is related instead to the classical mass dimension of the operator. Schematically, we thus have
\begin{equation}
    \eta_d = \frac{3}{d-2} (\text{\# of moduli}) \times (\text{mass dimension)} \ ,
\end{equation}
where it is also possible to rewrite the piece associated to the number of M-theory moduli using the length of the M-theory species vector as in \eqref{eigenotra}. Therefore, we generically expect the eigenvalue to depend both on the operator dimension as well as on the number of moduli. In subsequent parts of this paper, we will confirm that this is indeed the case.

\section{Setups with 8 Supercharges}\label{s:8supercharges}

In this section, we consider theories  exhibiting less amount of supersymmetry, and focus on those string-derived settings which preserve eight supercharges. In particular, we consider $\cN=(1,0)$ theories in 6d, $\cN=1$ theories in 5d, and 4d $\cN=2$ theories, which all share the same (locally) factorized moduli space structure $\cM=\cM_H \times \cM_{T,V}$, 
with $H,\,T$ and $V$ denoting the hyper, tensor and vector multiplet sectors, respectively. Having restricted supersymmetry implies that there are important modifications in the set of possible higher-derivative corrections, compared to the (half-)maximally supersymmetric case. First, these theories now allow for a non-trivial curvature-squared piece, which is moreover $\frac12$-BPS and depends only on the fields parametrizing $\cM_{T,V}$ \cite{Antoniadis:1993ze, Antoniadis:1995zn}. Secondly, most of the other higher-curvature terms---like the $\cR^4$-operator considered in Section \ref{s:maxsugra}---are not protected from receiving certain quantum corrections, making their systematic study significantly more challenging. For all these reasons, we will henceforth focus on the Wilson coefficient associated to the $\cR^2$-term, which is marginal (in the Wilsonian sense) in 4d and relevant in five and six dimensions. 

In the context of M-theory compactified on a Calabi--Yau threefold $Y_3$, the moduli dependence of the $\cR^2$-operator can be deduced by dimensionally reducing the analogous 11d fourth curvature invariant using (we neglect numerical factors, which can be found in \cite{Grimm:2013gma,Grimm:2017okk})\footnote{Similarly to the previous section, we will use the shorthand notation $\cR^2$ to refer to the four-derivative operator $\text{tr}\, \cR^2:= \mathrm{tr\,}\int \mathcal{R}\wedge \star \mathcal{R} =-\frac{1}{16}R_{\mu \nu \rho \sigma}R^{\mu \nu \rho \sigma}$.}
\begin{equation}\label{eq:r4tor2}
    t_8t_8\cR^4 \sim\, {\rm{tr}}\, \cR^2 \left(c_2\wedge J\right)\, +\, \text{(other terms)}\, ,
\end{equation}
where $J$ is the K\"ahler 2-form on $Y_3$ and $c_2(TY_3)$ denotes the second Chern class of its tangent bundle. Relatedly, in 6d a similar result follows from considering an elliptically-fibered Calabi--Yau and taking the F-theory limit of \eqref{eq:r4tor2} \cite{vandeHeisteeg:2023dlw}, while in four dimensions the same coefficient is determined by the genus-1 closed string topological free energy \cite{Witten:1988xj,Bershadsky:1993ta,Bershadsky:1993cx, Antoniadis:1993ze,Antoniadis:1995zn}, or rather by compactifying the previous expression on an extra circle. In all such cases, one finds that the corresponding Wilson coefficient reduces (asymptotically) to a linear function
\begin{equation}\label{eq:r2coeffasympt}
    \cF_{\cR^2} \sim c_{A}T^A \ ,
\end{equation}
with $T^A$ belonging to the vector/tensor multiplets, and $\{c_{A}\}$ being related to the topology of the underlying threefold. Let us remark though, that there could still be a priori further (non-)perturbative contributions to $\cF_{\cR^2}$,\footnote{For instance, in 4d a topological string computation at genus one reveals other perturbative contributions logarithmic in the moduli, as well as worldsheet-instanton corrections \cite{Bershadsky:1993ta,Bershadsky:1993cx,Grimm:2007tm}. Interestingly, the latter are seen to be absent in five and six dimensions.} and thus the expression above is only meant to capture the leading-order behavior close to the large volume point. 

In the rest of this chapter, we will argue that the aforementioned moduli-dependent function obeys an eigenvalue equation defined in terms of a suitable elliptic operator, which in four and six dimensions turns out being the corresponding scalar Laplacian. For the former case, we also provide more details on the link between the species convex hull and the constant term ansatz (cf. Section \ref{ss:4dN=2}), mirroring our previous discussion in maximal supergravity. Instead, in 5d (Section \ref{ss:5dMtheory}) and 6d (Section \ref{ss:6dFthy}) we focus on the top-down derivation of the differential condition, even though we also comment on the constraints imposed by the latter.

\subsection{Type IIA on a Calabi--Yau threefold}\label{ss:4dN=2}

Let us first consider 4d $\cN=2$ EFTs arising from Type IIA string theory compactified on a Calabi--Yau threefold $Y_3$. The bosonic part of the two-derivative action associated to the kinetic terms of the scalar and spin-2 fields in the Einstein frame reads
\begin{equation}\label{eq:4dIIAaction}
    S_{4\text{d}} = \frac{1}{2\kappa_4^2} \int \left( R\star 1 - 2g_{a\bar b} \, \td z^a \wedge \star \,\td \bar z^{\bar b} -2h_{uv }\, \td \xi^{u} \wedge \star \, \td \xi^v  \right)\, ,
\end{equation}
where $\{ z^a =b^a + t^a\} \, , \ a=1,\dots,h^{1,1}(Y_3)$, are complex coordinates parametrizing the K\"ahler deformations with $b^a\sim b^a+1$ being periodic, whilst the hypermultiplet moduli (including the complex structure fields) are described by $\{ \xi^u\}$. In the following, we focus on the vector multiplet sector, which exhibits a projective special K\"ahler metric $g_{a \bar b}=\partial_a \partial_{\bar b}K_{\text{ks}}$ and whose K\"ahler potential is determined at large volume by the triple intersection numbers $\cK_{abc}$ of $Y_3$
\begin{equation}
    K_\text{ks} = - \log\left( \frac{i}{6}\cK_{abc}(z-\bar z)^a (z-\bar z)^b (z-\bar z)^c\right) \, . 
\end{equation}
The latter can be written entirely in terms of the K\"ahler form as
\begin{equation}
    K_\text{ks} = - \log \left( \frac{1}{6}\int_{Y_3}J \wedge J \wedge J \right)\, ,
\end{equation}
by using the decomposition $J=t^a\omega_a$ in terms of an appropriate basis $\{\omega_a\}$ of $H^{1,1}(Y_3, \mathbb{Z})$.\footnote{For simplicity, in this work we always consider a basis of Nef divisors.} Being a complex manifold, the scalar Laplacian defined on this space adopts the simple form
\begin{equation}\label{eq:laplacian4d}
    \Delta_{\rm 4d} = 2g^{a\bar b} \partial_a \partial_{\bar b}\, .
\end{equation}
Here, we argue that the solutions of the Laplace equation constructed with $\Delta_{\rm 4d}$ capture the correct moduli dependence of the $\cR^2$-operator, normalized (in the Einstein frame) as
\begin{equation}
    S_{\cR^2} = \frac{1}{2} \int \td^4x \sqrt{-g} \, \cF(z,\bar z) \,\cR^2 \, ,
\end{equation}
Noticing that the aforementioned operator is marginal in four dimensions, and building on the intuition gained from the maximally supersymmetric setup (cf. eq. \eqref{eigenotra}), we are led to propose that the eigenvalue of the sought-after Laplace equation is zero in the present case. Hence, the equation thus obtained would read\footnote{This differential condition is related not only to the holomorphic anomaly equation as will be discussed in more detail below, but also resembles the `harmonicity equations' presented in \cite{Berkovits:1994vy, Berkovits:1998ex, Ooguri:1995cp}.}
\begin{equation}\label{eq:lapl4d}
    \Delta_{\rm 4d} \cF(z,\bar z) = 0\, .
\end{equation}
which is indeed satisfied in string theory up to  a subtle holomorphic anomaly that is related to threshold contributions associated to massless modes within the theory, see below. We show this first in full generality in Section \ref{sss:gencase4d}, paying special attention to various salient features. Subsequently, we illustrate the main points of the discussion by analyzing two detailed yet simple examples, namely a one modulus setup (Section \ref{sss:onemodexample}) as well as Type IIA string theory compactified on $\mathbb{P}^{1,1,1,6,9}[18]$ (Section \ref{sss:IIAonP11169}). 

\subsubsection{The general case}\label{sss:gencase4d}

Let us start by studying \eqref{eq:lapl4d} close to the large volume point, where we can organize our theory in terms of a classical (i.e., tree-level) contribution plus a series of quantum---in the worldsheet theory---stringy corrections. Hence, we consider a generic vector multiplet moduli space and we approximate the K\"ahler potential by its leading-order piece 
\begin{equation}\label{eq:polynomialkahlerpot}
    \cK = -\log \cG(t)\, ,
\end{equation}
with $\cG$ being an homogeneous function of degree $w$ in the saxions $\{t^a\}$ \cite{Grimm:2018cpv}. The metric can be then obtained by taking derivatives of \eqref{eq:polynomialkahlerpot} with respect to the K\"ahler fields, whereas its inverse may be written in general as follows
\begin{equation}\label{eq:inversemetricLV}
    g^{a \bar b} = \frac{4}{w-1}\left( t^a t^b - \frac{1}{w}\cG\cG^{ab}\right)\, ,
\end{equation}
where we introduced the quantity $\cG^{ab}$, which verifies $\cG^{ab}\partial_b\partial_c\cG/w(w-1) = \delta^a_c$. Thus, any function $\cF(z,\bar z)$ solving the Laplace equation on the \emph{full} complex moduli space must satisfy the differential condition
\begin{equation}\label{eq:laplace4dv2}
    \Delta_{\rm 4d} \cF(z,\bar z) = 2g^{a\bar b}\partial_a \partial_{\bar b}\cF(z,\bar z) =0\, .
\end{equation}
Notice that the metric \eqref{eq:inversemetricLV} does not depend explicitly on the compact directions $b^a$, such that $b^a \mapsto b^a+\lambda^a$ with $\lambda^a \in \bR$ becomes an approximate global symmetry of the theory. Indeed, in string compactifications it is a common feature that infinite distance limits restore some perturbative shift symmetry in terms of the axionic fields \cite{Corvilain:2018lgw}. We can thus parametrize asymptotic solutions to \eqref{eq:laplace4dv2} by imposing invariance under such shift transformations on the general form of $\cF$, mimicking the enhancement from a discrete to a continuous global symmetry as we reach infinite distance boundaries in K\"ahler moduli space. Furthermore, connecting with the logic followed in previous sections, we note that this is completely equivalent to imposing the existence of `constant terms' in the Fourier decomposition with respect to the periodic fields $b^a$. Taking this into account allows us to reduce the above equation to
\begin{equation}\label{eq:constantterm4d}
    \Delta_{\rm 4d} \cF_{\rm const} (t) = \frac12 g^{a \bar b}\partial_a \partial_b \cF_{\rm const} (t) = 0\, .
\end{equation}
To solve this, we propose a monomial ansatz of the form
\begin{equation}
    \cF_{\rm const}(t) \propto \prod_a \left( t^a \right)^{n_a}\, ,
\end{equation}
which means that $\cF_{\rm const}$ is homogeneous of degree $w_f = \sum_a n_a$. Using homogeneity to write
\begin{equation}
    \partial_a \partial_b\cF_{\rm const} (t) = \frac{1}{t^a t^b }(n_an_b - n_a \delta_{ab}) \cF_{\rm const}(t)\, ,
\end{equation}
such that
\begin{equation}
    \Delta_{\rm 4d}\cF_{\rm const}(t) = \frac{2}{w-1}\sum_{a,b}\left(1 - \frac{1}{w}\frac{\cG \cG^{ab}}{t^at^b}\right)\big(n_an_b - n_a \delta_{ab}\big)  \cF_{\rm const}\, ,
\end{equation}
leads us to the simpler condition
\begin{equation}\label{eq:lapl4dgeneralcase}
    \frac{\Delta \cF_{\rm const}}{\cF_{\rm const}} = \sum_{a,b}\cC_{ab}^{(0)} \big( n_a n_b - n_a\delta_{ab}\big)\, ,
\end{equation}
with the $\cC^{(0)}$ being homogeneous functions of degree zero in the saxions. One then immediately realizes that the only non-constant solutions are
\begin{equation}
    n_{\hat a} = 1, \quad n_{a\ne \hat a} =0 \quad \Longrightarrow \quad \cF_{\rm const}(t) = c_{2,a} t^a\, , \qquad  \frac{\LQG}{\Mpf} \sim \frac{1}{\sqrt{c_{2,a} t^a}}\, ,
\end{equation}
implying that the species function must be linear at large volume. This matches precisely the behavior observed in string theory \cite{Antoniadis:1993ze,Antoniadis:1995zn}, where the coefficients $c_{2,a}$ are related to the second Chern class of the underlying Calabi--Yau, and where such term provides the leading-order piece for the genus-1 topological free energy arising from the constant map contribution \cite{Bershadsky:1993ta}. 

More generally, one may notice that the previous discussion can be easily extended to the bulk of the moduli space, namely away from infinite distance loci. In fact, from \eqref{eq:laplace4dv2} it readily follows that the exact real function $\cF(z,\bar z)$ must be comprised by a purely holomorphic piece plus its complex conjugate. This result is intimately connected with the $\mathcal{N}= 2$ algebra of the 2d superconformal worldsheet theory \cite{Cecotti:1992vy}, which actually suffers from a subtle anomaly mixing the holomorphic and anti-holomorphic sectors \cite{Bershadsky:1993ta} according to the relation
\begin{equation}\label{eq:holan}
    \partial_a \partial_{\bar b}{\cF(z,\bar z)} = \tr(-1)^F C_a \bar C_{\bar b} + g_{a\bar b}\,\tr(-1)^F\, ,
\end{equation}
with $F$ denoting the fermion number and $C_a$ are the structure constants of the chiral ring. From the spacetime perspective, this anomaly accounts for threshold corrections associated to loops of massless fields within the theory, which may alter the Laplace equation considered herein. Indeed, solving \eqref{eq:holan} leads to \cite{Bershadsky:1993ta}
\begin{equation}
    \cF(z,\bar z) = \frac12 \left( 3+h^{1,1}-\frac{\chi_E(Y_3)}{12}\right) K_{\rm ks} +\frac12 \log \det g_{a \bar b} +\log |f|^2\, ,
\end{equation}
with $\chi_E(Y_3)= 2(h^{1,1}-h^{2,1})$ being the Euler characteristic of $Y_3$, and whose Laplacian yields 
\begin{equation}\label{eq:laplacian4dandcurvature}
    \Delta_{\rm 4d} \cF(z,\bar z) = h^{1,1} \left( 3+h^{1,1}-\frac{\chi_E(Y_3)}{12}\right)+\frac12 \mathsf{R}_{\rm mod}\, ,
\end{equation}
thereby introducing a source-term contribution in the right-hand side of \eqref{eq:lapl4d} that depends on the scalar curvature of moduli space $\mathsf{R}_{\rm mod}= 2 g^{a \bar b}\mathsf{R}_{a \bar b}$. Notice the resemblance with the situation in 8d maximal supergravity, where a similar---this time constant---term appears in the Laplace condition for the $\mathcal{R}^4$-operator (cf. eq. \eqref{eq:laplace8dcsource}). The latter is also linked with would-be logarithmic divergences induced by the massless fields and, as already argued, never provides for the leading-order piece in the species function when evaluated at infinity. Here we find an even more interesting result, since the source term may actually vary over the moduli space and can even diverge at certain special loci \cite{Strominger:1995cz,trenner2010, Marchesano:2023thx, Marchesano:2024tod,Castellano:2024gwi, Blanco:2025qom, CMP}. In any event, the important point for us is that, barring the anomaly captured by \eqref{eq:holan}, holomorphy together with the discrete periodicity in the axions $b^a$ implies that the function $\cF(z,\bar z)$ must admit a Fourier decomposition with a `zero mode' which can only be linear in the K\"ahler saxions, plus a series of exponential terms that are suppressed at large volume. From our perspective, the former property would follow from imposing \eqref{eq:lapl4d} in the Wilsonian effective action.

\subsubsection{One modulus example}\label{sss:onemodexample}

To illustrate some of the features highlighted in the previous discussion, we consider in the following a toy-example with a single modulus $z = b+it$ having complex periodicity one. Near the large radius point, the metric reduces to that of the hyperbolic upper-half plane $\mathfrak{h}$
\begin{equation}
    ds^2_{\cM_V} = \frac{3}{t^2}(db^2 + dt^2) \, .
\end{equation}
As mentioned before, equation \eqref{eq:lapl4d} admits as generic solution a sum of any holomorphic function with its complex conjugate. However, expanding in a Fourier series and imposing the presence of a `constant term' contribution---which dominates at infinite distance and thus gives rise to a continuous symmetry enhancement---yields a function of the form
\begin{equation}\label{eq:4donemod}
    \cF(z,\bar z) = -\frac{i}{2}c_2 z+\mathcal{O}\left( e^{2\pi i z}\right)+\text{c.c.}\, ,
\end{equation}
where by holomorphy the exponential terms are suppressed and hence subdominant at infinite distance. This coincides with the asymptotic expression of the topological string free energy discussed in \cite{Bershadsky:1993ta}, modulo logarithmic non-holomorphic terms which arise due to the massless threshold corrections that we have neglected. The fact that this Wilson coefficient should match the genus-1 computation was originally discussed in \cite{Antoniadis:1993ze,Antoniadis:1995zn}, and the statement that one can interpret this higher-curvature correction as being controlled by the species scale leads to the asymptotic approximation 
\begin{equation}\label{eq:4dspecies}
    \LQG = \cF^{-1/2} \sim \sqrt{t}\, ,
\end{equation}
in agreement with the generic expectations and microscopic counting arguments \cite{vandeHeisteeg:2022btw, Castellano:2023aum}.

To be more concrete, and in order to highlight certain similarities with the maximal supergravity case, let us consider an explicit stringy embedding of this toy model. One of the few compact Calabi--Yau manifolds on which topological string computations have been carried out is the Enriques threefold \cite{Ferrara:1995yx}. This space can be seen as a free $\bZ_2$ involution of the product space $K3\times \mathbf{T}^2$, yielding a $K3$-fibration with Enriques fibers over the fixed points of the torus factor.\footnote{Interestingly, this Calabi--Yau is also self-mirror-dual.} The involution enlarges the holonomy group from $SU(2)$ to $SU(2)\times \bZ_2$, hence breaking $\cN=4$ supersymmetry down to $\cN=2$. Given this minimal modification, however, many of the computational simplifications associated to having a higher amount of supersymmetry remain. Among those, one useful property is the factorization of the genus-1 amplitude between base and fiber \cite{Grimm:2007tm}
\begin{equation}
    \mathcal{F}^{(1)}(S,\bar S, z,\bar z) = \mathcal{F}^{(1)}_{\rm base}(S,\bar S) + \mathcal{F}^{(1)}_{E}(z,\bar z)\, ,
\end{equation}
where $\mathcal{F}^{(1)}_{E}$ refers to the genus-1 computation in the fiber-limit. We call $S$ the K\"ahler modulus of the torus, while $z$ refers to all other K\"ahler moduli. Focusing on the former dependence, the genus-1 free energy reads
\begin{equation}
    \mathcal{F}^{(1)}(S,\bar S,z,\bar z) = - 6\log\left( \text{Im}(S)|\eta(S)|^4\right) + f(z, \bar z)\, ,
\end{equation}
where $\eta$ is the Dedekind function and $f$ contains all $z, \bar z$ dependence. At large Im$(s)$, one can use eq. \eqref{eq:asymptotic behavior} to show that it agrees with \eqref{eq:4donemod} since, forgetting about $z$-dependence
\begin{equation}
    \mathcal{F}^{(1)}(S,\bar S) \sim 2\pi \text{Im}(S) + \cO(\log \text{Im}(S))\, .
\end{equation}
Notice that the leading corrections are logarithmic in Im$(S)$, which appears due to the holomorphic anomaly of the topological string \cite{Bershadsky:1993cx}, while at higher-orders one can find non-perturbative corrections due to worldsheet instantons of the form $e^{-\text{Im}(S)}$. This confirms that the terms responsible for upgrading \eqref{eq:laplace4dv2} to \eqref{eq:laplacian4dandcurvature} indeed capture subleading corrections to the Wilson coefficient at large volume. In more involved examples, however, these contributions might diverge along certain infinite-distance limits, since they are proportional to the moduli-space curvature. Even in this case, it is consistent to neglect them for the computation of the quantum gravity cutoff, as the topological string is capturing field-theoretic degrees of freedom due to a rigid subsector of the theory decoupling from gravity. 

Finally, and connecting again with maximal supergravity, we notice that the exact function appearing in the full genus-1 computation is simply the regularized $SL(2,\bZ)$ Eisenstein series $\hat E_{1}^{sl_2}$, which is further discussed in Appendix \ref{ap:Massform}. Thus, the Laplace equation can be also related in this context to the $SL(2,\bZ)$ duality symmetry appearing in the torus factor of the compactification manifold.

\subsubsection{Two moduli example: $\mathbb{P}^{1,1,1,6,9}[18]$}\label{sss:IIAonP11169}

We study now a two-moduli theory obtained by reducing Type IIA on the threefold ${\bP}^{1,1,1,6,9}[18]$, which will allow us to make the discussion in Section \ref{sss:gencase4d} more concrete and connect with the species polytope, as done in the (half-)maximally supersymmetric setting. The compactification manifold can be seen as an elliptic fibration over a ${\bP^2}$ base with $h^{1,1}(Y_3) = 2$ K\"ahler moduli $z^1,z^2,$ giving rise to a K\"ahler potential of the form (see e.g., \cite{Castellano:2023jjt})
\beq 
e^{-K_\text{ks}} = 12(t^1)^3  + 12(t^1)^2t^2 + 4t^1(t^2)^2+\ldots\, .
\eeq
Restricting to the cubic piece of $K_\text{ks}$ leads to the inverse moduli space metric
\beq
 g^{-1}\, =\, \begin{pmatrix}
     \frac{(t^1)^2 \left(3 (t^1)^2+6 t^1 t^2+2 (t^2)^2\right)}{t^2 (3 t^1+t^2)} & -\frac{3 (t^1)^2 (t^1+t^2)}{t^2} \\
 -\frac{3 (t^1)^2 (t^1+t^2)}{t^2} & \frac{9 (t^1)^3}{t^2}+9 (t^1)^2+3 t^1 t^2+(t^2)^2 
 \end{pmatrix}\, .
\eeq
If we now impose the Laplace equation \eqref{eq:lapl4d} on the Wilson coefficient and follow the procedure outlined in Section \ref{sss:gencase4d}, one is able to write explicitly the degree-zero $\cC^{(0)}(t^1,t^2)$ functions appearing in \eqref{eq:lapl4dgeneralcase} and look for constraints on the pair of exponents $\{p,q\}$ defining the shift-symmetric ansatz $\cF_{\rm const} \sim (t^1)^p (t^2)^q$. Indeed, condition \eqref{eq:lapl4dgeneralcase} reads
\begin{equation}
    \frac{\Delta \cF_{\rm const}}{\cF_{\rm const}} = 2{pq}\, \cC_{pq}^{(0)}(t^1,t^2) + p(p-1)\,\cC_{pp}^{(0)}(t^1,t^2) + q(q-1)\,\cC_{qq}^{(0)}(t^1,t^2) \stackrel{!}{=}0\, ,
\end{equation}
where we defined
\begin{equation}\label{eq:twomodfun}
    \begin{cases}
        &\cC_{pq}^{(0)} = -\frac{1}{2(t^2)^2} \left[t^1 (t^1+t^2)\right]\\
        &\cC_{pp}^{(0)} =\frac{1}{t^2(3 t^1+t^2)}\,  \left[3 (t^1)^2+6 t^1 t^2+2 (t^2)^2\right]\\
        &\cC_{qq}^{(0)} = \frac{1}{(t^2)^2}\left[\frac{9 (t^1)^3}{t^2}+9 (t^1)^2+3 t^1 t^2+(t^2)^2\right] 
    \end{cases}\, ,
\end{equation}
which can be seen to be generically non-vanishing functions over moduli space.

Therefore, the only non-trivial dependence of $\cF_{\rm const}$ on the K\"ahler saxions is constrained to be linear, and the species scale agrees with the top-down computation\footnote{We remind the reader that the coefficients $c_{2,i}$ are microscopically related to the topology of $Y_3$, in particular to its second Chern class. For the present example, one has $c_2(\mathbb{P}^{1,1,1,6,9}[18])=102 [D_1]+36 [D_2]$ \cite{Morrison:1996pp}.}
\begin{equation}
    \frac{\LQG}{\Mpf} \sim \left( \frac{1}{\sqrt{c_{2,1} t^1 + c_{2,2}t^2}}\right)\, .
\end{equation}
The two possible limits precisely correspond to the $t^1 \gg t^2$ regime of decompactification to M-theory or $t^2 \gg t^1$ decompactification to F-theory \cite{Castellano:2023jjt}, which give, respectively 
\begin{equation}
    \frac{\LQG^{\text{(M-th)}}}{\Mpf} \sim \frac{1}{\sqrt{t^1}}\, , \qquad \frac{\LQG^{\text{(F-th)}}}{\Mpf} \sim \frac{1}{\sqrt{t^2}}\, .
\end{equation}
Correspondingly, the towers are a $p=1$ tower of D0-branes in the first case, i.e.,  the KK modes of the M-theory circle, and a $p=2$ effective tower of D0-D2 bound states \cite{Castellano:2021mmx,Castellano:2022bvr}. It is worth noticing, however, that the towers themselves do not satisfy in general the Laplace equation, since
\begin{equation}
    \frac{m_{\text{D0}}^{\text{(M-th)}}}{\Mpf} \sim \frac{1}{(t^1)^{3/2}}\,, \qquad \frac{m_{\text{D0}}^{\text{(F-th)}}}{\Mpf} \sim \frac{1}{\sqrt{t^1}t^2}\,, \qquad \frac{m_{\text{D2}}^{\text{(F-th)}}}{\Mpf} \sim \frac{\sqrt{t^1}}{t^2}\,.
\end{equation}
Next, the saxionic slices of moduli space are Riemann flat, and one can introduce a set of  globally-defined canonically normalized coordinates as (see \cite{Castellano:2024bna} for the exact expressions)
\begin{equation}
    \hat t_1 = \log(t^1)  + \cO\left( \frac{t^1}{t^2}\right), \quad \hat t_2= \frac{1}{\sqrt{2}}\log(t^2)+ \cO\left( \frac{t^1}{t^2}\right)\, ,
\end{equation}
to rewrite the $\cF$ function as a sum of exponentials, i.e.,
\begin{equation}
    \cF \sim c_{2,1}\, e^{\hat t_1} + c_{2,2}\, e^{{\sqrt{2}\hat t_2}}\, ,
\end{equation}
so that the Wilson coefficient diverges (and the species scale goes to zero) exponentially fast, as required by the Distance Conjecture. Given the form of the metric and the canonically normalized moduli, one can then compute the species hull vectors in the two limits discussed above, which yields
\begin{equation}
    \vec\cZ_{\text{M-th}} = \left( \frac{1}{3\sqrt{2}},\frac13\right)\, , \qquad \vec \cZ_{\text{F-th}} = \left(0,\frac12 \right)\, ,
\end{equation}
fully recovering the top-down result from the Laplace equation.

\subsubsection{Laplacian and scalar Weak Gravity Conjecture}

To conclude our discussion in four dimensions, we highlight a direct connection between the Laplace equation and certain scalar formulations of the Weak Gravity Conjecture (SWGC) that have been proposed in the literature \cite{Palti:2017elp,Lee:2018spm,Gonzalo:2019gjp,Gonzalo:2020kke,DallAgata:2020ino,Benakli:2020pkm,Andriot:2020lea,Etheredge:2022opl,Benakli:2022shq, Dudas:2023mmr, Etheredge:2023usk}.
For simplicity, let us consider the case of a single complex modulus $S= c+is$ in $4d,\  \cN=1$ supergravity with kinetic term
\beq
S_{\text{kin}}  =  \frac{1}{2\kappa_4^2} \int \td^4 x \sqrt{-g}\,\, \frac{\partial S \cdot \partial \bar S}{a^2 \, \rm{Im}(S)^2}\, ,
\eeq
with $a \in \bR$, such that one can introduce its associated canonically normalized saxion $\phi$ as $s=e^{a\phi}$. The Laplace constraint discussed above can be turned into a condition for the species scale through the relation \eqref{eq:4dspecies}, namely
\beq
\frac {|\partial_S \LQG|^2}{\LQG^2}  =  \frac {1}{d-1}  \frac {\Delta \LQG}{\LQG}\, ,
\label{species}
\eeq
where from this point onwards, $d$ is left explicit. Consider now a tower of states with characteristic mass $m_{\rm t}$ (in $4d$ Planck units), which generates this scale via species counting. The relation between the $m_{\rm t}$ and $\LQG$ is given by \cite{Castellano:2021mmx,Castellano:2022bvr}
\beq
\LQG = m_{\rm t}^{\frac{p}{d-2+p}}\, ,
\label{relacionlambda}
\eeq 
where $p$ parametrizes the effective tower degeneracies, so that e.g., for $p=1\  (\infty)$ one has a single Kaluza--Klein (string) tower.  Recall that, as already seen in Section \ref{sss:IIAonP11169}, towers do not, by themselves, satisfy the Laplace equation. However, knowledge of the states' degeneracies can be used to construct a suitable differential equation for their moduli-dependence. Indeed, plugging definition \eqref{relacionlambda} into \eqref{species}, one obtains 
\beq
\left(1+\frac {1}{\lambda}\right) \frac{|{\partial}_S m_{\rm t}|^2}{m_{\rm t}^2} - \frac {\Delta m_{\rm t}}{m_{\rm t}} = 0\, ,\qquad \lambda  =  \frac {(d-2+p)}{p(d-2)}\, .
\label{paramasas}
\eeq
Following the logic of the rest of the section, it is easy to check that this equation admits an asymptotic shift-symmetric solution of the form  $m_{\rm t} \sim  s^{-\lambda}$ which, when expressed in terms of the canonical field $\phi$, yields the typical exponential behavior $m_{\rm t}= e^{-a\lambda \phi}$ for the mass scale of the tower, compatibly with the Distance Conjecture. Furthermore, by using the purportedly universal pattern put forward in \cite{Castellano:2023stg, Castellano:2023jjt}, it is possible to constrain the parameters $a$ and $\lambda$ as
\begin{equation}\label{eq:constr}
   \vec \cZ_{\text{sp}} \cdot \vec \zeta_t = \, \frac{a^2 \lambda}{d-2} \stackrel{!}=\frac{1}{d-2} \ \ \implies  \ \   a^2 \lambda = 1 \ \Rightarrow \lambda_{\rm dec} = \sqrt{\lambda} \ ,
\end{equation}
where we defined $\lambda_{\text{dec}} \equiv a\lambda$ as the decay rate of the tower in the canonical frame. This agrees with the expected rate given in \cite{vandeHeisteeg:2023ubh} (see also \cite{Etheredge:2022opl,Calderon-Infante:2023ler}). Notice that with the above ingredients, one is able to saturate the SWGC proposed in \cite{Gonzalo:2020kke}, which for a single field in 4d reads\footnote{The factor of 1/2 is needed as the metric in \cite{Gonzalo:2020kke} corresponds to the K\"ahler line element.}
\beq
\frac{1}{2}g^{ss} \left| (\partial_s m_{\rm t}^2)^2   - m_{\rm t}^2 \partial_s^2 m_{\rm t}^2 \right| \geq  m_{\rm t}^4 \ .
\label{PPSWGC}
\eeq
Indeed, applying the constraint \eqref{eq:constr} to the asymptotic solution of \eqref{paramasas}, one is able to rewrite this differential equation in a purely algebraic fashion as
\beq 
a^2\left| 2\lambda^2 - \lambda(2\lambda-1) \right|= a^2 \lambda \stackrel{\eqref{eq:constr}}= 1 \ .
\eeq
To conclude, we point out that the combination of \eqref{PPSWGC} with the differential condition \eqref{paramasas} brings us to a bound on the norm of the gradient of the mass scale of the tower, namely
\beq
\frac{|\partial_S m_{\rm t}|^2}{m_{\rm t}^2} \ge \lambda \ ,
\eeq
which reproduces the `sharpened' SWGC proposed in \cite{Etheredge:2022opl}. Analogously to the preceding case, using constraint \eqref{eq:constr} saturates the bound for the solution to the Laplace equation.

\subsection{F-theory on an elliptic Calabi--Yau threefold}\label{ss:6dFthy}

In six dimensions, the field content of $\cN=(1,0)$ supergravity EFTs consists of the gravity multiplet, $H$ hypermultiplets, $T$ tensor multiplets and $V$ vector multiplets (see e.g., \cite{Taylor:2011wt}). In the following, we consider those theories that are obtained from compactifying F-theory on an elliptically fibered Calabi--Yau threefold \cite{Vafa:1996xn,Morrison:1996na,Morrison:1996pp}, and we focus on their tensor branch. The latter can be described entirely in terms of the K\"ahler 2-form $J_{B_2}$ associated to the base $B_2$ of the elliptic threefold. This object may be naturally regarded as a vector in $\bR^{1,T}$ satisfying the non-linear constraint
\begin{equation}\label{eq:6dconstraint}
    J_{B_2} \cdot J_{B_2} = \Omega_{IJ} J^I J^J=  1\, , 
\end{equation}
where $\Omega_{IJ}$ is a matrix of signature $(1,T)$ that defines a symmetric inner product in $\bR^{1,T}$. Introducing a basis of $h^{1,1}(B_2)$ 2-cycles $[C_I]$ within $B_2$, we can write the K\"ahler form as follows
\begin{equation}
    J_{B_2} = J^I [C_I] = J^0 [C_0] + J^i [C_i]\, ,\qquad i=1, \ldots, T=h^{1,1}(B_2)-1\, ,
\end{equation}
where $[C_0]$ is a distinguished shrinkable curve in the base which singles out the only kind of infinite distance degenerations that can occur in these theories, namely emergent string limits \cite{Lee:2018urn,Lee:2018spm,Lee:2019xtm}. Using these variables, the moduli space is given by the hypersurface 
\begin{equation}\label{eq:6dconstr}
     \mathfrak{F}_6 = \Omega_{I K}J^IJ^K = 1\, , \qquad \mathfrak{F}_6 \subset \frac{SO(1,T)}{SO(T)}\, .
\end{equation}
Notice that the constraint is expressed in terms of a $SO(1,T)$-invariant quadratic form, thereby preserving the isometries of the embedding space. Using a set $\{ \psi^\a\}$ of $h^{1,1}(B_2)-1$ independent fields that parametrize the hypersurface \eqref{eq:6dconstraint}, the relevant two-derivative action becomes \cite{Nishino:1997ff,Ferrara:1996wv,Bonetti:2011mw}
\begin{equation}\label{eq:6daction}
    S_{\text{6d}} = \frac{1}{2 \kappa_6^2} \int \left( R \star 1 - \frac12 g_{\a \b} \, d\psi^\a\wedge \star  d\psi^\b + \dots \right) \ , \quad \text{with}\ \ g_{\a \b} = \frac{\partial J^I}{\partial \psi^\a} \frac{\partial J^J}{\partial \psi^\b} \left( 2{J_I J_J}  - \Omega_{IJ}\right)\, ,
\end{equation}
with the indices lowered by $\Omega_{IJ}$ and where the kinetic term of the scalars have been pulled-back to the constrained moduli space. The ambient space, which lacks any such constraint, has a metric which can be written as
\begin{equation}\label{eq:ambientspacemetric6d}
    g_{IJ} = -\frac12 \partial_I \partial_J \log \fF_6\, . 
\end{equation}
Unlike in the four-dimensional case, for 6d $\cN=(1,0)$ supergravity the $\cR^2$-operator is not marginal. Therefore, according to the intuition gained from maximal supergravity in $d\geq8$, we are left with a choice not only of operator but also of eigenvalue, which in general may depend on the number of moduli, as per the discussion in Section \ref{ss:maxsugracomments}. To settle this, we adopt a top-down approach by focusing on the leading, higher-curvature Wilson coefficient. In particular, the relevant operator has been computed by considering first M-theory on the fibration $\pi: T_2 \hookrightarrow Y_3\to B_2$ and then uplifting the resulting theory to six dimensions \cite{vandeHeisteeg:2022btw,vandeHeisteeg:2023dlw}. Ignoring numerical prefactors, one obtains
\begin{equation}\label{eq:R26dFtheory}
    S_{\cR^2} = \frac{\Mps^2}{2} \int \td^6 x \sqrt{-g} \, \left( \int_{B_2}c_1(B_2) \wedge J\right)\, \mathrm{tr}\, \cR^2 .
\end{equation}
In the following, we will always assume $\int c_1(B_2) \wedge C_0  \ne 0$ to avoid finely tuned cases where the asymptotically tensionless string that emerges is identified with a Type II string with enhanced $\cN=4$ supersymmetry, which naturally provides stronger protection to the $\cR^2$-interaction. Additionally, we will leave the values of the second Chern class unspecified, taking the Wilson coefficient to be given by some linear function of the form
\begin{equation}
    \cF = c_{1,I} J^I\, .
\end{equation}
This will suffice to extract the corresponding eigenvalue. For simplicity, we consider here two classes of simple examples, which were also studied in \cite{vandeHeisteeg:2023dlw} within the species scale context.

\subsubsection{Laplacian on constrained manifolds}

The moduli space of 6d supergravity theories exhibits a coset structure with locally hyperbolic metric, defined by some slice of $SO(1,T)/SO(T)$ as per \eqref{eq:6dconstr}. However, instead of finding an explicit parametrization of this hypersurface and computing its associated Laplacian, it turns out to be easier to relate the Laplace operator on the ambient space with the one induced on the constrained submanifold. Therefore, given a function $f$, it is a standard result that \cite{ecker2004regularity}
\begin{equation}\label{eq:laplonconstr}
    \Delta_t f = \Delta_Xf - \text{Hess}_f(\mathbf{n},\mathbf{n}) + \mathbf{H}_{\mathscr{M}} \cdot df\, ,
\end{equation}
where $\mathbf{n}$ is any unit normal vector to the codim-1 submanifold $\mathscr{M}$, $\Delta_t$ denotes its Laplacian, $\Delta_X$ the corresponding one defined in the ambient space $\mathscr{N}$, $\text{Hess}_f = \nabla_I \partial_J f dX^I dX^J$ is the Hessian 2-form, and $\mathbf{H}_{\mathscr{M}}$ the mean curvature vector of the submanifold with respect to $\mathscr{N}$. The latter is defined as
\begin{equation}\label{eq:meancuarvature}
    (\mathbf{H}_{\mathscr{M}})^L = -(\nabla_K n^K)n^L\, .
\end{equation}
The above result can be easily recovered by rewriting the Laplacian using the projector onto the hypersurface $\Pi^I_J=\delta^I_J -n^In_J$. In particular, upon applying this operator to the 1-form $\partial_If$ and taking into account that both projected and induced metrics agree (imposing \eqref{eq:6dconstraint}), one arrives at eq. \eqref{eq:laplonconstr}. In addition, a simplifying feature of the type of submanifolds considered herein is that their mean curvature vanishes identically, as shown in Appendix \ref{ap:meancurv}. This relies on the fact that the vector $\mathbf{n}$ can be simply expressed as
\begin{equation}
     n^L = g^{LK}\frac{\partial_K \mathfrak{F}_6}{|\partial \mathfrak{F}_6|}\, , \qquad \text{with}\quad |\partial \mathfrak{F}_6|= \sqrt{g^{IJ} \partial_I \mathfrak{F}_6 \partial_J \mathfrak{F}_6}\, .
\end{equation}
Note that from here it also follows that the scalar curvature of the ambient space is identified with that of the submanifold. Therefore, for the case at hand eq. \eqref{eq:laplonconstr} reduces to
\begin{equation}\label{eq:laplacianrelation}
     \Delta_t f = \Delta_X f - \,\text{Hess}_f(\mathbf{n},\mathbf{n})\, .
\end{equation}
Moreover, in the 6d models of interest, the inverse metric on ambient space reads
\begin{equation}\label{eq:invmetr6d}
    g^{IJ} = 2J^I J^J -\mathfrak{F}_6 \,\Omega^{I J}\, ,
\end{equation}
where $\Omega^{IJ}$ denotes the inverse matrix of the pairing form appearing in $\mathfrak{F}_6 =  \Omega_{IK}J^I J^K$. Consequently, we find that the extra contribution relating the Laplacian on the full space and the one defined on the constrained submanifold is (cf. Appendix \ref{ap:meancurv})
\begin{equation}
    \text{Hess}_f(\mathbf{n},\mathbf{n}) =f\, ,
\end{equation}
for any \emph{linear} function in ambient space, thereby lowering the eigenvalues of $\Delta_X$ by one unit. We thus conclude that indeed the higher-curvature correction \eqref{eq:R26dFtheory} obeys an equation in the ambient space of the kind
\begin{equation}
    \Delta_J \cF = (\lambda + 1) \cF\, ,\qquad \text{with}\quad \Delta_\psi \cF = \lambda \cF,\ \lambda \in \mathbb{Z}_{\geq 0}\, .
\end{equation}

\subsubsection{From examples to the general case}\label{sss:examples6d}

To compute the eigenvalue of the Laplacian acting on $\cF(J)$, we first consider two families of concrete examples, following the analysis in \cite{vandeHeisteeg:2023dlw,Morrison:1996na}. Subsequently, we generalize the results for any 6d model obtained from compactifying F-theory on a Calabi--Yau threefold.

\paragraph{F-theory on del Pezzo surfaces.} If $B_2 = dP_r$, i.e.,  blowups of $\mathbb{P}^2$ in $r$ generic points, then the prepotential can be written as
\begin{equation}\label{eq:6dprepotconstraint}
    \fF_6 = (J^0)^2 - \sum_{i=1}^r (J^i)^2 \stackrel{!}{=}1\, ,
\end{equation}
where we used as basis of curves the hyperplane class $[C_0]$ inherited from $\bP^2$ and the exceptional divisors $[C_i]$ \cite{Donagi:2004ia}. Taking $r=1$ as an illustrative example, the metric \eqref{eq:ambientspacemetric6d} reads
\begin{equation}
   g =  \frac{1}{\left[(J^0)^2-(J^1)^2\right]^2}\left(
\begin{array}{cc}
 (J^0)^2+(J^1)^2 & -2 J^0 J^1 \\
 -2 J^0 J^1 & (J^0)^2+(J^1)^2 \\
\end{array}
\right)\, ,
\end{equation}
such that upon computing the Laplacian, we find that any linear function in $J^I$ obeys an eigenvalue equation in ambient space of the type\footnote{\label{fnote:chernclassdP}For this class of examples, we have $c_1(dP_r)= 3 [C_0]-\sum_i [C_i]$, thus leading to $\cF_{dP_r}=3J^0-\sum_i J^i$ \cite{vandeHeisteeg:2023dlw}.} 
\begin{equation}
    \Delta_J \cF = 2 \cF \ \ \Longrightarrow \ \  \lambda=1\, .
\end{equation}
Repeating the same calculation for other surfaces characterized by the rank, we simply have
\begin{equation}
    \Delta _J \cF = (\lambda+1) \cF = (r+1) \cF\, .
\end{equation}

\paragraph{F-theory on Hirzeburch surfaces.} In this case, $B_2 = \mathbb{F}_n$, where $n=1,\dots,8,12$ \cite{Morrison:1996pp}. The prepotential for the two moduli $f,h$ obtained by decomposing the K\"ahler form in the basis of Mori-cone generators $H,F$, has the form
\begin{equation}
    \fF_6 = 2h f+ nh^2 \stackrel{!}{=}1\, ,
\end{equation}
whereas the metric on ambient space is very similar to that of $dP_2$. The latter reads
\begin{equation}
   g= \left(
\begin{array}{cc}
 \frac{2}{(2 f+h n)^2} & \frac{n}{(2 f+h n)^2} \\
 \frac{n}{(2 f+h n)^2} & \frac{2 f^2+2 f h n+h^2 n^2}{h^2 (2 f+h n)^2} \\
\end{array}
\right)\, .
\end{equation}
Once again, the eigenvalue equation for $\cF_{\mathbb{F}_n}=2f+(2+n)h$ is satisfied with
\begin{equation}
    \Delta_J \cF = 2 \cF \ \ \Longrightarrow \ \  \lambda=1\, ,
\end{equation}
analogously to the aforementioned del Pezzo example.

\paragraph{The general case.} We have seen that in these simple examples, the higher-derivative operator satisfies a Laplace equation on moduli space with eigenvalue given by the total number of tensor multiplets, i.e.,
\begin{equation}\label{eq:6dlapl}
    \Delta_j \cF = T \cF \, .
\end{equation}
In the following, we argue that this should always be the case, since the ambient metric is locally hyperbolic and takes the form
\begin{equation}
    g_{IJ} = \frac{1}{\fF_6} \hat g_{IJ}\, , \qquad \hat g_{IJ} = -\Omega_{IJ} + \frac{2}{\fF_6}\Omega_{IA}\Omega_{JB}J^AJ^B\, .
\end{equation}
Notice that, for a generic linear function, the only relevant piece of the Laplacian is the one involving linear derivatives
\begin{equation}
     \Delta_X \to \Delta_{\text{linear}} = \frac{1}{\sqrt{|g|}}\partial_I \left(\sqrt{|g|}g^{IJ}\right)\partial_J\, ,
\end{equation}
whilst the inverse metric is given by \eqref{eq:invmetr6d}, and the determinant can be computed explicitly to be $\sqrt{|g|} = \fF_6^{-\frac{T+1}{2}}$.\footnote{This follows from a simple application of the determinant lemma, i.e., $|\hat g |= |\det(-\Omega)(1 - \frac{2}{\fF_6}J_I\Omega^{IJ} J_J)| = 1$.} Moreover, a direct computation reveals that $\partial_I g^{IJ} =2(T+1)J^J$, resulting in the linear part of the Laplacian being 
\begin{equation}
    \fF_6^{\frac{T+1}{2}} \partial_I\left( \fF_6^{-\frac{T+1}{2}} g^{IJ}\right)\partial_J = (T+1) J^I \partial_I\, ,
\end{equation}
which, when acting on linear functions in $J^I$, yields precisely $\lambda = T$. 

Now that we confirmed explicitly that the Wilson coefficient associated to the $\cR^2$-operator in 6d $\cN=(0,1)$ supergravity EFTs arising from F-theory obeys a Laplace equation with eigenvalue $\lambda=T$, we would like to follow the inverse logic. Namely, we want to show how to recover the precise asymptotic form of the function $\cF(J)$ by assuming the Laplace equation to hold, similarly to what we did in Section \ref{s:maxsugra}. For concreteness, we illustrate this in two simple del Pezzo cases with $T=1,2$. Let us start with the former, where \eqref{eq:6dlapl} reads
\begin{equation}\label{eq:LaplacedP1}
    \partial_{x}^2 \cF = \cF\, ,
\end{equation}
when written in terms of the canonically normalized scalar $x$. In this setup, the solutions to the above equation are given by 
\begin{equation}\label{eq:solsdP1canonical}
    \cF= c_1 e^{x} + c_2 e^{-x} = \frac{c_1+c_2}{2}\cosh x + \frac{c_1 - c_2}{2}\sinh x\, ,
\end{equation}
with $\{c_1,c_2\}$ being some integration constants (cf. footnote \ref{fnote:chernclassdP}). Furthermore, one can readily see that taking $J^0 = \cosh x\, ,\, J^1 =\sinh x$ is consistent with \eqref{eq:6dprepotconstraint}---as well as the K\"ahler cone constraint if we also impose $x\geq 0$ \cite{Donagi:2004ia}, namely
\begin{equation}\label{eq:constraintdP1}
    (J^0)^2  - (J^1)^2 = 1\, ,
\end{equation}
and, at the same time, yields a canonical metric $g_{x x}= 1$. Therefore, we conclude that \eqref{eq:solsdP1canonical} recovers globally---as opposed to only asymptotically---the linear dependence of $\cF$ in ambient space, since
\begin{equation}
    \cF = \frac{c_1+c_2}{2}J^0 + \frac{c_1 - c_2}{2}J^1\, .
\end{equation}
One can repeat the same steps for the $T=2$ case. In particular, the embedding submanifold can be parametrized now in terms of $(x,\theta)$ as
\begin{equation}\label{eq:parametrizationdP2}
    J^0 = \cosh x, \quad J^1 = \sinh x  \cos \theta , \quad  J^2 = \sinh x  \sin \theta\, ,
\end{equation}
so that the metric on the constrained manifold reads
\begin{equation}
    \td s^2 = \td x^2 + \sinh^2 x\,  \td \theta^2\, , 
\end{equation}
with $x\geq0$ and $\theta \in [0,\pi/2]$,\footnote{This restriction, together with the condition $\cosh x +\sinh x(\cos \theta - \sin \theta) \geq 0$, is required in order to stay inside the K\"ahler cone \cite{Donagi:2004ia,vandeHeisteeg:2023dlw}.} whilst the Laplace equation takes the form
\begin{equation}\label{eq:laplaceeqdP2}
    \Delta_{\rm 6d} \cF = 2 \cF\, .
\end{equation}
Here, $\Delta_{\rm 6d}$ is the Laplacian on the upper half-plane in hyperbolic polar coordinates, i.e., 
\begin{equation}
 \Delta_{\rm 6d} = \partial^2_x \ + \coth x \, \partial_x + \frac{1}{\sinh^2 x}\partial_\theta^2\, .
\end{equation}
The space of solutions to this differential equation is richer than that of the one-modulus case. Following the discussion in maximal supergravity, we first expand the Wilson coefficient in a Fourier series with respect to the angular variable
\begin{equation}\label{eq:Fourier6ddP2}
    \cF (x,\theta) = \sum_{n \in \mathbb{Z}} f_n(x) e^{in\theta }\, ,
\end{equation}
and solve for each mode independently. Moreover, as explained in Appendix \ref{ap:delPezzolaplace}, finding regular solutions in moduli space away from (infinite distance) singularities selects both the `constant' (i.e., $\theta$-independent) and the $n=1$ terms as the unique possible contributions, thus leading to the linear function
\begin{equation}
    \cF (x, \theta) =c_0 \cosh x+c_1 \sinh x \cos \theta +c_2 \sinh x \sin \theta =  c_0\, J^0+c_1\, J^1 +c_2\, J^2\, .
\end{equation}
Therefore, once again we are able to recover the global form---up to numerical prefactors---of the relevant Wilson coefficient by simply imposing the Laplace constraint \eqref{eq:laplaceeqdP2}. In fact, one can argue that this should be the case for any del Pezzo model upon using that a convenient parametrization for the constrained hypersurface \eqref{eq:6dprepotconstraint} is
\begin{equation}
   J^0=\cosh x\, ,\quad J^i=x^i \sinh x\, ,\qquad \text{with}\quad \sum_i (x^i)^2=1\, ,
\end{equation}
which gives rise to a pulled-back metric that can be written as
\begin{equation}
    \td s^2 = \td x^2 + \sinh^2 x\, \td\Omega_{T-1}^2\, ,
\end{equation}
where $\td\Omega^2_{T-1}$ is the line element on the unit $(T-1)$-dimensional sphere. From here, one indeed finds (see Appendix \ref{ap:delPezzolaplace} for details)
\begin{equation}
    \left[ \partial_x^2 + (T-1) \coth x \partial_x + \frac{1}{\sinh^2 x}\Delta_{\mathbf{S}^{T-1}}\right] \cF = T\cF \ \ \implies \ \ \cF = c_0\, J^0+c_i\, J^i\, ,
\end{equation}
which generalizes straightforwardly the $T=2$ discussion above.

To conclude this section, let us briefly comment on the duality group exhibited by the theory, under which the linear ansatz for the Wilson coefficient should be invariant. In general, for 6d $\cN = (0,1)$ supergravity EFTs one expects this to be a discrete subgroup with definite signature $G^{1,T} \subset SO(1,T)$ \cite{Taylor:2011wt}.\footnote{The signature is inherited from the coset structure of the classical theory and is required to leave invariant the charge lattice $\Gamma$ of objects that couple to the $B$-field, which has the same signature.} From the geometric point of view, one can see this duality acting on the lattice $H_2(d P_2, \bZ)$ endowed with the intersection product $\Omega^{IJ}$ \cite{Iqbal:2001ye}. As discussed in more detail in \cite{Iqbal:2001ye}, following F/M-theory duality, this corresponds to the Weyl group of the maximal supergravity theory in 9d, e.g., the $\bZ_2$ discussed in Section \ref{ss:maxsugra9d}. Here, in the two-moduli example, it acts as global diffeomorphisms of the $dP_2$ which preserve the canonical class $K = -c_1(dP_2)$ by exchanging $J^1 \leftrightarrow  J^2$ (i.e., the two blowup points) through $\theta \to \pi/2-\theta$. The full $SL(2,\bZ)$ duality of the maximal supergravity setup, instead, is expected to map topologically different backgrounds in F-theory by also changing this quantity. Given all of this, from the EFT point of view, only the $\bZ_2$-duality factor is expected to be present in its moduli-dependence. Notice that this acts always by a discrete action on the compact moduli, meaning the asymptotic form of the coefficient is identical in all duality frames. This is consistent with the fact that all infinite-distance limits in such theories are given by emergent string limits in the same duality orbit.

\subsection{M-theory on a Calabi--Yau threefold}\label{ss:5dMtheory}

In this section, we finally consider our last set of examples exhibiting lower supersymmetry. These correspond to 5d $\cN=1$ supergravity theories characterized by containing, besides the gravity multiplet, an arbitrary number of vector and hypermultiplets. From the top-down perspective, they can be obtained by compactifying M-theory on a Calabi--Yau threefold $Y_3$. The relevant two-derivative action reads \cite{Cadavid:1995bk,Ferrara:1996hh,Ferrara:1996wv,Bergshoeff:2004kh,Lauria:2020rhc}
\begin{equation}
    S_{5\rm{d}} = \frac{1}{2 \kappa_5^2}\int \left( R \star 1  -  g_{ij}(\phi) \,d\phi^i \wedge \star d\phi^j + \dots\right)\, ,
\end{equation}
where $\phi^i$ are $n_V=h^{1,1}(Y_3)-1$ real fields parameterizing the vector multiplet moduli space. Its geometry, together with the interactions entering the Lagrangian, are dictated by a single cubic function depending on the variables $M^I$
\begin{equation}\label{eq:constraintMth}
    \fF_5 = \frac{1}{3!}C_{IJK} M^I M^J M^K\, ,\qquad I=1, \ldots, n_V+1\, ,
\end{equation}
with $C_{IJK} \in \bZ$ being some constant parameters. Indeed, these projective coordinates define a ($n_V + 1$)-dimensional real manifold, where the metric is given by
\begin{equation}\label{eq:5dmetric}
    g_{IJ} = -\frac{1}{2}\partial_I \partial_K \log \fF_5\, , 
\end{equation}
whereas the actual vector multiplet moduli space spanned by the $\phi^i$ corresponds to the hypersurface determined via the \emph{very special geometry} constraint 
\begin{equation}\label{eq:5dconstraint&metric}
    \fF_5 = 1\, , \qquad g_{ij} = \partial_i M^I \partial_j M^J g_{IJ}\, ,
\end{equation}
with $g_{ij}$ being the pull-back of the ambient metric \eqref{eq:5dmetric}.

By further compactifying this theory on a circle, we can relate M-theory on $Y_3 \times \mathbf{S}^1$ with a strongly coupled limit of Type IIA supergravity on $Y_3$. From the 11-dimensional perspective, the resulting moduli can be related to the corresponding string theory quantities as \cite{Cadavid:1995bk, Witten:1995ex}
\begin{equation}
    2\pi R_{\mathbf{S}^1} = \frac{\cV^{1/3}}{\cV_M^{1/3}}\, , \qquad M^I = t^I \frac{\cV_M^{1/3}}{\cV^{1/3}}\, ,
\end{equation}
where $R_{\mathbf{S}^1}$ is the $\mathbf{S}^1$ radius, $X^M$ the K\"ahler moduli, and $\cV_{M}$ the total volume of the Calabi--Yau, all of them measured in eleven-dimensional Planck units. Instead, $\cV$ and $t^I$ are measured in string units. To make even more explicit the relation with the 5d description above, we introduce the circle radius in 5d Planck units $R_5$,\footnote{The relation between $\ell_5$ and $\ell_{11}$ is $\ell_{11} = \ell_5 \cV_M^{-1/3}$.} which can be identified with $\cV$ by
\begin{equation}
    2\pi R_5 = \cV^{1/3}\, . 
\end{equation}
The 5d ambient coordinates, $M^I$, can thus be related to the $h^{1,1}(Y_3)$ K\"ahler moduli $t^I$ appearing in the 4d description (see discussion around eq. \eqref{eq:4dIIAaction}) via the rescaling 
\begin{equation}\label{eq:5dMthy4dIIAmap}
    M^I = \frac{t^I}{\cV^{1/3}}\, .
\end{equation}
In this language, the $C_{IJK}$ constants become the triple-intersection numbers associated to $Y_3$---which were previously denoted by $\mathcal{K}_{IJK}$, and the hypersurface $\fF_5=1$ can be reinterpreted as fixing the overall threefold volume in 11d M-theory units to one, consistently with the fact that the latter belongs to a 5d hypermultiplet \cite{Bergshoeff:2004kh, Lauria:2020rhc}. 

As mentioned at the beginning of this chapter, one can compute exactly certain higher-derivative corrections within this class of theories. Particularly interesting is the $\cR^2$-operator, which is obtained from 11d by dimensionally reducing the analogous $\cR^4$-term term on $Y_3$. Using \eqref{eq:r4tor2} and integrating over the Calabi--Yau, one has (neglecting numerical prefactors) 
\begin{equation}\label{eq:5dhigherder}
    \frac{\Mpe}{2} \int \td^{11}x \sqrt{-g_{\small{\rm 11d}}} \ t_8 t_8 \cR^4 = \frac{\Mpfive}{2}\int \td^5x \sqrt{-g_{\small{\rm 5d}}} \left( \frac{1}{\cV_M^{1/3}}\int_{Y_3} c_2(Y_3) \wedge J\right) \cR^2\, ,
\end{equation}
where by decomposing $J$ in terms of the $M^I$ coordinates we can eliminate the dependence on the overall volume factor. Thus, the Wilson coefficient ends up being a linear function with respect to the ambient space fields
\begin{equation}\label{eq:R25dsugra}
    \cF (M) = c_{2,I}M^I\, ,
\end{equation}
similarly to the 6d case. However, in contrast to F-theory, we will show in the remainder of this section that the Laplacian is not the appropriate elliptic operator with respect to which $\cF$ is an eigenfunction—much like in ten-dimensional Type IIA. We will first illustrate this issue through a couple of simple examples, leaving a more detailed explanation for Chapter \ref{s:Laplace&Symm}.

\subsubsection{One modulus example: $Y_{2,86}$}\label{sss:onemod5d}

We consider first a simple Calabi--Yau threefold with Hodge numbers $(h^{1,1},h^{2,1})=(2,86)$, which was initially studied in \cite{Greene:1995hu,Greene:1996dh} and recently reviewed by \cite{Alim:2021vhs,vandeHeisteeg:2023dlw} in the context of the Swampland program. This theory exhibits two different geometric phases connected by a flop transition at $M^2=0$. The first one is characterized by the following cubic prepotential
\begin{equation}
    \fF_5 = \frac56 (M^1)^3 + 2(M^1)^2M^2\, ,
\end{equation}
with a K\"ahler cone spanned by $M^1,M^2 \geq 0$, and for the following analysis we will only consider this phase, which contains the unique infinite distance locus within the vector multiplet moduli space (see \cite{vandeHeisteeg:2023dlw} for details). The Wilson coefficient, when written in terms of the two non-vanishing Chern numbers $c_{2,I}$, has the simple expression
\begin{equation}
    \cF = \frac{1}{12}( 50M^1 + 24M^2)\, .
\end{equation}
Since the constrained manifold defined by $\fF_5=1$ is one-dimensional, we can introduce a flat\footnote{This can be straightforwardly verified using the fact that the ambient space metric takes the form:
\begin{equation}
   g= \frac{1}{\fF_5^2}\left(
\begin{array}{cc}
 \frac{25}{24} (M^1)^4+4(M^1)^2(M^2)^2+\frac{10}{3} (M^1)^3M^2 & \frac56 (M^1)^4 \\
 \frac56 (M^1)^4 & 2 (M^1)^4 \\
\end{array}
\right)\, .
\end{equation}
such that $g_{xx}=1$ once we pull back to the actual moduli space via \eqref{eq:5dconstraint&metric}.} coordinate $x \in [-\log (6/5)/\sqrt{3},\, \infty)$ which parametrizes $M^1$ and $M^2$ as follows
\begin{equation}
    M^1 = e^{-x/\sqrt{3}}\, , \qquad M^2 = -\frac{5}{12}e^{-x/\sqrt{3}} + \frac12e^{2x/\sqrt{3}}\, ,
\end{equation}
so that the species function extracted from the higher-derivative coefficient reads
\begin{equation}
    \cF \sim e^{2x/\sqrt{3}} \iff  \LQG \sim e^{-x/\sqrt{3}}\, ,
\end{equation}
when taking the limit $x\rightarrow\infty$. As already remarked, within this model the latter is the only possible infinite distance limit, since moving in the other direction we first hit the flop wall located at $M^2=0$ and, continuing towards the second geometric phase, a SCFT boundary at $M^2=-1/(2 \cdot 3^{1/3})$ is found also at finite distance. From here, we already see that in the most simple setup with one-modulus and using flat coordinates---where the Laplacian is simply given by $\partial_{x}^2$, the Wilson coefficient of the $\mathcal{R}^2$-term does not obey an eigenvalue equation with respect to $\Delta_{\rm 5d}$ in a \emph{global} fashion, but only does so \emph{asymptotically}
\begin{equation}
    \partial_x^2 \cF \sim \frac43 \cF\, ,\qquad \text{as}\quad x\to \infty\, ,
\end{equation}
which is precisely where the exponential behavior of the species scale is expected to arise. 

However, we may readily recognize essentially the same structure as for the 10d Type IIA case (cf. Section \ref{ss:10dIIA}) appearing in here, and thus the way in which the Laplace equation is violated is completely analogous. In particular, in this very simple case, we can easily work out the form of the appropriate elliptic operator by a simple deformation
\begin{equation}\label{eq:twomodelliptic}
    \mathcal{D}^2_x \cF =\left(\partial_x^2 -\frac{1}{\sqrt{3}}\partial_x \right) \cF = \frac23\cF\, ,
\end{equation}
where we see that $\mathcal{D}^2_x$ differs from the Laplacian by a linear derivative term, exactly like in the 10d Type IIA theory (cf. eq. \eqref{eq:typeIIAeq}).

\subsubsection{Two moduli example: $\mathbb{P}^{1,1,2,8,12}[24]$}\label{sss:twomod5d}

Let us now study a more complicated example involving three (ambient) moduli fields, which arises from compactifying M-theory on $\mathbb{P}^{1,1,2,8,12}[24]$ (cf. e.g., in \cite{Hosono:1993qy,Hosono:1994ax,Klemm:1996bj}). In this case, determining the suitable elliptic operator is not as simple, and in fact constructing the latter explicitly turns out to be hard in general, as will be discussed during the rest of this chapter. Similarly to the 6d case, the form of the metric and the Laplacian defined on the constrained submanifold is fairly cumbersome, and for this reason it becomes convenient to work instead with the one associated to the ambient space, using \eqref{eq:laplonconstr} and the results of Appendix \ref{ap:meancurv}.

For the case at hand, the prepotential reads \cite{Hosono:1993qy}
\begin{equation}\label{eq:solM1}
    \fF_5 = \frac43(M^2)^3 +(M^2)^2M^1 + 2 (M^2)^2 M^3 + M^2(M^3)^2+M^1M^2M^3\, ,
\end{equation}
with $M^I \geq 0$, $I=1,2,3$, and the constraint $\fF_5=1$ can be simply solved by setting 
\begin{equation}
    M^1 = \frac{3-4(M^2)^3 -6(M^2)^2M^3 -3M^2(M^3)^2}{3(M^2)^2 + 3M^2M^3}\, .
\end{equation}
The Wilson coefficient for $\mathcal{R}^2$-term, involving the second Chern numbers, reads
\begin{equation}\label{eq:5d3mhigherder}
    \cF = c_{2,1}M^1 + c_{2,2} M^2 +  c_{2,3} M^3\, ,
\end{equation}
whilst the boundaries of moduli space are given by the representatives
\begin{equation}
    \big\{ (M^1,M^2,M^3)\big\} = \bigg\{(\infty,0,0),\, (0,\left( 3/4\right)^{\frac13},0),\, (0,0,\infty) \bigg\}\, .
\end{equation}
The second one lies at finite distance and we include it for completeness, while the other two are infinite distance emergent and F-theory limits, respectively \cite{Marchesano:2023thx}.

Let us first consider the case $M^1\gg M^2,M^3$, which can be achieved by taking $M^2 \rightarrow 0$ with $M^3$ fixed on the restricted submanifold. In this case, $\cF$ diverges linearly in $M^1$ (considered as a function of $M^2,M^3$) and the Laplace equation would give asymptotically
\begin{equation}\label{eq:asymptoticeigenvalueP112812}
    \frac{\Delta_{\rm 5d}\cF}{\cF} = \frac{\Delta_M\cF}{\cF} - \frac{\rm{Hess}_\cF(\mathbf{n},\mathbf{n})}{\cF} \stackrel{M^2\ll M^3}{\sim} \frac43\, ,
\end{equation}
as in the previous model. To arrive at the last result, we also replaced the Hessian contribution computed in Appendix \ref{ss:Hessian5d} (see in particular eq. \eqref{eq:5dhess}). This limit corresponds to an emergent string degeneration lying at infinite distance. Again, we notice that the Laplace equation is solved but only asymptotically, in line with the fact that the damping of the Wilson coefficient needs to be exponential at least therein. 

Similarly, one can probe the regime $M^2 \gg M^3$ on the restricted submanifold. In this limit, one finds (cf. eq. \eqref{eq:solM1}) 
\begin{equation}
    M^1 \sim \frac{3-4 (M^2)^3}{3(M^2)^2}\, ,
\end{equation}
which means that, in order to stay inside the K\"ahler cone, $M^2 \in \left[0,(\frac{3}{4})^{\frac13}\right]$, as anticipated. Moreover, we find that in the limit $M^3 \rightarrow 0, \ M^2 \rightarrow (\frac{3}{4})^{\frac13}$, the Laplace equation has a non-trivial solution
\begin{equation}
      \frac{\Delta_M \cF}{\cF}\, \stackrel{M^2\gg M^3}{\sim}\, \frac{2(2c_{2,1} + c_{2,2}+c_{2,3})}{3c_{2,2}}\, ,
\end{equation}
where the $\cF$ function does not diverge but rather stays finite, signaling that the aforementioned boundary is indeed at finite proper distance.

The last limit that can be probed corresponds instead to a decompactification to 6d F-theory. This arises by taking $M^1, \, M^2 \,\to \,0,\ M^3 \to \infty$, which can be achieved when $M^3 \sim (M^2)^{-1/2}$ on the restricted submanifold. Applying the same procedure, we find that 
\begin{equation}
    \frac{\Delta \cF}{\cF}\, \stackrel{M^3\to \infty}{\sim}\, \frac{4}{3}\, ,
\end{equation}
which is remarkably the same eigenvalue that was found before. One could then hope that, by using the scalar Laplacian $\Delta_{\rm 5d}$, even if $\cF$ is not a global eigenfunction, it asymptotes to one with a unique eigenvalue depending only on the number of moduli. However, in Appendix \ref{app:more5dmodels} we show that this is not the case, since there are examples exhibiting different eigenvalues for different limits if the Chern class is not tuned appropriately. Our findings are summarized in Table \ref{tab:5d}.
\begin{table}\renewcommand{\arraystretch}{1.1}
\centering
\begin{tabular}[h]{ |c|c|c|c|c|c|c|} 
\hline
Model & $\lambda$ & Smooth? & Unique $\lambda$?& $h^{1,1}$  \\ 
\hline
\hline
 $Y_{2,86}$ & $\frac43 $ &\cmark & \cmark& 2 \\
 $\mathbb{P}^{1,1,2,8,12}[24]$ & $\frac43$ & \cmark &  \cmark & 3 \\
Fibration over $\mathbb{P}^2$--I &  &\xmark & \xmark* & 3  \\
Fibration over $\mathbb{P}^2$--II & $\frac56$ &\xmark & \cmark & 3  \\
Fibration over $\mathbb{F}_1$ & $\frac43$ & \xmark & \cmark & 4  \\
Extra section over $\mathbb{P}^2$ & \ \  & \xmark & \xmark* & 3  \\
 \hline
\end{tabular}
\caption{\small Summary of results obtained by studying several 5d Calabi--Yau models (see Appendix \ref{app:more5dmodels}). The third column refers to whether the fibration structure is smooth or not. The * refers to the fact that the eigenvalues can be tuned to be unique only for a specific choices of the second Chern class. Since in general we do not have access to its computation, we leave it as an interesting possibility.}
\label{tab:5d}
\end{table}
Therefore, in the 5d setup it does not seem straightforward to obtain $g_{ij}(\phi)$ in general nor to engineer a suitable differential equation satisfied by $\cF$. In what follows, we will argue that by compactifying the theory on a circle and using the four-dimensional results discussed in Section \ref{ss:4dN=2}, one can extract the operator we seek for.

\subsubsection{Finding the appropriate elliptic operator}\label{sss:uplift}

As we have seen by inspecting simple one- and two-moduli examples, the higher-derivative Wilson coefficient of interest does not obey an eigenvalue equation involving the Laplacian. Consequently, in order to construct a suitable elliptic operator that does so, it is useful to compactify the theory on circle, in the same spirit of Section \ref{sss:typeIIAop}. In particular, denoting respectively by $\cF_5$ and $\cF_4$ the 5d and 4d $\cR^2$ Wilson coefficients, we have 
\begin{equation}
    \frac{\Mpfive}{2}\int \td^5x\, \left(\cF_5 \, \cR^2_5\right) \star_5 1 = \int \td^4x \, \left(\cF_5 R_5\, \cR_4^2\right)\star_4 1\, ,
\end{equation}
where we recall that $R_5$ denotes the $\mathbf{S}^1$ radius measured in 5d Planck units. Thus, the relation between the two functions is simply given by
\begin{equation}\label{eq:4d5dwilson}
    \cF_5 = \frac{\cF_4}{R_5}\, .
\end{equation}
Following the discussion in Section \ref{ss:4dN=2}, and using the fact that the five-dimensional ambient coordinates are defined in terms of the 4d saxionic $t^I$ as $M^I = t^I/R_5$, we are lead to the expectation that
\begin{equation}
    \cF_5  = c_I M^I\, ,
\end{equation}
reconstructing the explicit computation. Notice that the relation between 5d and 4d moduli, usually framed in terms of top-down quantities (e.g., volumes of $Y_3$ in 11d and string units), actually involves only geometrical quantities defined already in the 5d supergeavity theory. Therefore, knowing that the 4d coefficient satisfies\footnote{Similarly to how we derived the 10d Type IIA operator from the 9d equation in Section \ref{sss:typeIIAop}, we truncate all axion-dependence from the 4d Laplacian.}
\begin{equation}\label{eq:4dlaplforupl}
    \frac{1}{2}g^{I \bar J}\partial_{t^I} \partial_{t^J} \cF_4 = 0\, ,
\end{equation}
with $g_{I \bar J}$ denoting the 4d K\"ahler metric, one can extract some analogous equation that should hold for the 5d $\mathcal{R}^2$-coefficient. In order to keep the discussion concise, we relegate the explicit computation to Appendix \ref{app:uplift}, summarizing here the main points. Indeed, from eq. \eqref{eq:4dlaplforupl} we find that the 5d Wilson coefficient satisfies the following condition in ambient space
\begin{equation}
    \cD^2_M \cF_5 = \frac23\left(h^{1,1} - 1\right) \cF_5\, , 
\end{equation}
where 
\begin{equation}
    \cD^2_M  =\frac{2}{3}\left( h^{1,1}-1\right)  M^I\partial_I + \frac14\left( g^{IJ}  - \frac{8}{3}M^IM^J\right) \partial_I \partial_J\, .
\end{equation}
Of course, we are ultimately interested in deriving an eigenvalue equation on the constraint hypersurface \eqref{eq:5dconstraint&metric}. To this end, let us compute the restriction of this operator on the $\fF = 1$ slice by employing the projector \cite{ecker2004regularity}
\begin{equation}\label{eq:5dprojector}
    \Pi^{I}_J = \delta^I_J - n^In_J\, ,
\end{equation}
where $n^I= \sqrt{\frac23}M^I$ is the normal vector to the hypersurface (cf. eq. \eqref{eq:meancurvvanishes}). The $\cD_M^2$ ambient space operator can be rewritten in a covariant fashion as in \eqref{eq:5dcovop}:
\begin{equation}
  \cD_{M}^2 = \left( \frac{1}{4}g^{IJ}-n^In^J \right) \nabla_I \partial_J - \frac{1}{24}\left[P^I - \frac{13}{2}\sqrt6\,h^{1,1}\,n^I\right]\partial_I\, ,
\end{equation}
with the vector field $\mathbf{P}$ reading explicitly $P^I = C C^{JK}C_{JKL}C^{LI}$. By acting on the 1-form $\partial_I\cF_5$ with the projector \eqref{eq:5dprojector}, and using \eqref{eq:meancurvvanishes}, one is able to write
\begin{equation}\label{eq:5doperator}
    \cD^2_{\fF_5=1} = \frac14(g^{IJ}-n^In^J) \nabla_I \partial_J -\frac{1}{24}\left( P^I - \sqrt{\frac32}h^{1,1}n^I\right)\partial_I\, .
\end{equation}
where the terms that have been removed contribute to the difference
\begin{equation}
    \cD^2_M - \cD^2_{\fF_5=1} = -\frac34n^In^J \nabla_I \partial_J + \frac{1}{\sqrt{6}}h^{1,1}n^I\partial_I\, .
\end{equation}
which, when evaluated on linear functions in ambient space, yields a constant times the function itself, as per eq. \eqref{eq:5dhess}. Thus, we have that on the constrained manifold, the $\mathcal{R}^2$-coefficient obeys
\begin{equation}\label{eq:5deigconstraint}
    \cD^2_{\fF_5=1} \cF_5 = \frac16\left( h^{1,1}-1 \right)\cF_5\, ,
\end{equation}
where the eigenvalue is written in terms of the number $n_V = h^{1,1}-1$ of vector multiplets. For concreteness, and to check consistency with previous results, let us compute this explicitly for the $Y_{2,86}$ example analyzed in Section \ref{sss:onemod5d}. Remarkably, the restricted operator can be rewritten as
\begin{equation}\label{eq:twomodrestrop}
    \cD^2_{\fF=1}  =\frac{1}{4}\frac{\partial X^I}{\partial \Delta}\frac{\partial X^J}{\partial \Delta} \nabla_I \partial_J - \frac{1}{4\sqrt{3}}\frac{\partial X^I}{\partial \Delta}\partial_I = \frac{1}{4}\left( \partial_\Delta^2 - \frac{1}{\sqrt{3}}\partial_\Delta\right)\cF_5\, ,
\end{equation}
which is proportional to the guessed operator, namely \eqref{eq:twomodelliptic}. This yields a non-trivial check of the picture advocated in the present work. As a final remark, notice that given the explicit form of $\cD_M^2$ and the eigenvalue $\lambda$, the differential equation will generally be solved by a linear function in the ambient coordinates $M^I$, leading to a Wilson coefficient with the form $\cF_5 = c_IM^I$, consistently with the 4d$\to$5d uplift \eqref{eq:4d5dwilson}. 

\subsubsection{Connecting with six-dimensional supergravity}

Lastly, to bridge the remaining gap between the five- and six-dimensional descriptions, and to elucidate how the six-dimensional equation written in terms of the Laplacian is spoiled in 5d, we consider circle compactifications of F-theory on an elliptically fibered Calabi--Yau. Indeed, the recurring algorithm we used in relating elliptic operators in $d$ and $d+1$ dimensions relies on the triviality of the circle fibration, which applies in the relevant cases of maximal supergravity as well as in going from 5d $\cN=1$ to 4d $\cN=2$ (at large volume). In this section, we aim to show how the presence of quantum corrections spoil the six-dimensional Laplace equation when reducing down to five dimensions. Applied to the F-theory case, this leads to M-theory on an elliptic Calabi--Yau threefold, whose constraint polynomial---splitting the coordinates in $M^I=\{M^0,M^\a\}$ and identifying $M^0$ as the cubic contribution \eqref{eq:constraintMth}---reads 
\begin{align}
 \fF_5 & = 2\Omega_{\a\b}c_1^\a c_1^\b (M^0)^3 + 6 \Omega_{\a\b}c_1^\a M^\b (M^0)^2 + 6 \Omega_{\a\b} M^\a M^\b M^0 \notag\\
& = 6 \tilde{M}^0\left(  \Omega_{\a\b}\tilde{M}^\a\tilde{M}^\b + \frac{1}{12}\Omega_{\a\b} c_1^\a c_1^\b \tilde{M}^0 \tilde{M}^0\right)\, ,
\label{CKF}   
\end{align}
where in the second step we have defined $\{ \tilde{M}^0=M^0,\, \tilde{M}^\a = M^\a + \frac{1}{2} c_1^\a M^0 \}$, with $c^\a_1 \in \bZ$ being constant parameters denoting the expansion coefficients of the first Chern class of $B_2$ \cite{Corvilain:2018lgw} in top-down constructions. For instance, in the case of the threefold $\mathbb{P}^{1,1,1,6,9}[18]$ \cite{Candelas:1994hw}, which has two K\"ahler moduli $\{ M^0, M^1\}$, one can simply identify 
\begin{equation}
  \tilde{M}^0=M^0\, , \qquad \tilde{M}^1 = M^1+\frac32 M^0\, ,
\end{equation}
with $\Omega_{11}= 1$, $c^1_1=3$. The F-theory limit consists of shrinking the fiber, identified with $M^0$, while keeping the overall volume of the base fixed. To see how this comes about, one can define a new set of variables
\begin{equation}\label{eq:defj}
  J^0=M^0\, ,\qquad J^\a = \sqrt{M^0} M^\a\, ,
\end{equation}
which in the limit of shrinking fibre, i.e.,  $M^0 \to 0$, verify the relation
\begin{equation}\label{eq:constraintFthy}
	\fF_6=\Omega_{\a \b} J^\a J^\b \stackrel{!}{=} 1\, ,
\end{equation}
thereby reproducing the 6d constraint \eqref{eq:6dconstr} which defines the 6d $\cN=(1,0)$ moduli space. 

Using the coordinates \eqref{eq:defj}, one can reexpress the five-dimensional metric \eqref{eq:5dmetric}, dubbed $g^M_{IJ}$, as follows
\begin{equation}
\begin{split}\label{eq:metricjcoords}
g^M_{\a \b} &= J_\a J_\b - \Omega_{\a \b}\, ,\\
g_{0 0}^M &= \frac{1}{(J^0)^2} - {\mathfrak{a}J^0} + {\mathfrak{a}^2 (J^0)^4} -{\mathfrak{a} J^0}\left( J_\b J^\b\right) + \frac{1}{4 (J^0)^2}\left(J_\a J_\b - \Omega_{\a \b}\right)J^\a J^\b\, ,\\
g_{0 \a}^M &= -\frac{J_\a}{J^0} \left(J_\b J^\b - 1\right) + {(J^0)^2} \mathfrak{a} J_\a\, ,
\end{split}
\end{equation}
where $\mathfrak{a}= \frac{1}{12} \Omega_{\a \b} c^\a_1 c^\b_1 = \frac{1}{12}(10-h^{1,1}(B_2))$ \cite{Bonetti:2011mw}. The contributions proportional to $\mathfrak{a}$ can be traced back to the second term in the prepotential \eqref{CKF}, which in turn may be interpreted as a 1-loop effect induced by the tower of M2-branes wrapping the elliptic fibre \cite{Corvilain:2018lgw}. On the other hand, in the special case where $\mathfrak{a}=0$ one obtains decoupling between the coordinates, leading to a (sub-)metric $g_{\a \b}^M$ of the form
\begin{equation}
g^M = 
\begin{pmatrix}
\frac{3}{4 (J^0)^2} & -\frac{J_\a}{2J^0}\\
-\frac{J_\a}{2J^0} & \frac{1}{2} \left(J_\a J_\b - \Omega_{\a \b}\right)
\end{pmatrix}\, ,
\end{equation}
which is block diagonal when evaluated on the constrained surface \eqref{eq:constraintFthy}. In the following, we analyze a simple example to understand how the five-dimensional Laplace equation \eqref{eq:6dlapl} gets quantum-corrected by powers of the fiber coordinate within the 5d bulk moduli space.

\subsubsection*{Application to the two moduli example: $\mathbb{P}^{1,1,2,8,12}[24]$}

We can try to reexpress the Laplacian we found in the Section \ref{sss:twomod5d} for the coordinates $\{ M^1,M^2,M^3\}$ in terms of 6d variables $\{J^0,J^1,J^2\}$. First, notice that via some change of coordinates of the kind
\begin{equation}
    \tilde M^1 = 2M^3 - M^1 +2M^2\, , \qquad \tilde M^2 = M^2\,,\qquad  \tilde M^3 = M^1- M^3 +M^2\, ,
\end{equation}
the prepotential takes the form
\begin{equation}
    \fF_5 = M^2\left[ 6(\tilde M^1)^2 + 15\tilde M^1 \tilde M^3 + 9 (\tilde M^3)^2 + (\tilde M^2)^2\right]\, ,
\end{equation}
which agrees with \eqref{CKF}. Then, defining the $J$-coordinates as 
\begin{equation}
    \{J^0,J^1,J^2\} = \left\{\tilde M^2, \sqrt{\tilde M^2}\tilde M^1, \sqrt{\tilde M^2}\tilde M^3\right\}\, ,
\end{equation}
we have that in the limit of vanishing fiber volume $J^0 \to 0$ one gets
\begin{equation}
    \fF_5 = \frac{(J^0)^3}{3} + 2(J^1)^2 + 5J^1 J^2 + 3(J^2)^2\ \stackrel{J^0 \rightarrow 0}{\implies}\ \fF_6 =  2(J^1)^2 + 5J^1 J^2 + 3(J^2)^2 \stackrel{!}{=} 1\,,
\end{equation}
as required. It is instructive to perform the change of coordinates and express the Laplacian on the constrained manifold as a power expansion in $J^0$. The linear higher-derivative correction \eqref{eq:5d3mhigherder} in the new coordinate system becomes $\cF = [(c_{2,1}+c_{2,3}) J^1 +(2c_{2,1}+c_{2,3})J^2]/\sqrt{J^0} + (c_{2,2}-c_{2,3})J^0$. Upon doing so, one finds
\begin{equation}
    \Delta_{\rm 5d} \cF  = \frac{4}{3}\underbrace{\left[\frac{(c_{2,1}+c_{2,3}) j^1 +(2c_{2,1}+c_{2,3})j^2}{\sqrt{j^0}} + (c_{2,2}-c_{2,3})j^0 \right]}_{\cF}\, +\, \cO \left((j^0)^{5/2}\right).
\end{equation}
It is also useful to split which contribution comes from the ambient Laplacian and which one arises from the Hessian contraction with the normal hypersurface vectors. By looking at \eqref{eq:5dhess} we deduce that the ambient space Laplacian is such that
\begin{equation}\label{eq:ambient5dexpansion}
    \Delta_M \cF = 2 \left[\frac{(c_{2,1}+c_{2,3}) j^1 +(2c_{2,1}+c_{2,3})j^2}{\sqrt{j^0}} + (c_{2,2}-c_{2,3})j^0 \right]\, +\, \cO((j^0)^{5/2})\, .
\end{equation}
This can be interpreted as the fact that the quantum corrections spoil the Laplace equation in the bulk, in terms of powers of $J^0 \to 0$. Asymptotically, however, we recover the correct condition with a unique eigenvalue in the smooth fibration case. For the non-smooth fibration examples (cf. Appendix \ref{app:more5dmodels}), where the eigenvalue is different along different limits, this kind of uplift cannot be performed. Indeed, in this case, one cannot put $\cF$ in the form \eqref{CKF}.

On the other hand, if we take instead $\fF_6$ as defined above, and set up the 6d Laplace equation for the linear ansatz $\cF_{6}= c_{2,1} J^1 +c_{2,2}J^2$, we have that the ambient space Laplacian acts as
\begin{equation}
    \Delta^{(\rm 6d)}_J \cF_{6} = 2\cF_{6}\, .
\end{equation}
This matches what we get from the 5d ambient space computation along the $J^0 \to 0$ limit. The difference in eigenvalues is instead brought by the restriction to the submanifold. In the 5d case $\text{Hess}_\cF(\mathbf{n},\mathbf{n})^{(6d)} = \frac23 \cF$ while in 6d $\text{Hess}_\cF(\mathbf{n},\mathbf{n})^{(6d)} =  \cF$, as discussed in Appendix \ref{ap:meancurv}.

\section{Laplacian and Symmetries of Moduli Space}\label{s:Laplace&Symm}

Let us summarize our findings so far. For the case of maximal supergravity in $d \geq 8$, we find that the Wilson coefficient $\cF$ for the relevant higher-derivative operator obeys the eigenvalue equation \eqref{eq:eigenvalueeq} written in terms of the Laplacian in moduli space, with the exception of 10d Type IIA, where the proper differential equation \eqref{eq:typeIIAeq} can be thought of as inherited from Type IIA/B duality. With less supersymmetries, we find more exceptions, such as all cases with 16 supercharges or, in general, 5d $\cN=1$ supergravity modes. In this section, we will argue that the common thread between all these theories is that the naive isometries of the supergravity sigma-model are broken by other terms in the two-derivative effective action.

When referring to \textit{naive} (continuous) isometries, we specifically mean those associated to the kinetic terms for the scalar fields of the theory. In particular, for the dilaton in 10d it would correspond to a shift symmetry acting as $\phi \to \phi + \lambda, \ \lambda \in \bR$, while for the axio-dilaton of Type IIB it would be the standard $SL(2,\bR)$ action 
\begin{equation}
    \tau \to \frac{a \tau + b}{c \tau+d} \ , \quad \begin{pmatrix}
        a & b \\ c& d
    \end{pmatrix} \in SL(2,\bR) \ .
\end{equation}
To illustrate the link between these isometries and the Laplace-Beltrami operator, it is useful to reframe automorphic forms in representation-theoretic terms. Let us pick, for simplicity, the Maass forms of $SL(2,\bZ)$ given by the Eisentein series $E_s(\tau,\bar \tau)$ (cf. Appendix \ref{ss:SL2forms}). Recall that they are automorphic in the sense of being non-holomorphic modular forms, invariant under $SL(2,\bZ)$ \cite{DHoker:2022dxx}. Additionally, they satisfy a Laplace equation which schematically reads
\begin{equation}
    \Delta_{\rm sl_2} E_s(\tau,\bar \tau) = s(s-1) E_s(\tau,\bar \tau)\, ,
\end{equation}
and are at most of polynomial growth at the cusp $\tau_2 \to \infty$. Here, $\Delta_{\rm sl_2}$ is the usual hyperbolic Laplacian \eqref{eq:IIboperator} on the upper half-plane $\mathfrak{h}$. From the group-theoretic perspective, this space admits a differentiable action in terms of elements of $SL(2,\bR) \ni g$, which act on functions as
\begin{equation}
    f(g \cdot m) = (g \cdot f) (m)\, ,
\end{equation}
where we denote by $m$ any point on $\mathfrak{h}$.
This, in turn, allows one to relate to the Lie algebra element $x \in sl(2,\bR)$ the vector field
\begin{equation}\label{eq:liealgvecfield}
    (X_x f)(m) = \frac{d}{dt}f( e^{tx}\cdot m)\bigg|_{t=0}\, .
\end{equation}
Quadratic differential operators thus correspond to products of Lie algebra elements, which belong to the universal enveloping algebra. In this context, one distinguished element is the Casimir operator, commuting with the $sl(2,\bR)$ action. By the previous dictionary, and using the fact that the $G$-invariant operator descends through the quotient map $q: SL(2,\bR) \to SL(2,\bR)/SO(2)$ \cite{Garrett_2018}, this is identified with an invariant differential operator on $\mathfrak{h}$ of degree-two. To determine what this is, and following \cite{Garrett_2018}, we choose the generators
\begin{equation}
    H = \begin{pmatrix}
        1 & 0 \\
        0 & -1
    \end{pmatrix}\,, \quad X = \begin{pmatrix}
        0 & 1 \\
        0 & 0
    \end{pmatrix}\,, \quad Y =\begin{pmatrix}
        0 & 0 \\
        1 & 0
    \end{pmatrix}\, .
\end{equation}
The Killing form in this basis is 
\begin{equation}
    \langle w,v \rangle = \tr(vw)\, ,
\end{equation}
which means that the (unique) Casimir operator corresponds to 
\begin{equation}
    \cC^2 = \frac{1}{2}H^2 + XY +YX = \frac{1}{2}H^2 + 2YX -H\, ,
\end{equation}
where we also used the commutation relation $[X,Y] = H$. The computation can be further simplified by the fact that functions on the upper half plane are, by construction, right $SO(2)$-invariant. The generator of this $so(2)$ inside $sl(2,\bR)$ is given by $X-Y$, which means that acting with the corresponding differential operator $X_{X-Y}$ annihilates $f$. Then, the Casimir can be written as
\begin{equation}
    \cC^2 = \frac{1}{2}H^2+2X^2 - H\, .
\end{equation}
Mapping the generators to vector fields using \eqref{eq:liealgvecfield}, we thus identify
\begin{equation}
    X = y\partial_x\, , \quad H =2y\partial_y \quad \Longrightarrow \quad 
    \cC^2 = 2y^2 (\partial_x^2 + \partial_y^2) = 2\Delta_{\text{sl}_2}\, .
\end{equation}
Therefore, the Laplacian corresponds to the Casimir, and a Wilson coefficient sitting in an irreducible representation of $SL(2,\bR)$ will satisfy an eigenvalue equation. For the case of Maass forms, they are said to be \textit{automorphic representations} of $SL(2,\bR)$ since they are also invariant under the duality group \cite{Garrett_2018}. The same happens for all maximal supergravity theories studied in Chapter \ref{s:maxsugra}. At the quantum level, the classical group of symmetries is thus broken \textit{covariantly} by the higher-derivative operator, since its coefficient transforms in a specific (automorphic) representation.\footnote{Since there could be other Casimir operators that need to be diagonalized, we cannot argue for irreducibility in general. At most, we can say the representation should be an eigenspace of the only Casimir we inspect.} Of course, since we compute the Laplacian starting from the sigma-model metric, it is imperative that its naive isometry group is not broken by other interactions in the two-derivative action, as this spoils the identification with the classical group of symmetries of the theory. We will see that this is precisely what happens in all cases where the Laplacian is replaced by a more general elliptic second order operator.

Let us then turn to the exceptions we found, which are of two types. The first one is due to the moduli space being one-dimensional, with modulus $\phi$---i.e., the dilaton. When this is the case, the moduli space has only one possible candidate symmetry, which corresponds to a shift of the field $\phi \to \phi + \lambda$, as already mentioned. However, due to the presence of $e^{\phi}$ factors in the action, the interactions with other fields break this symmetry. Indeed, this is the statement that changing the vev of the dilaton modifies the low-energy physics. Focusing on Type IIA, this is evident as the strong-coupling and weak-coupling limits bring us to very different low-energy descriptions. Precisely the same holds for the 10d heterotic cases. This asymmetry is thus deeply related with the existence of a strongly-coupled M-theory description. The second type occurs in models with more than one moduli, where only a part of the naive isometries are broken. In particular, in the 9d cases with 16 supercharges, the symmetric space factor \eqref{eq:9d16modulisp} indeed describes proper symmetries of the two-derivative action, while the $\bR$ factor suffers the same fate as in the 10d Type IIA case. The 5d setup is more subtle, and the breaking of the sigma-model isometries has been discussed in \cite{de_Wit_1992}. Indeed, the upshot is that the cubic constraint that needs to be imposed on the coordinates $M^I$, i.e., $C_{IJK}M^IM^JM^K =6$, is such that the naive isometries associated to the ambient space are broken. Following \cite{de_Wit_1992}, we illustrate this point in the simple case where the sigma-model is given by
\begin{equation}
    \frac{SO(1,2)}{SO(2)}\, ,
\end{equation}
where the triple intersection numbers in $\{ M^1,M^2,M^3\}$-coordinates are $C_{122}=1,\ C_{233}=-1$, while all other vanish. Naively, the isometries of this model are three\footnote{In particular, since $so(1,2) \cong sl(2,\bR)$, they are the same as the ones in the upper-half plane $\mathfrak{h}$.} but the ones preserving the $C_{IJK}$ tensor and the $\fF_5=1$ constraint are just given by the following matrices acting linearly on $M^I$
\begin{equation}
    \tilde B_{(2)} = \begin{pmatrix}
        \frac{4}{3} & & \\
        & -\frac23 & \\
    & & \frac13
    \end{pmatrix}\, , \qquad  \tilde B_{(3)} = \begin{pmatrix}
        0& 0& 2 \\
        0& 0& 0 \\
        0& 1 &0   \\
    \end{pmatrix}\, ,
\end{equation}
which form a non-abelian, solvable subalgebra of the full naive isometry algebra. In particular, since the action is transitive, the moduli space is still homogenous, but it is not symmetric anymore, since one of the isometries of the sigma-model is absent. It is instructive to compare this with what happens in six dimensions. There, the action is written in terms of the intersection form $\Omega_{IJ}$, which is by construction $SO(1,T)$ invariant and also generates the constraint. Since this is the case, all naive isometries of the $SO(1,T)/SO(T)$ F-theory sigma-model will correspond to classical continuous symmetries of the 6d supergravity theory. Note that this also happens in four dimensions, where one can match the naive isometry group with the supergravity symmetry group by inspecting the 4d $\cN=2$ action.\footnote{This is discussed also in \cite{de_Wit_1992}, where the authors dimensionally reduce the above 5d example on a circle.}

\section{Summary and Discussion}\label{s:conclusions}

In this work, we have studied the leading higher-derivative corrections to the effective gravitational action in theories with 32, 16 and 8 supercharges and we have argued that they obey an eigenvalue equation in terms of an elliptic differential operator defined over moduli space. To this end, we examined operators that possess some degree of protection due to supersymmetry (or anomaly cancellation) and for which the Wilson coefficients---in particular their moduli-dependence---are known exactly. Specifically, we focused on the $\cR^4$-operator in setups with $d=10,9,8,$ preserving 32 or 16 supercharges, as well as on the $\cR^2$-term in $d=6,5,4,$ theories with $8$ supercharges. For all such cases, the eigenvalue turned out being dependent just on the dimensionality of the operator and the number of moduli in the low-energy EFT. Among the various possibilities, the ones which are marginal (in the Wilsonian sense) deserve special attention. In principle, whenever this occurs, the full coefficient (e.g., in maximal supergravity in 8d, cf. \eqref{eq:eigenvalueeq}) obeys a non-homogeneous differential equation with a (possibly constant) source term. However, as we commented in the main text, the most relevant moduli-dependence for the infinite distance limits---and for the computation of the quantum gravity cutoff $\Lambda_{\rm QG}$---is captured instead by the homogeneous zero-eigenvalue equation, given that the additional contributions associated to the source terms reflect thresholds corrections due to massless modes or, in 4d $\cN=2$, field-theoretic states decoupling from gravity. In many instances, the form of the elliptic operator was shown to agree with Laplace-Beltrami. For the remaining ones, string dualities allowed us to motivate its form by linking the latter to Laplacians across various dimensions. These include 10d Type IIA (cf. Section \ref{ss:10dIIA}), the $d=10,9,$ heterotic cases discussed in Section \ref{ss:16sugra}, and 5d $\cN=1$ supergravity (Section \ref{ss:5dMtheory}).

We have further argued that this Laplace-like equation encodes non-trivial information about the behavior of the quantum gravity cutoff $\Lambda_{\rm QG}$ close to asymptotic limits in moduli space. Specifically, we have shown in Section \ref{s:maxsugra} that solving the equation by an exponential ansatz in the canonically normalized saxions allows one to put bounds on the species vectors as well as their related polytope in maximal supergravity. In the same spirit, we related the explicit connection between the $d$ and $d+1$ eigenvalue equations upon circle compactification to the relationship between their respective convex hulls. The same has been moreover discussed in cases with less supersymmetries in Section \ref{s:8supercharges}, where the $\cR^2$ Wilson coefficient depends only on the tensor/vector-branch of the moduli space. The way to motivate this ansatz from the bottom-up is two-fold. First, in maximal supergravity, it reproduces the typical asymptotic behaviour of Eisenstein series (i.e., their `constant terms'), which are required by duality to appear in the Wilson coefficients. On the other hand, this is also the characteristic form one expects from the Distance Conjecture \cite{Ooguri:2006in}. In the case of Type IIA on a Calabi--Yau, we remarked that this ansatz captures the ubiquitous shift-invariance that arises in infinite distance limits in K\"ahler moduli space \cite{Corvilain:2018lgw}. Notice that, by finding an independent bottom-up rationale for this ansatz, one can also motivate the Distance Conjecture. 

It is interesting that the elliptic operator does not always correspond to the Laplacian, which is naturally defined in moduli space. As discussed in Section \ref{s:Laplace&Symm}, a common thread between the exceptions is that their naive continuous isometries---i.e., those associated to the scalar kinetic terms of the theory---are broken by other terms in the two-derivative action. One prototypical example of this is 10d Type IIA, where the naive shift symmetry associated to the dilaton $\phi \to \phi + \lambda$, with  $\lambda \in \bR$, is broken by $e^\phi$-interactions with other sectors, even at string tree-level. Physically, this is related to the fact that the low-energy physics is sensitive to its vev and, in particular, it informs us about the existence of a strong coupling limit different from the weak-coupling one. In the 5d M-theory case, the breaking of the supergravity sigma-model isometries has been already discussed in \cite{de_Wit_1992}, and it is due to the constant volume constraint imposed on the projective (real) special K\"ahler geometry.

As a closing remark, one could try to speculate as to why such an eigenvalue equation should hold, in the first place. Taken at face value, one is reminded of a steady-state heat equation, which could encode asymptotically non-trivial information about the entropy of towers of sates becoming light \cite{Cribiori:2023ffn,Herraez:2024kux}. This is very suggestive, as in four dimensions the $\cR^2$-coefficient coincides with the entropy of species-sized (i.e., `small' in the sense of \cite{Sen:1994eb,Sen:1995in,Hamada:2021yxy}) black holes. From the worldsheet perspective instead, the maximal supergravity constraint for the Wilson coefficient in $d\le9$ can be associated with the remarkable analogous differential equation satisfied by the partition function of Narain CFTs \cite{Obers_2000, Maloney:2020nni,benjamin2021harmonic}. Moreover, as argued in \cite{Aoufia:2024awo}, the equation remains true asymptotically in moduli space, regardless of the amount of supersymmetry or the nature of the internal CFT. This corroborates the fact that the latter should be closely related with the counting of states and their entropy. As another possible interesting link with Swampland conjectures, notice that the form of the equations that we found resembles fixed points of the geometric flow discussed e.g., in \cite{Kehagias:2019akr, DeBiasio:2023hzo} and more recently in \cite{Demulder:2024glx}, which aim to generalize the Distance Conjecture beyond exact moduli spaces. Lastly, the differential equations we obtained can be of practical use in defining an appropriate measure of complexity of the species scale function in the sense of \cite{Grimm:2024elq}. From the perspective of the effective theory, this aligns with the Tameness Conjecture of \cite{Grimm:2021vpn}, as these higher-derivative corrections should be more appropriately defined as tame functions.

\vspace{0.05cm}
		
\textbf{Acknowledgments.} We are indebted to I. Basile, J. Calderón-Infante, D. van de Heisteeg, F. Marchesano, L. Melotti, M. Montero, A. Uranga, C. Vafa, and I. Valenzuela for discussions, and to A. Herráez for collaboration during the initial stages. A.C. thanks the Aspen Center for Physics, funded by the NSF grant PHY-2210452, for hospitality. C.A. and L.I. are supported through the grants CEX2020-001007-S and PID2021-123017NB-I00, funded by MCIN/AEI/
13039/501100011033 and by ERDF `A way of making Europe'. The work of A.C. is supported by a Kadanoff, an Associate KICP fellowships, and through the NSF grants PHY-2014195 and PHY-2412985. A.C. is grateful to T. Lobo for her continuous encouragement and support.


\appendix

\section{Relevant Automorphic Forms}
\label{ap:Massform}

This appendix provides a brief overview of the relevant literature on automorphic functions, focusing mostly on the discrete groups $SL(2, \mathbb{Z})$ and $SL(3, \mathbb{Z})$ capturing the duality symmetries in 10d, 9d and 8d maximal supergravity (see Section \ref{s:maxsugra}). A similar analysis can be performed for the (bigger) duality groups that arise upon reducing the number of non-compact spacetime dimensions, but we refrain from reviewing those in the present work. 

For the purposes of this paper, an automorphic function with respect to a given continuous group $G$ and a discrete subgroup $\Gamma \subset G$ is defined as a map from a space $\mathcal{M}$ admitting a differentiable $G$-action to $\mathbb{R}$ (or more generally $\mathbb{C}$), such that it is left invariant under the corresponding $\Gamma$-action. In the following, we will particularize to those automorphic functions of $G$ which are moreover real analytic since, as remarked in the main text, they appear as Wilson coefficients in the EFT expansion of the gravitational effective Lagrangian in maximal supergravity theories. More specifically, we will mainly be interested in their rich interplay with the spectral theory of $\cM$. In fact, there is an economic way to generate such analytic functions as eigenfunctions of some appropriate $\Gamma$-invariant\footnote{Since the space admits a $G$-action, which defines a $\Gamma$-action on functions, by $\Gamma$-invariant we mean that the action commutes with the $\Gamma$-action on said functions.} elliptic operator. Lastly, their definition requires  them to be (at most) of polynomial growth at the cusp points\footnote{These can be more easily understood as the infinite distance loci of $\Gamma\backslash\cM$.} of $\cM$. In the remainder of the appendix, we specialize to the cases where $\Gamma=\{SL(2, \bZ), \, SL(3,\bZ)\}$, focusing on the asymptotic behaviour of the respective automorphic functions, closely following \cite{Kiritsis:1997em,Green:2010wi,benjamin2021harmonic,Castellano:2023aum,Castellano:2024bna}.

\subsection{$SL(2, \mathbb{Z})$ Maass waveforms}\label{ss:SL2forms}

The real-analytic Eisenstein series $E_\ell^{sl_2}(\tau,\bar \tau)$ constitute, for generic values of $\ell \in \bC$, the set $\{\mathcal{E}_\ell\}$ of (non-holomorphic) automorphic forms of $SL(2,\bZ)$ on the upper-half plane, also called \textit{Maass waveforms}.\footnote{For specific values of $\ell$, one also needs to consider \textit{Maass cusp forms}, which vanish as $\tau_2 \to \infty$ and are associated to the discrete part of the spectrum of $\Delta_2$. As such, we will not be interested in them in what follows. The interested reader can find more details in \cite{DHoker:2022dxx}.} The value of $\ell$ labels the eigenvalue with respect to the hyperbolic Laplacian operator \eqref{eq:IIboperator} 
\begin{equation}
   \Delta_{\rm sl_2} E_\ell^{sl_2} =\tau_2^2 \left(\partial_{\tau_1}^2 + \partial_{ \tau_2}^2\right)E_\ell^{sl_2} =\ell(\ell-1)E_\ell^{sl_2} \ ,
\end{equation}
where once again $\tau=\tau_1 +i\tau_2$. Eisenstein series can be formally defined through a Poincarè series over images of the modular group as
\begin{equation}\label{eq:nonholoEisenstein}
    E_\ell^{sl_2}(\tau,\bar \tau) \equiv \zeta(2\ell)\sum_{\gamma \in \Gamma_\infty\backslash PSL(2,\bZ)} (\mathrm{Im}\gamma \tau)^\ell = \sum_{(m, n) \in \mathbb{Z}^2 \setminus \lbrace (0,0) \rbrace} \frac{\tau_2^\ell}{\left| m+n\tau\right|^{2\ell}}\ ,
\end{equation}
where $\Gamma_\infty$ is the subgroup of $PSL(2,\bZ)$ fixing $\tau_2$. The previous expression is modular-invariant due to the sum, and although it converges absolutely only if $\text{Re}\, \ell >1$, it admits a meromorphic continuation to the entire complex plane. The polynomial growth of the Eisenstein series is apparent, since upon taking the limit $\tau_2 \to \infty$, the infinite series is clearly dominated by the terms with $n=0$. More precisely, the functions $E_{\ell}^{sl_2}(\tau,\bar \tau)$ have an alternative Fourier expansion in $\tau_1$, which can be obtained upon Poisson resumming on the integer $n$, yielding
\begin{align}\label{eq:nonpertexpansion}
	\notag E_{\ell}^{sl_2}(\tau , \bar \tau) =\, & \bigg[ 2\zeta(2\ell) \tau_2^{\ell} + 2\pi^{1/2}\frac{\Gamma(\ell-1/2)}{\Gamma(\ell)} \zeta(2\ell-1) \tau_2^{1-\ell}\\
 &+ \frac{8 \pi^\ell \tau_2^{1/2}}{\Gamma(\ell)} \sum_{m=1}^{\infty} m^{\ell-1/2} \sigma_{1-2\ell} (m)\, \cos(2\pi m \tau_1)\, K_{\ell-1/2} (2\pi m \tau_2)\bigg]\, ,
\end{align}
where $\sigma_{1-2\ell} (m) = \sum_{d|m} d^\ell$ runs over all divisors $d$ of $m$, and $K_\ell(y)$ is the modified Bessel function of second kind, which is defined as follows
\begin{equation}
    K_\ell(y)=\frac{1}{2} \int_0^{\infty} dx\, x^{\ell-1} \exp \left[ -\frac{y}{2} \left( x + \frac{1}{x}\right)\right]\, ,
    \end{equation}
and decays asymptotically as $ K_\ell(y) \sim y^{-1/2} e^{-y}$ for $y \to \infty$.
Thus we see that, as outlined in the main text, the large $\tau_2$ behaviour can be extracted by taking the zero mode---also referred to as \textit{constant term}---with respect to this Fourier decomposition. Similarly, for larger groups possessing a greater number of compact variables, the behaviour of automorphic functions near cusp points can be extracted as above by inspecting their constant terms.

The meromorphic continuation of the Eisenstein series, as a function of $\ell$, has simple poles at $\ell=0,1$. Nonetheless, one can extract the regular part of $E_{\ell=1}^{sl_2}(\tau,\bar \tau)$ by taking $s=1+\epsilon$ and expanding in a Laurent series \cite{Green:2010wi} as
\begin{equation}\label{eq:sl2eisregularized}
     E_{1+\epsilon}^{sl_2}(\tau,\bar \tau) = \frac{\pi}{\epsilon} - 2\pi\left(\gamma_{\text{e}} - \log 2\right) - \pi \log \left( \tau_2\,|\eta(\tau)|^4 \right) + \cO(\epsilon)
\end{equation}
where $\gamma_{\text{e}}$ is the Euler-Mascheroni constant and $\eta(\tau)$ denotes the Dedekind eta function, which may be defined as
\beq \label{eq:Dedekind}
\eta(\tau) = q^{\frac{1}{24}} \prod_{k=1}^{\infty} \left( 1-q^k\right)\, , \qquad q=e^{2\pi i \tau}\, .
\eeq
It is thus useful to define for this particular value of $\ell$ a \textit{regularized} Eisenstein series by subtracting the pole and a constant, getting to
\begin{equation}
    \hat E_{1}^{sl_2} (\tau,\bar \tau)= -\pi \log \left( \tau_2\,|\eta(\tau)|^4 \right) \ .
\end{equation}
To conclude, let us note that even though the function $\hat{E}_{1}^{sl_2}(\tau)$ arises as the regularization of $E_{1}^{sl_2}(\tau)$, it is actually not strictly speaking a Maass form, since 
\begin{equation}
\Delta_2 \hat{E}_{1}^{sl_2}(\tau) = \pi \ ,
\end{equation}
and in particular is not proportional to $\hat{E}_{1}^{sl_2}(\tau)$ itself.
In any event, what remains true is that the large modulus behaviour of $\hat{E}_{1}^{sl_2}(\tau)$ matches with that expected for $E_{\ell=1}^{sl_2}(\tau)$, since upon using the Fourier series expansion for $\eta(\tau)$
\beq
\eta(\tau) = q^{\frac{1}{24}} \left( 1-q-q^2+q^5 + \mathcal{O}(q^7) \right)\, ,
\eeq
one finds the following relevant asymptotic expression 
\beq \label{eq:asymptotic behavior}
-\pi \text{log} \left(\tau_2\,|\eta(\tau)|^4\right)\, \sim\, -\pi \text{log} \left(\tau_2\,e^{-\frac{\pi \tau_2}{3}}\right)\, \sim\, \frac{\pi^2}{3} \tau_2 - \pi \text{log} (\tau_2)\, ,
\eeq
whose first term precisely is $2 \zeta(2) \tau_2$ (cf. eq. \eqref{eq:nonpertexpansion}). The log-terms that are subleading in this large-$\tau_2$ expansion and contribute to $\hat{E}_{1}^{sl_2}(\tau)$, as discussed in the main text, are related to the massless threshold log-corrections that appear e.g., when dimensionally regularizing the field-theoretic divergence of the amplitude in $d=8$. In fact, one can show that this regularization procedure is equivalent to the pole subtraction in \eqref{eq:sl2eisregularized} \cite{Green:2008uj,Green:2010wi}.

\subsection{$SL(3, \mathbb{Z})$ Maass waveforms}\label{ss:SL3forms}

We now turn to the case of $SL(3,\bZ)$-automorphic functions on the coset space $\cM = SL(3, \mathbb{R})/$ $SO(3)$. Adopting the parametrization $\{ \nu, \tau,b,c\}$ defined in Section \ref{ss:8dmaxsugra} (see also \cite{Castellano:2023aum,Castellano:2024bna}), let us first introduce the appropriate $SL(3, \mathbb{Z})$-invariant Laplace operator \eqref{eq:8dlapl} on $\cM$
\beq \label{eq:laplacianSL3}
\Delta_{\rm sl_3} = 4\tau_2^2 \partial_{\tau} \partial_{\bar \tau} + \frac{1}{\nu \tau_2} \left| \partial_b -\tau \partial_c \right|^2 + 3 \partial_{\nu} \left( \nu^2 \partial_{\nu}\right)\, .
\eeq
It is useful to group the previous coordinates into the following $3\times3$ matrix (see e.g., \cite{Kiritsis:1997em})
\beq
 \mathcal{B}= \nu^{1/3} \begin{pmatrix}
		\frac{1}{\tau_2} \quad  \frac{\tau_1}{\tau_2} \quad \frac{c+\tau_1 b}{\tau_2}\\ \frac{\tau_1}{\tau_2} \quad  \frac{|\tau|^2}{\tau_2} \quad \frac{\tau_1 c+|\tau|^2 b}{\tau_2}\\ \frac{c+\tau_1 b}{\tau_2} \quad  \frac{\tau_1 c+|\tau|^2 b}{\tau_2} \quad \frac{1}{\nu} + \frac{|c+\tau b|^2}{\tau_2}
	\end{pmatrix}\, ,
\eeq
which satisfies $\mathcal{B}=\mathcal{B}^{\text{T}}$ as well as $\det \mathcal{B}=1$. The matrix $\mathcal{B}$ is introduced due to its transformation properties under $SL(3,\bZ)$. Specifically, upon performing some transformation $\mathcal{A} \in SL(3, \mathbb{Z})$, one finds that it transforms in the adjoint as
\beq \label{eq:Btransf}
  \mathcal{B} \rightarrow \mathcal{A}^{\text{T}}\, \mathcal{B}\, \mathcal{A}\, .
\eeq
With this, we are now ready to define the Eisenstein $SL(3, \mathbb{Z})$ series of order $\ell$:
\beq\label{eq:SL3Eisenstein}
E_{\ell}^{sl_3} = \sum_{\mathbf{n}\, \in\, \mathbb{Z}^3 \setminus \lbrace \vec{0} \rbrace} \left( \sum_{i, j =1}^3 n_i\, \mathcal{B}^{ij}\, n_j\right)^{-\ell} = \sum_{\mathbf{n}\, \in\, \mathbb{Z}^3 \setminus \lbrace \vec{0} \rbrace} \nu^{-\ell/3} \left[ \frac{\left| n_1 + n_2 \tau + n_3 \left( c+\tau b\right)\right|^2}{\tau_2} + \frac{n_3^2}{\nu}\right]^{-\ell}\, ,
\eeq
with $\mathcal{B}^{ij}$ denoting the components of the inverse matrix of $\mathcal{B}$. Note that the above expression is manifestly $SL(3, \mathbb{Z})$-invariant, since the vector $\textbf{n} = \left( n_1, n_2, n_3\right)$ transforms as $\textbf{n} \rightarrow \mathcal{A}^{\text{T}}\, \textbf{n}$ under the duality group. As also happened with the non-holomorphic Eisenstein series defined in eq. \eqref{eq:nonholoEisenstein} above, the functions $E_{\ell}^{sl_3}$ are eigenvectors of the Laplacian $\Delta_3$, satisfying
\beq 
  \Delta_{\rm sl_3} E_{\ell}^{sl_3} = \frac{2\ell (2\ell-3)}{3} E_{\ell}^{sl_3}\, .
\eeq
Once again, the series is absolutely convergent if $\ell>3/2$ but can be meromorphically continued over the complex plane, with simple poles in $\ell=0,\frac32$. Following the procedure outlined in the $SL(2,\bZ)$ case, one may define a \textit{regularized} Eisenstein series for $\ell=3/2$ by employing the Kronecker limit formula \cite{terras2013harmonic} 
\beq \label{eq:regularisation}
  \hat{E}_{3/2}^{sl_3} \equiv \lim_{\ell\to 3/2} \left( E_{\ell}^{sl_3} - \frac{2\pi}{\ell-3/2} - 4\pi(\gamma_{\text{e}}-1)\right)\, ,
\eeq
where again $\gamma_{\text{e}}$ denotes the Euler-Mascheroni constant. Such newly defined function is no longer singular and remains invariant under $SL(3, \mathbb{Z})$ transformations, with the price of not being a zero-mode of the Laplacian \eqref{eq:laplacianSL3} anymore.

Following suit the $SL(2,\bZ)$ discussion, we are now ready to perform a Fourier expansion in the compact directions $\{\tau_1,b,c\}$ of $\cM$. This will allow us to rewrite the $SL(3,\mathbb{Z})$ Eisenstein series in a way which makes manifest the perturbative and non-perturbative origin of the different terms that appear in the expansion. We closely follow Appendix A of \cite{Kiritsis:1997em, Castellano:2023aum}. First, let us introduce the following integral representation
\begin{align}\label{eq:integralrep}
	\notag E_{\ell}^{sl_3} &= \frac{\pi^\ell}{\Gamma(\ell)} \int_0^{\infty} \frac{dx}{x^{1+\ell}} \sum_{\mathbf{n}\, \in\, \mathbb{Z}^3 \setminus \lbrace \vec{0} \rbrace} \exp \left[-\frac{\pi}{x} \left( \sum_{i, j =1}^3 n_i\, \mathcal{B}^{ij}\, n_j\right) \right]\\
 &= \nu^{-\ell/3} \frac{\pi^\ell}{\Gamma(\ell)} \int_0^{\infty} \frac{dx}{x^{1+\ell}} \sum_{\mathbf{n}\, \in\, \mathbb{Z}^3 \setminus \lbrace \vec{0} \rbrace} \exp \left[-\frac{\pi}{x} \left( \frac{\left| n_1 + n_2 \tau + n_3 \left( c+ \tau b\right) \right|^2}{\tau_2} + \frac{n_3^2}{\nu}\right) \right]\, ,
\end{align}
which can be shown to coincide with the defining series \eqref{eq:SL3Eisenstein} after performing the change of variables $y=x^{-1}$ and using the definition of the $\Gamma$-function 
\begin{align}
	\Gamma(z) = \int_0^{\infty} dy\, y^{z-1} e^{-y}\, .
\end{align}
After carefully separating the sum in the integers $n_i$ and performing a series of Poisson resummations, one arrives at a Fourier series expansion of the form \cite{Kiritsis:1997em,Basu:2007ru,Basu:2007ck}
\begin{align}\label{eq:instexpSL3}
	\notag E_{\ell}^{sl_3} &= 2\nu^{-\ell/3} \tau_2^\ell \zeta(2\ell) + 2 \sqrt{\pi} T_2 \left( \tau_2 \nu^{1/3}\right)^{3/2-\ell} \frac{\Gamma(\ell-1/2)}{\Gamma(\ell)} \zeta(2\ell-1) + 2\pi \nu^{2\ell/3-1} \frac{\zeta(2\ell-2)}{\ell-1}\\
 &+  2 \frac{\pi^\ell \sqrt{\tau_2}}{\Gamma(\ell) \nu^{\ell/3}} \sum_{m,n \neq 0} \left| \frac{m}{n}\right|^{\ell-1/2} e^{2\pi \text{i} m n\tau_1}\, K_{\ell-1/2} (2\pi |m n| \tau_2)\, +\, \sum_{m, n \in \mathbb{Z} \setminus \lbrace (0,0) \rbrace} \mathcal{I}^\ell_{m, n}\, ,
\end{align}
where we have defined $T_2 \equiv \text{Im}\, T$, with $T= b+ \text{i} \left( \nu \tau_2\right)^{-1/2}$, and 
\begin{align}
	\mathcal{I}^\ell_{m, n} = 2\frac{\pi^\ell \nu^{\ell/6-1/2}}{\Gamma(\ell) \tau_2^{\ell/2-1/2}} \sum_{k \neq 0} \left| \frac{m+n\tau}{k}\right|^{\ell-1} e^{2\pi \text{i} k \left[n(c+\tau_1 b)- (m+n\tau_1)b \right]}\, K_{\ell-1} \left(2\pi |k|\frac{\left| m+n\tau \right|}{\sqrt{\nu \tau_2}}\right)\, .
\end{align}
Notice that upon using the Fourier expansion for the $SL(2,\mathbb{Z})$ series in eq. \eqref{eq:nonpertexpansion}, one can group the terms which depend on $\nu^{-\ell/3}$ into the following expression
\begin{align}\label{eq:SL3&SL2}
	 E_{\ell}^{sl_3} &= \nu^{-\ell/3} E_{\ell}^{sl_2}(\tau) + 2\pi \nu^{2\ell/3-1} \frac{\zeta(2\ell-2)}{\ell-1}\, +\, \sum_{m, n \in \mathbb{Z} \setminus \lbrace (0,0) \rbrace} \mathcal{I}^\ell_{m, n}\, .
\end{align}
It should be emphasized that the fact that the same modular Eisenstein series appear also in eq. \eqref{eq:SL3&SL2} can be understood from dimensional reduction arguments, since the 8d theory inherits certain perturbative and non-perturbative corrections from its 10d and 9d analogues, where the $SL(2,\mathbb{Z})$ duality group is all there is. Additionally, the sum over the pair of integers $(m,n)$ corresponds to certain instanton corrections arising from bound states of $(p,q)$-strings wrapping some $T^2$ of the internal geometry.

To close the appendix, let us briefly discuss the particular case of the regularized $SL(3,\mathbb{Z})$ Eisenstein series with $\ell=3/2$, given its role in Section \ref{ss:8dmaxsugra} from the main text.\footnote{This is easy to see from eq. \eqref{eq:instexpSL3} above, since the functions $\zeta(1+x)$ as well as $\Gamma(x)$ present simple poles at $x=0$. Indeed, one obtains the following expansions around the pole:
\beq
\notag \zeta(1+\epsilon)= \frac{1}{\epsilon} + \gamma_{\text{e}} + \mathcal{O}(\epsilon)\, , \qquad \Gamma(\epsilon)=\frac{1}{\epsilon} - \gamma_{\text{e}} + \mathcal{O}(\epsilon)\, .
\eeq} Upon using \eqref{eq:regularisation} in the above expansion, one finds the following expression \cite{Kiritsis:1997em}
\begin{align}\label{eq:Eisenstein3/2}
	\notag \hat{E}_{3/2}^{sl_3} &= 2\zeta(3) \frac{\tau_2^{3/2}}{\nu^{1/2}} + \frac{2 \pi^2}{3} T_2 + \frac{4\pi}{3} \log \nu\\
 &+  4\pi \sqrt{\frac{\tau_2}{\nu}} \sum_{m,n \neq 0} \left| \frac{m}{n}\right| e^{2\pi \text{i} m n\tau_1}\, K_{1} (2\pi |m n| \tau_2)\, +\, \sum_{m, n \in \mathbb{Z} \setminus \lbrace (0,0) \rbrace} \mathcal{I}^{3/2}_{m, n}\, ,
\end{align}
which rewritten in terms of the M-theory coordinates $\{\cV,\tau,R_3 \}$ discussed in Section \ref{ss:8dmaxsugra} becomes \cite{Green:2010wi,Castellano:2023aum,Castellano:2024bna}
\begin{equation}\label{eq:Eisenstein3/2-2}
\begin{split}
	\hat{E}_{3/2}^{sl_3} & = 2 \zeta(3) \mathcal{V}_2^{9/7} R_3^{-2} + \frac{2\pi^2}{3} \tau_2 - 2\pi \log(\tau_2) - \frac{2\pi}{3} \log\left( \mathcal{V}_2^{9/7} R_3^{-2} \right) + \cdots .
\end{split}
\end{equation}
Once again, it is important to stress that the leading asymptotic behaviour of each automorphic function is capture by the zero mode of its Fourier expansion---i.e., its constant term---with respect to the compact directions within $\cM$. Moreover, the leading log-divergence brought about by massless thresholds of the $\ell=3/2$ Eisenstein series can again be understood from the dimensional regularization of the four-graviton amplitude \cite{Green:2008uj,Green:2010wi}.

\section{Details on Mean Curvature and Hessian Analysis}\label{ap:meancurv}

In this appendix, we provide some details concerning the determination of the the mean curvature and Hessian differential forms corresponding to the submanifolds that define the moduli spaces of 5d and 6d (minimal) supergravity theories. This calculation is relevant since the Laplace operators defined both in the ambient space and on the constrained manifold are related to each other precisely by the aforementioned quantities, as per \eqref{eq:laplonconstr}. In particular, the hypersurfaces are implicitly defined in terms of a constant volume slice $\fF_ =1$ of either the whole Calabi--Yau or the base $B_2$ of the elliptic fibration, respectively\footnote{Recall that the indices $I,J,$ in \eqref{eq:constantvolhypersurface} run over $1,\ldots, n_{V,T},$ for the 5d and 6d supergravity cases, respectively (see Sections \ref{ss:6dFthy} and \ref{ss:5dMtheory} for details).}
\begin{equation}\label{eq:constantvolhypersurface}
\begin{split}
    &\fF_5 = \frac{1}{6}C_{IJK} M^I M^J M^K\, , \\
    &\fF_6 = \Omega_{IJ}J^I J^J\, ,
\end{split}
\end{equation}
Furthermore, recall that in both cases the metric tensor in the ambient space is given by $g_{IJ} = -\frac12\partial_I \partial_J \log \fF$, such that the normal vector to the hypersurface can be written as 
\begin{equation}
    n^L = g^{LK}\frac{\partial_K \mathfrak{F}}{|\partial \mathfrak{F}|}\, .
\end{equation}
Therefore, using the properties of the K\"ahler moduli, we prove in the following that the mean curvature with respect to the ambient space, namely the quantity 
\begin{equation}
    (\mathbf{H}_{\mathscr{M}})^L = -(\nabla_K n^K)n^L\, ,
\end{equation}
vanishes for both five- and six-dimensional theories. For the Hessian 2-form, we will show instead that, when restricting ourselves to linear functions in ambient space, the latter becomes proportional to the function itself.

\subsection{Five-dimensional case}\label{ss:Hessian5d}

In 5d $\mathcal{N}=1$ supergravity, the inverse moduli space metric (in ambient space) takes the explicit form
\begin{equation}
    g^{KL} = M^KM^L -2C^{KL}\fF_5\, ,
\end{equation}
where $C^{KL}$ is the inverse of $C_{KL} \equiv C_{KLM}M^M$. With this, one may easily prove the following identities by direct computation
\begin{equation}\label{eq:identities5d}
    g^{IJ}C_I C_J  = \frac23 C^2\, ,\quad C^{IJ}C_J = M^I\, ,\quad  g^{IJ}C_J = \frac23 C M^I\, ,\quad   M^I g_{IJ} = \frac32 \frac{C_J}{C}\, ,
\end{equation}
where we introduced the notation $C_I = C_{IJK}M^JM^K$ and $C=6\fF_5$. Equation \eqref{eq:identities5d} implies that the norm $|\partial \fF_5| = \sqrt{g^{KL}\partial_K \fF_5 \partial_L \fF_5} = \sqrt{6}\, \fF_5$, from which it follows that
\begin{equation}\label{eq:meancurvvanishes}
    n^L = \sqrt{\frac23}M^L \quad \Longrightarrow \quad \nabla_K n^L = 0\, ,
\end{equation}
where we used the relation $g^{KL}\partial_L\fF_5 = 2\fF_5 M^K$. In other words, the normal vector to the constrained hypersurface is covariantly constant and thus the constrained manifold is said to be \textit{totally geodesic} \cite{KobayashiNomizu}. This implies that the mean curvature of the hypersurface vanishes identically and that the normal vector field $n^I$ can be integrated to give a global section of the normal bundle associated to the submanifold, which becomes then automatically trivial. 

On the other hand, assuming that the function of interest $f(M)$ is linear in ambient space---motivated by the dependence of the $\mathcal{R}^2$ Wilson coefficient \eqref{eq:R25dsugra} in 5d M-theory, one computes the Hessian contribution $\text{Hess}_f(\mathbf{n},\mathbf{n})$ as follows
\begin{equation}\label{eq:hess5d}
    \text{Hess}_f(\mathbf{n},\mathbf{n}) =\frac23 M^IM^J \nabla_I \partial_Jf \stackrel{\text{linear }f}{=}  -\frac{2}{3}M^IM^J\Gamma^K_{IJ}\partial_K f\, .
\end{equation}
Imposing now that the Christoffel symbols are determined by
\begin{equation}
    \Gamma_{IJ}^K = -\frac{1}{4}g^{KP}\partial_{I}\partial_{J}\partial_{P}\log \fF_5\, ,
\end{equation}
we can rewrite \eqref{eq:hess5d} using the relation $\Gamma_{IJ}^KM^I = -\delta^K_J$ as follows
\begin{equation}\label{eq:5dhess}
    \text{Hess}_f(\mathbf{n},\mathbf{n}) \stackrel{\text{linear }f}{=} \frac{2}{3}f\, .
\end{equation}

\subsection{Six-dimensional case}\label{ss:Hessian6d}

Similarly to the five-dimensional case, in 6d $\mathcal{N}=(1,0)$ supergravity theories the inverse metric reads
\begin{equation}
    g^{IJ} = 2J^IJ^J - \fF_6 \Omega^{IJ}\, ,
\end{equation}
where $\Omega^{IJ}$ denotes the inverse of $\Omega_{IJ}$. Analogous properties to the ones shown in \eqref{eq:identities5d} hold for the 6d theories, namely
\begin{equation}\label{eq:identities6d}
    g^{IJ}J_I J_J = \fF_6^2\, ,\quad \Omega^{IJ}J_J = J^I\, ,\quad  g^{IJ}J_J = \fF_6 J^I\, ,\quad J^I g_{IJ} = \frac{J_J}{\fF_6}\, ,
\end{equation}
where we have defined $J_J = \Omega_{IJ}J^J$ above. Using these identities, it is easy to verify that, once again, we have
\begin{equation}
    n^L = J^L \quad \Longrightarrow \quad \nabla_K n^L
 = 0\, ,
\end{equation} 
implying that the mean curvature of the volume hypersurface defined by $\fF_6 =1$ is zero. The Hessian computation proceeds very much in the same way, where by using $\Gamma_{IJ}^K J^J =-\delta^K_I$ one arrives at
\begin{equation}
    \text{Hess}_f (\mathbf{n},\mathbf{n}) \stackrel{\text{linear }f}= -J^LJ^M\Gamma_{ML}^K \partial_K f  = f\, ,
\end{equation}
with the first equality being understood as valid for any linear function $f(J)=c_I J^I$ in ambient space.

\section{Solutions to Laplace Equation in del Pezzo Surfaces}\label{ap:delPezzolaplace}

In this appendix, we provide more details concerning the study and determination of Laplace-Beltrami eigenfunctions in 6d $\mathcal{N}=(1,0)$ EFTs obtained from compactifying F-theory on del Pezzo surfaces. We consider first the simpler case of $dP_2$, where it is shown how the Laplace condition singles out the known moduli structure of the $\mathcal{R}^2$-operator within these theories. Subsequently, in Section \ref{ss:FthydPr}, we explain how to extend this result for any such 6d model. The material presented here is complementary to the discussion in Section \ref{sss:examples6d}.

\subsection{F-theory on $dP_2$}\label{ss:FthydP2}

As already discussed around eq. \eqref{eq:parametrizationdP2}, the hypersurface constraint \eqref{eq:6dprepotconstraint} can be solved in the present two-dimensional example by introducing a pair of real coordinates $(x \geq 0,\theta \in [0,\pi/2])$ leading to the following line element
\begin{equation}
    \td s^2 = \td x^2 + \sinh^2 x\,  \td \theta^2\, , 
\end{equation}
which thus parametrize (a slice) of the hyperbolic 2d plane. The corresponding Laplace operator reads as
\begin{equation}\label{eq:laplacedP2}
 \Delta_{\rm 6d} = \partial^2_x \ + \coth x \, \partial_x + \frac{1}{\sinh^2 x}\partial_\theta^2\, .
\end{equation}
and it severely constrains the moduli dependence of a certain protected curvature-squared operator entering the resulting EFT. The latter is seen to satisfy the differential condition
\begin{equation}\label{eqap:laplaceeqdP2}
    \Delta_{\rm 6d} \cF(x,\theta) = 2 \cF(x,\theta)\, .
\end{equation}
Here we aim to show how, starting directly from \eqref{eqap:laplaceeqdP2}, we can recover the linear dependence on the K\"ahler moduli $J^I$. To do so, and thanks to the periodicity of the angular variable $\theta$, we first expand the function $\cF(x, \theta)$ in a Fourier series
\begin{equation}\label{eqap:Fourier6ddP2}
    \cF (x,\theta) = \sum_{n \in \mathbb{Z}} f_n(x) e^{in\theta }\, ,\qquad \text{with}\quad f_{-n}(x)=f_n^*\, .
\end{equation}
Therefore, imposing the Laplace condition above results in the following differential equation for each mode
\begin{equation}\label{eq:diffeqfn}
    \frac{d^2f_n}{dx^2}+\coth x \frac{df_n}{dx}-\frac{1}{\sinh^2 x}\left( n^2+2\sinh^2 x\right) f_n=0\, .
\end{equation}
To exhibit explicitly its solutions, it is convenient to first perform the change of coordinate $\rho=\cosh x$, which transforms the previous equation into
\begin{equation}\label{eq:Legendre}
    (1-\rho^2)\frac{d^2f_n}{d\rho^2} -2\rho \frac{df_n}{d\rho} +\left( 2+\frac{n^2}{\rho^2-1}\right) f_n=0\, .
\end{equation}
Notice that this is nothing but the Legendre differential equation of degree $\ell=1$ and order $n$. The solutions to the most general form of this equation are well known, and they are usually referred to as associated Legendre polynomials of first and second kind, i.e., $\mathsf{P}_{\ell}^n(\rho),\, \mathsf{Q}_{\ell}^n(\rho)$ respectively. In general, they can be obtained from the expressions \cite{bateman_2023}\footnote{More generally, one may define these solutions in terms of hypergeometric functions as follows \cite{bateman_2023}
\begin{equation}
\begin{aligned}
    \mathsf{P}_\mu^\nu(z)&= \frac{1}{\Gamma(1-\nu)} \left(\frac{1+z}{1-z}\right)^{\nu/2} {}_2 F_1\left( -\mu,\mu+1;1-\nu;\frac{1-z}{2}\right)\, ,\\
    \mathsf{Q}_\mu^\nu(z)&= \frac{\sqrt{\pi}\, \Gamma(1+\mu+\nu)}{\Gamma(\mu + 3/2)} \frac{1}{z^{1+\mu +\nu}} \left(1-z^2\right)^{\nu/2} {}_2 F_1\left( \frac{1+\mu +\nu}{2},\frac{\mu +\nu}{2}+1;\mu +\frac32;\frac{1}{z^2}\right)\, ,
\end{aligned}
\end{equation}
which can be extended to general complex values of the argument by analytic continuation.}
\begin{equation}\label{eq:Legendrepolynomials}
\begin{aligned}
    \mathsf{P}_\ell^n(\rho)&= (\rho^2-1)^{n/2} \frac{d^{n}}{d\rho^{n}}\mathsf{P}_\ell(\rho)= \frac{1}{2^\ell \ell!}(\rho^2-1)^{n/2} \frac{d^{\ell+n}}{d\rho^{\ell+n}}(\rho^2-1)^\ell\, ,\\
    \mathsf{Q}_{\ell}^n(\rho)&= \frac12 \mathsf{P}_\ell(\rho) \log \left( \frac{1+\rho}{1-\rho}\right)-\mathsf{R}_\ell(\rho)\, ,
\end{aligned}
\end{equation}
where $\mathsf{R}_\ell(\rho)$ denotes some rational function of $\rho$. The two are moreover related by the equation \cite{Bailey_1932,magnus2013}
\begin{equation}
    \mathsf{Q}_{\ell}^n(\rho)=\frac{\pi}{2 \sin(n \pi)} \left( \mathsf{P}_\ell^n(\rho)\cos(n \pi)-\frac{\Gamma(1+\ell+n)}{\Gamma(1+\ell-n)} \mathsf{P}_\ell^{-n}(\rho)\right)\, ,
\end{equation}
with
\begin{equation}
    \mathsf{P}_\ell^{-n}(\rho)= (-1)^n \frac{(\ell-n)!}{(\ell+n)!}\mathsf{P}_\ell^n(\rho)\, .
\end{equation}
However, as is familiar to us from e.g., the study of angular momentum in quantum mechanics, the only \emph{regular} solutions in the interval $\rho \in [1,\infty)$ correspond to the functions $\mathsf{P}_\ell^n(\rho)$ with $|n|\leq \ell$, whereas the rest all diverge---with an order that increases by steps of $1/2$ as $n$ is raised---at $\rho=1$. Hence, since this point corresponds to the centre of the hyperboloid and is completely regular (namely there is no singularity of any kind at this locus, as seen from e.g., its moduli space metric), we conclude that the only physically acceptable functions that can contribute to $\cF (x, \theta)$ in \eqref{eqap:Fourier6ddP2} are $\mathsf{P}_1^{0}(\rho) = \rho$ and $\mathsf{P}_1^{\pm 1}(\rho)=\pm \sqrt{\rho^2-1}$, leading to
\begin{equation}
    \cF (x, \theta) =c_0 \cosh x+c_1 \sinh x \cos \theta +c_2 \sinh x \sin \theta\, ,
\end{equation}
as the most general solution of \eqref{eqap:laplaceeqdP2} in the present case, thereby reproducing the linear behavior observed in string theory.

\subsection{General del Pezzo surface}\label{ss:FthydPr}

For the remaining $dP_r$ models, with $r=1, \ldots, 9$, one can use a similar parametrization as in the two-dimensional example above so as to solve the constraint
\begin{equation}
    (J^0)^2 - \sum_{i=1}^r (J^i)^2 \stackrel{!}{=}1\, ,
\end{equation}
by taking
\begin{equation}\label{eq:dPrparametrization}
   J^0=\cosh x\, ,\quad J^i=x^i \sinh x\, ,
\end{equation}
where $\{ x^i\}$ parametrize a $(r-1)$-dimensional sphere with unit radius, such that $\sum_i (x^i)^2=1$.\footnote{Imposing the K\"ahler cone constraints in $dP_{r<9}$ amounts to taking $x\geq 0,\, \theta_i \in [0, \pi/2],\, \phi \in [0, \pi/2]$, as well as restricting the surface of the hyperboloid such that certain linear relations (e.g., $J^0-J^i-J^j \geq 0$, $\forall i\neq j \in \{1, \ldots, r\}$, $2J^0-J^i-J^j-J^k-J^l-J^m \geq 0$, $\forall i\neq j \neq\ k \neq l \neq m$, etc.) are satisfied. The case of $dP_9$ requires special care, see e.g., \cite{Donagi:2004ia} and references therein.} The resulting metric reads
\begin{equation}
    \td s^2 = \td x^2 + \sinh^2 x\, \td\Omega_{r-1}^2\, ,
\end{equation}
whose associated Laplace-Beltrami operator can be readily computed to give
\begin{equation}\label{eq:laplacedPr}
 \Delta_{\rm 6d} = \partial^2_x \ + (r-1)\coth x \, \partial_x + \frac{1}{\sinh^2 x}\Delta_{\mathbf{S}^{r-1}}\, .
\end{equation}
Here, $\Delta_{\mathbf{S}^{r-1}}$ denotes the Laplacian on the unit $(r-1)$-sphere which, in hyperspherical coordinates $(\theta_1, \ldots, \theta_{r-2},\phi)$ with $\theta_i \in [0, \pi]$ and $\phi \in [0, 2\pi)$, can be obtained from the recursive relation
\begin{equation}\label{eq:laplacedSr-1}
 \Delta_{\mathbf{S}^{r-1}}= \frac{1}{\sin^{r-2} \theta_1} \partial_{\theta_1} \left( \sin^{r-2} \theta\, \partial_{\theta_1}\right)+ \frac{1}{\sin^2 \theta_1} \Delta_{\mathbf{S}^{r-2}}\, .
\end{equation}
Therefore, following the same strategy as in Section \ref{ss:FthydP2}, we can try to solve the relevant Laplace equation (cf. eq. \eqref{eq:6dlapl})
\begin{equation}\label{eqap:laplacedPr}
    \Delta_{\rm 6d} \cF (x,\theta_i, \phi) = r \cF(x,\theta_i, \phi) \, ,
\end{equation}
by expanding the Wilson coefficient in (hyper)spherical harmonics, namely
\begin{equation}
    \cF (x,\theta_i, \phi) = \sum_{\ell_{r-2}\leq \ell_{r-3}\leq \ldots \leq \ell_1}\sum_{|m|\leq \ell_{r-2}} f_{\ell_i,m} (x)\, \mathsf{Y}^m_{\ell_1,\ldots,\ell_{r-2}} (\theta_i, \phi)\, ,
\end{equation}
where $\mathsf{Y}^m_{\ell_1,\ldots,\ell_{r-2}} (\theta_i, \phi)$ is defined as \cite{Higuchi:1986wu}
\begin{equation}
 \mathsf{Y}^m_{\ell_1,\ldots,\ell_{r-2}}(\theta_i, \phi ) = \frac{1}{\sqrt{2\pi}} e^{i m \phi} \prod_{j=1}^{r-2} {}_{r-j}\bar{P}_{\ell_j}^{\ell_{j+1}}(\theta_j)\, ,\qquad \text{with}\quad \ell_{r-1} \equiv m\, ,
\end{equation}
which is itself comprised by products of Legendre functions
\begin{equation}
 {}_{j}\bar{P}_{n}^{\ell}(\theta) = \sqrt{\frac{2n + j - 1}{2}  \frac{(n + \ell + j - 2)!}{(n - \ell)!}} \, \sin^{\frac{2 - j}{2}}(\theta) \, \mathsf{P}_{n + \frac{j - 2}{2}}^{-\left(\ell + \frac{j - 2}{2}\right)}(\cos \theta)\, ,
\end{equation}
and moreover satisfy \cite{Higuchi:1986wu}
\begin{equation}
 \Delta_{\mathbf{S}^{r-1}}\mathsf{Y}^m_{\ell_1,\ldots,\ell_{r-2}}  = -\ell_1(\ell_1 +r-2)\, \mathsf{Y}^m_{\ell_1,\ldots,\ell_{r-2}}\, .
\end{equation}
This leads to the following differential equation for the expansion coefficients $f_{\ell_i, m} (x)$
\begin{equation}\label{eq:genLegendre}
    (1-\rho^2)\frac{d^2f_{\ell_i, m}}{d\rho^2} -r\rho \frac{df_{\ell_i, m}}{d\rho} +\left( r+\frac{\ell_1(\ell_1 +r-2)}{\rho^2-1}\right) f_{\ell_i, m}=0\, ,
\end{equation}
where we have already substituted $\cosh x =\rho$ in eq. \eqref{eq:laplacedPr} above. This equation can be regarded as a simple deformation/generalization of the associated Legendre one (cf. \eqref{eq:Legendre}), and their solutions can be obtained in a straightforward manner order by order. We refrain from displaying them here and just state that, similarly to what happens in the two-dimensional del Pezzo case, the only regular solutions at $\rho=1$ are those for which $\ell_1=0,1$. These precisely span the ambient space coordinates \eqref{eq:dPrparametrization}, thus leading to generic linear functions on $\{J^I\}$ solving \eqref{eqap:laplacedPr}, as advertised in Section \ref{sss:examples6d}.

\section{Further 5d $\mathcal{N}= 1$ Examples}\label{app:more5dmodels}

In this appendix, we consider a pair of non-smooth two moduli examples which were recently studied in \cite{Marchesano:2023thx} in the context of the moduli space curvature, a model with extra sections discussed in \cite{FierroCota:2023bsp}, a non-smooth three moduli case over a $\mathbb{F}_1$ base, and a smooth three moduli example in five-dimensional minimal supergravity. The analysis of Section \ref{ss:5dMtheory} suggests that the Laplace equation is violated in the bulk of moduli space. However, one could still wonder whether the eigenvalue that we find asymptotically is unique throughout all possible infinite distance limits. In what follows, we argue that this is indeed not the case, and the pattern of eigenvalues gets rather cumbersome. Our findings are summarized in Table \ref{tab:5d}. In all the examples below, we will always work in a Nef basis of divisors.

\subsection*{Two moduli example with a non-smooth fibration over $\mathbb{P}^2$\,--\, I}

We consider a non-smooth fibration over a $\bP^2$ base with two exceptional divisors. Using the triple intersection numbers given e.g., in \cite{Marchesano:2023thx} we have that the prepotential takes the form 
\begin{equation}
   \fF_5 =  M^1 (3 M^2+M^3)^2+\frac{3}{2} M^2 \left[7 (M^2)^2+5 M^2 M^3+(M^3)^2\right]\, .
\end{equation}
We follow the same exact procedure as in Section \ref{ss:5dMtheory}, knowing that the infinite distance loci are $\mathsf{L}_{\text{dec}} \equiv (M^1,M^2,M^3) = (0,0,\infty)$ and $\mathsf{L}_{\text{string}} \equiv (M^1,M^2,M^3) = (\infty,0,0)$, which correspond to a decompactification to 6d F-theory with $T= h^{1,1}(\mathbb{P}_2) - 1= 0$ and an emergent string limit, respectively. The relevant Wilson coefficient has the linear form 
\begin{equation}
    \cF = c_{2,1}M^1 +c_{2,2}M^2 + c_{2,3}M^3\, ,
\end{equation}
and we compute the (asymptotic) Laplace equation, eventually arriving at
\begin{equation}
    \Delta_{\rm 5d} \cF\, \stackrel{ \mathsf{L}_{\text{string}}}{\sim}\, \frac{5c_{2,1} + 6c_{2,2} -18c_{2,3}}{6c_{2,1}}\,\cF\, , \qquad \Delta_{\rm 5d} F\, \stackrel{\mathsf{L}_{\text{dec}}}{\sim}\, \frac{5}{6}\cF\, .
\end{equation}
Notice that there is a slight dependence on the $c_{2,i}$ coefficients. However, since $M^1$ diverges, it is consistent in the first string limit case to take $c_{2,2}=c_{2,3}=0$, recovering $\frac56$ for all limits. Let us stress, though, that from the perspective of the top-down computation these eigenvalues will not agree if the second Chern class does not turn out to be precisely tuned as above.

\subsection*{Two moduli example with a non-smooth fibration over $\mathbb{P}^2$\,--\,II}

As in the previous example, the triple intersection numbers can be found in e.g., \cite{Marchesano:2023thx}. Using those, one readily computes the prepotential 
\begin{equation}
\begin{split}
    \fF_5 = &\frac{1}{6} \bigg[ 50 (M^1)^3+30 (M^1)^2 M^2+240 (M^1)^2 M^3+6 M^1 (M^2)^2+ \\ & 96 M^1 M^2 M^3+384 M^1 (M^3)^2+9 (M^2)^2 M^3+75 M^2 (M^3)^2+203 (M^3)^3\bigg]\, .
\end{split}
\end{equation}
This time, there is only one infinite distance locus at $\mathsf{L}_{\text{dec}} \equiv (M^1,M^2,M^3) = (0,\infty,0)$, which corresponds to a decompactification to 6d F-theory with $T=0$. Similarly to the previous case, the (asymptotic) Laplace equation for a linear Wilson coefficient reads
\begin{equation}
     \Delta_{\rm 5d} \cF\, \stackrel{\mathsf{L}_{\text{dec}}}{\sim}\, \frac{5}{6}\cF\, .
\end{equation}

\subsection*{Three moduli example with a non-smooth fibration over $\mathbb{F}_1$}

We now consider a Calabi--Yau threefold with $h^{1,1}=4$, given by a non-smooth fibration over a Hirzebruch surface. The cubic prepotential for this model can be computed from the toric data in \cite[Section 4.2]{Hayashi:2023hqa}, and reads
\begin{equation}
\begin{split}
   \fF_5=\frac16 \bigg[&8 (M^1)^3 + 9 (M^1)^2 M^2 + 3 M^1 (M^2)^2 + 6 (M^1)^2 M^3 + 6 M^1 M^2 M^3 + 
 \\ &72 (M^1)^2 M^4 + 54 M^1 M^2 M^4 + 9 (M^2)^2 M^4 + 36 M^1 M^3 M^4 + 18 M^2 M^3 M^4 + 
 \\&204 M^1 (M^4)^2 + 129 M^3 (M^4)^2 + 188 (M^4)^3\bigg]\, ,
 \end{split}
\end{equation}
where we readily recognize the $M^1\to \infty$ and $M^4\to \infty$ asymptotic directions corresponding to finite distance singularities bounded by the fixed volume constraint. We can then take the following parametrization of the Wilson coefficient 
\begin{equation}
    \cF = c_{2,1}M^1 +c_{2,2}M^2 + c_{2,3}M^3 + c_{2,4}M^4\, ,
\end{equation}
and compute the Laplace equation in the $M^4\sim M^1\to0$ subspace. We get
\begin{equation}
    \Delta \cF= \frac43 \cF + \cO (M^4,M^1)\, .
\end{equation}
Notice that, in this case, where the 6d limit would be with $T =1$, we recover the result of the smooth two-moduli example $\mathbb{P}^{1,1,2,8,12}[24]$ (cf. eq. \eqref{eq:asymptoticeigenvalueP112812}). In all cases, these values seems to strongly depend not only on the number of moduli but also on the specific 6d uplift.

\subsection*{Two moduli example with an extra section over $\mathbb{P}_2$}

This model, together with its geometric realization, are discussed more in detail in e.g., \cite{FierroCota:2023bsp}. The corresponding prepotential is given by 
\begin{equation}
    \cF = \frac16\bigg[54 M^1 (M^2)^2 + 9 (M^2)^3 + 36 M^1 M^2 M^3 + 9 (M^2)^2 M^3 + 6 M^1 (M^3)^2 + 3 M^2 (M^3)^2 \bigg]\, .
\end{equation}
Notice that the 6d uplift would correspond to a model with $T=0$. We have again two infinite distance boundaries at $\{ (M^1,M^2,M^3)\}=\{(\infty,0,0), \,(0,0,\infty)\}$. We can compute the Laplacian with an ansatz
\begin{equation}
    \cF = c_{2,1}M^1 +c_{2,2}M^2 + c_{2,3}M^3\, ,
\end{equation}
getting us to, respectively for the two limits:
\begin{equation}
    \Delta_{\rm 5d} \cF\ \stackrel{M^1\to \infty}\sim\ \frac{11c_{2,1} + 6c_{2,2}-18c_{2,3}}{6c_{2,1}} \cF , \qquad  \Delta_{\rm 5d} \cF\ \stackrel{M^3\to \infty}\sim\ \frac56 \cF\, .
\end{equation}
If we take again consistently with the case above $c_{2,2}=c_{2,3}=0$ for the first limit, we get an eigenvalue of $\frac{11}{6}$. Once again, if we do not tune the Chern class such that $c_{2,2}-3c_{2,3} = -c_{2,1}$, then the eigenvalues along different limits will disagree.

\section{Uplifting the 4d Laplacian to Five Dimensions}\label{app:uplift}

In Section \ref{sss:uplift} from the main text, we explained how the correct functional dependence of the five-dimensional $\mathcal{R}^2$ Wilson coefficient can be derived by solving the four-dimensional Laplace equation and subsequently uplifting the result using the familiar dictionary under circle reductions. In this appendix, we show that the four-dimensional Laplacian 
\begin{equation}\label{eq:4dsaxop}
    \Delta_{\rm 4d}  = 2 g^{a \bar b}\partial_a \partial_{\bar b}\, ,
\end{equation}
when truncated to the saxionic fields $t^a = \text{Im}(z^a)$, can be uplifted to a well-defined five-dimensional operator under which the corresponding 5d coefficient becomes an eigenfunction. To begin with, it is useful to list some K\"ahler identities involving the 4d fields for later use:
\begin{equation}
     g^{a \bar b}\cK_a \cK_b = \frac43\cK^2 \,,\quad 
     \cK^{ab} \cK_b = t^a \, ,\quad 
     g^{a \bar b} \cK_b = \frac{4}{3}\cK t^a \, , \quad 
     t^a g_{a \bar b} = \frac34\frac{\cK_b}{\cK}\, , 
\end{equation}
where we introduced the notation $\cK =\cK_{abc}t^at^bt^c, \ \cK_a = \cK_{abc}t^b t^c, \ \cK_{ab} = \cK_{abc}t^c$.

Recalling now that the 4d and 5d moduli are related by (cf. eq. \eqref{eq:5dMthy4dIIAmap})
\begin{equation}
    M^I = \frac{t^I}{2\pi R_5(t)} = \frac{t^I}{\left( \frac{1}{6}\cK\right)^{1/3}}\, ,  
\end{equation}
we can simply rewrite the operator \eqref{eq:4dsaxop} in terms of 5d (ambient space) coordinates as follows
\begin{equation}\label{eq:5dellipticop}
    \cD^2_M = \frac{1}{2}g^{a \bar b}\frac{\partial M^I}{\partial t^a \partial t^b}\partial_I + \frac12 g^{a \bar b}\frac{\partial M^I}{\partial t^a}\frac{\partial M^J}{\partial t^b}\partial_{I}\partial_J\, ,
\end{equation}
where via explicit computation we find
\begin{equation}
\begin{split}
     &\frac{1}{2}g^{a \bar b}\frac{\partial M^I}{\partial t^a \partial t^b} = \frac{2}{3}(h^{1,1}-1)M^I \, ,\\
     & \frac12 g^{a \bar b}\frac{\partial M^I}{\partial t^a}\frac{\partial M^J}{\partial t^b} = \frac{1}{4}g^{IJ}-\frac{2}{3}M^IM^J\, .
\end{split}
\end{equation}
Note that, in order to reach the right hand side, one needs to take into account that the 5d and 4d (large volume) metrics satisfy $g_{\rm \small 5d} = 2g_{\rm \small 4d}$. On the other hand, we have that the four-dimensional Wilson coefficient satisfies $\Delta_{\rm 4d}\cF_4 = 0$, which implies---upon using \eqref{eq:4d5dwilson}---that
\begin{equation}
    0 = \Delta_{\rm 4d} \left(R_5 \cF_5 \right) = R_5 \left(\cD^2_M\cF_5 \right) + \frac{1}{2}g^{a \bar b}\partial_a\partial_bR_5(t) \cF_5+ g^{a \bar b}\partial_a R_5 \partial_b \cF_5\, ,
\end{equation}
where $(2\pi R_5)^3 = \frac16\cK$. Remarkably, the last term vanishes since $t^b\frac{\partial M^I}{\partial t^b}=0$, due to the $M^I$-coordinates being homogeneous of degree zero in $t^a$. Thus, the eigenvalue condition reads
\begin{equation}
    \cD_M^2 \cF_5 = -\frac{1}{R_5}g^{a \bar b}\partial_a \partial_b R_5 \cF_5= \frac{2}{3}(h^{1,1}-1) \cF_5\, .
\end{equation}
This equation admits, given the explicit form $\cD_M^2$, an obvious linear solution in the moduli, $\cF_5 = c_I M^I$, as already mentioned in the main text. It is interesting to note that the operator thus found can be expressed in a manifestly covariant way. Indeed, using \eqref{eq:identities5d} it follows that
\begin{equation}
    -\frac14g^{IJ} \Gamma_{IJ}^K = \frac{C}{24}C^{IJ}C_{IJK}C^{KL} + \frac{1}{8}h^{1,1} M^L\, ,
\end{equation}
which, in turn, allows us to rearrange the operator \eqref{eq:5dellipticop} as
\begin{equation}\label{eq:5dcovop}
\begin{split}
    \cD^2_M &= \left( \frac{1}{4}g^{IJ}-\frac{2}{3}M^IM^J \right) \nabla_I \partial_J - \frac{1}{24}\left[P^I - 13h^{1,1}M^I\right]\partial_I\, ,\\
    &=\left( \frac{1}{4}g^{IJ}-n^In^J \right) \nabla_I \partial_J - \frac{1}{24}\left[P^I - \frac{13}{2}\sqrt6\,h^{1,1}\,n^I\right]\partial_I\, ,
\end{split}
\end{equation}
where we used the normal vector field $n^I=\sqrt{\frac23}M^I$ introduced in \eqref{eq:meancurvvanishes}, as well as a new one $P^I \equiv C C^{JK}C_{JKL}C^{LI}$. Notice that since the coordinates can be rewritten in terms of $n^I$, and $C_{IJK}$ defines a tensor of $\cN=1$ 5d supergravity \cite{Bergshoeff:2004kh,Lauria:2020rhc}, $P^I$ constitutes some properly defined vector field whose physical interpretation remains unclear.

\bibliography{papers}
\bibliographystyle{JHEP2015}

\end{document}